\documentclass[iop,preprint,10pt]{emulateapj}
\usepackage{natbib}
\usepackage{graphicx}
\usepackage[dvipsnames]{color}
\usepackage{float}
\usepackage{longtable}
\usepackage{amsmath}
\usepackage{fancyref}
\definecolor{orange}{RGB}{255,69,0}
\definecolor{green}{RGB}{0,255,0}
\definecolor{darkred}{RGB}{139,0,0}
\usepackage{amssymb}
\usepackage{lineno}
\usepackage[colorlinks=true,
            linkcolor=orange,
            urlcolor=magenta,
            citecolor=blue]{hyperref}
\usepackage{nameref}
\usepackage{array,multirow,makecell}
\usepackage{tabularx}
\usepackage{morefloats}
\usepackage{gensymb}

\begin{document}
\title
{ Long term study of the light curve of PKS 1510-089 in GeV energies}

\author{Raj Prince$^{1}$, Pratik Majumdar$^{2}$, Nayantara Gupta$^{1}$ }
\affil{$^{1}$Raman Research Institute, Sadashivanagar, Bangalore 560080, India\\
 $^{2}$Saha Institute of Nuclear Physics, HBNI, Kolkata, West Bengal 700064, India }
\email{rajprince@rri.res.in}
\begin{abstract}
We have analyzed data from the Flat Spectrum Radio Quasar PKS 1510-089 collected over a period of 8 years 
from August 2008 to December 2016 with the Fermi-LAT. We have identified several flares of this highly variable
source, studied their temporal and spectral properties in detail and compared with previous works on flares of PKS 1510-089.
Five major flares and few sub-flares/sub-structures have been identified in our study. The fastest variability time is found to be
1.30$\pm$0.18 hr between MJD 55852.063 and 55852.188 where we estimate the minimum size of the emission region to be 
$4.85 \times 10^{15}$ cm.
In most of the flares the spectral energy distributions are better fitted with Logparabolic distribution compared to 
simple Power law or Power law with exponential cut-offs. This has strong physics implications regarding the nature of the 
high energy gamma-ray emission region. 
\end{abstract}
\section{\textbf{Introduction}}
Understanding the physics of  blazar flares is one of the most intriguing topics of research in high energy gamma ray 
astronomy. The origin of flux variability or flares could be in internal shocks in the blazar jets as discussed in earlier 
studies (\citealt{Spada et al.(2001)}).
It could also be from perturbation in accretion rate, variations in activity of the central engine (\citealt{Kelly et al.(2011)}) or 
fluctuations in the local magnetic field and particle densities.
The observed emissions of different frequencies could be from a single zone or 
multiple-zones. Depending on the spectral properties of the sources the underlying mechanism of variable emission may vary 
from one source to another.
\par
PKS 1510-089, located at a redshift of 0.361 (\citealt{Burbidge et al.(1966)}; \citealt{Thompson et al.(1990)}), is highly
variable and has been observed in gamma ray energies upto 400 GeV 
(\citealt{Abramowski et al.(2013)}, HESS Collaboration; \citealt{Aleksic et al.(2014)}, MAGIC Collaboration). 
This highly variable Flat Spectrum Radio Quasar (FSRQ) has been monitored by Fermi-LAT over a period of eight years (2008-2016).
The multi-wavelength data from flares of PKS 1510-089 during its high state between September 2008 and June 2009 
showed variabilities in timescales of 6 to 12 hours (\citealt{Abdo et al.2010a}, Fermi LAT Collaboration). Fermi LAT collaboration recorded 
isotropic luminosity in gamma rays of  approximately $2\times 10^{48}$ erg/s on March 26 in 2009.
This luminosity exceeds the estimated Eddington's luminosity L$_{Edd}$ = $6.86\times 10^{46}$ erg/s. This is 
calculated by using the black hole mass given in \citealt{Abdo et al.2010a}.
It is hard to find correlation in emissions in different frequencies (\citealt{Ahnen et al.(2016)}, MAGIC Collaboration).
\par
The multi-wavelength emission of PKS 1510-089 has been modeled previously in the framework of both leptonic and hadronic 
models.

In the leptonic model, the low energy component of the spectral energy distribution (SED) 
is produced by synchrotron radiation of relativistic electrons in the jet. The high energy component is produced
by inverse Compton (IC) process where the seed photons can be due to synchrotron radiation (commonly 
called the Synchrotron Self Compton process) or photons from the Broad Line region (BLR) or 
dusty torus (DT) (commonly known as the External Compton radiation (EC)). For more details on EC modeling, 
see \citet{Barnacka et al.(2014)}, \citet{Aleksic et al.(2014)}, \citet{Bottcher et al.(1998)}, \citet{Tavecchio et al.(2008)}.


In hadronic models, the required jet luminosity is high (\citealt{Bottcher et al.(2013)}) because proton cooling is much more inefficient 
than electron cooling. Proton synchrotron origin of X-ray and gamma-ray emission has been considered recently 
(\citealt{Pratim et al.(2016)}) 
to explain the week scale flares (during March-April 2009 and February-April 2012) of PKS 1510-089. In this study they have fitted the radio
to gamma ray data with a single zone model of synchrotron emission from electrons and protons  for jet luminosity comparable 
to \citet{Bottcher et al.(2013)}.
\par
The study of light curve and the identification of the location of flares (\citet{Tavecchio et al.(2010)}) are of much interest 
due to the wealth of flare data observed by Fermi-LAT.
It has been suggested before that there could be multiple simultaneously active gamma ray emission regions along the 
jet of PKS 1510-089 (\citealt{Brown(2013)}).
\citet{Dotson et al.(2012)} discussed about locating the distances of the emission regions of flares from the black hole with the cooling time 
scales of the energetic electrons.
 The temperature and density of the seed photons are different in the BLR and molecular torus (MT) regions which determine the 
 inverse Compton cooling regime (Klein - Nishina or Thomson) and time scale of the electrons. 
 The maximum decay time difference of the flares could impose an upper limit on the location of the flares.
 For luminosity of seed photons in the MT region $10^{45}$ erg/s and Lorentz factor of the jet $\Gamma=10$,
they found the distance of the flare to be within $2.3\times 10^{18}$ cm  for the Fermi LAT observed flares of energy between 
100 MeV to 1 GeV. In a more recent study \citet{Dotson et al.(2015)} have discussed that the emission
regions of 2009 GeV flares of PKS 1510-089 are distributed  over a large distance along the length of the jet ranging from the 
BLR to the MT and to the VLBI radio core zone 10pc away from the black hole. 
\par
Due to the extreme nature of variability of the source, the light curve of PKS 1510-089 has shown many interesting results 
and has been studied by various authors (\citealt{Abdo et al.2010a,Foschini et al.(2013), Zacharias et al.(2016),
Ahnen et al.(2016)}).  
However, most of the work that can be found in literature has been focused on the variability studies on short timescales
($\sim$ few hours to few tens of minutes). As of now, not much effort has been concentrated on study of the 
long term light curve of the source. In this paper, we aim to address the long term light curve of PKS 1510-089 in the gamma-ray
band using the data collected over a period of 8 years with the Fermi-LAT detector.  

\par 
From the Fermi-LAT data collected over a period of eight years we have selected the high states/flares of PKS 1510-089 to 
compare their spectral and temporal properties. Although some of these high states of PKS 1510-089 have been studied before 
by other authors, a comprehensive study  including all the high states observed by Fermi-LAT Collaboration till December, 2016 
and a comparison of their spectral and temporal characteristics is not available in literature.
Thus our work provides a detailed, complete and updated analysis of the flares of PKS 1510-089 detected by Fermi-LAT.
The rest of the paper is organized as follows : in section 2, we describe the data analysis procedures, conduct a
detailed
study of various flares and construct the spectral energy distributions of the various states of the source. 
In section 3 we discuss the results and draw conclusions from our analysis.
\par
Throughout the paper we noted the flux in units of $10^{-6}$ ph/cm$^2$/s unless otherwise mentioned.

\section{\textbf{Fermi-LAT Data Analysis of PKS 1510-089}}
The Fermi-LAT is a pair conversion $\gamma$-ray telescope sensitive to photon energies greater than 20 MeV with a 
field of view of about 2.4 sr (\citealt{Atwood et al.(2009)}). The primary observation mode of Fermi-LAT is survey mode in which the 
LAT (Large Area Telescope) scans the entire sky every 3 hours.
PKS 1510-089 has been continuously monitored by Fermi-LAT since Aug 2008.
We consider here the Fermi-LAT data for PKS 1510-089 from 05 Aug 2008--31 Dec 2016 (MJD 54683--57753).
The data analysis has been done with the help of \textit{gtlike/pyLikelihood} method, as implemented in the latest version (v10r0p5) of \textit{Fermi 
Science Tools} software package.
In this analysis we have considered photons of energy greater than 100 MeV.

\par
Gamma ($\gamma$) rays are also produced, in the upper atmosphere by the interaction of cosmic rays with ambient medium matter/radiation.
In order to reduce the contribution from these $\gamma$ rays (also called Earth limb $\gamma$ rays), our analysis is 
restricted to a maximum zenith angle of 105$\degree$.
The latest \textit{Fermi Science Tools} include the Instrument Response Function (IRF) ``P8R2\_SOURCE\_V6'' which has been used in the analysis.
The photons are extracted from a circular region of 10$\degree$ around the source, 
which is also called the region 
of interest (ROI). To include all the sources lying within the ROI we have used the third Fermi-LAT catalog 
(3FGL; \citealt{Acero et al.(2015)}). 
The spectral parameters were left free for sources lying within the 10$\degree$.
It must be noted that several other sources are also present in the 10$\degree$ to 20$\degree$ ROI. 
In the model file their spectral parameters have been kept fixed to the 3FGL catalog value.
To gauge the significance of $\gamma$-ray signal we have done the Maximum Likelihood (ML) test which is defined by 
TS=2$\Delta$ $\log(L)$, where L is the likelihood function between models with and without a point source at the 
position of source of interest (\citealt{Paliya(2015)}).
We have first performed the Maximum Likelihood analysis over the period of interest and for further analysis we have removed the 
sources of TS$<$9 (TS = 9, corresponds to $\sim$ 3 $\sigma$ detection; for details see \citealt{Mattox et al.(1996)}).
The standard background model was used to extract the spectral information. In our analysis, we have also used 
the latest isotropic background model, 
``iso\_P8R2\_SOURCE\_V6\_v06'', and the Galactic diffuse emission model, ``gll\_iem\_v06''. 
(available on \textit{Fermi Science Tools} website\footnote{https://fermi.gsfc.nasa.gov/ssc/data/access/lat/ Background Models.html}).
The variability of the source can be clearly seen by producing light curves with different time-bins (7 days, 1 day, 
12 hr, 6 hr and 3 hr). In Fig.1, we show the weekly light curve which clearly reveals that the source is highly 
variable. In addition we performed the spectral analysis in the energy range 0.1--300 GeV over several periods of the flaring states 
by using the \textit{unbinned likelihood analysis}. 
\par
The differential photon spectra have been fitted with three different functions whose forms are presented below.

\begin{itemize}
\item A power law (PL), defined as 
\begin{equation} \label{1}
 dN(E)/dE = N_p (E/E_p)^{-\Gamma},
 \end{equation}
\end{itemize}
with $E_p$ = 100 MeV (constant for all the SEDs) 
\begin{itemize}
 \item A log-parabola (LP), defined as
 \begin{equation} \label{2}
 dN(E)/dE = N_0 (E/E_0)^{-\alpha-\beta\ln(E/E_0)},
 \end{equation}
\end{itemize}
with $E_0$ = 300 MeV (constant for all the SEDs), 
where $\alpha$ is the photon index at $E_0$, $\beta$ is the curvature index and 
where ``ln" is the natural logarithm;
\begin{itemize}
 \item A power law with an exponential cut-off (PLEC), defined as
 \begin{equation} \label{3}
  dN(E)/dE = N_0 (E/E_p)^{-\Gamma} \exp(-E/E_c),
  \end{equation}
\end{itemize}
with $E_p$ = 200 MeV (constant for all the SEDs) 

\subsection{\textbf{Identifying the Flares of PKS 1510-089}}
PKS 1510-089 is one of the most variable (variability index = 11014.00) blazars in the 3FGL catalog. 
The variability can be seen in Fig.1, which shows the weekly light curve history of the source observed by the 
Fermi-LAT during Aug 2008--Dec 2016. Most of the time PKS 1510-089 is in the quiescent state accompanied by 
occasional periods of high activity where the flux greatly surpasses the quiescent state flux. These episodes of high 
activity are also referred to as flaring states. The duration of the flaring state is very short (ranging from a few days 
to a couple of weeks) after which the source returns to its pre-flare quiescent state.
\par
The light curve history of PKS 1510-089 shows that so far there have been five major flaring states (see Fig.1). 
We refer to these 
states in our work as Flare-1, Flare-2, Flare-3, Flare-4 and Flare-5 which happened during MJD 54825--55050, MJD 55732--56015,
MJD 56505--56626, MJD 57082--57265 and MJD 57657--57753 respectively.
We have zoomed out these major flares in bins of 1 day (not shown here) where sub-structures are not clearly seen, 6 hr 
(primarily for light curve study) and 3 hr (for variability time scale study). The 6 hour binning 
clearly reveals that there are sub-structures and various phases
(pre-flare, plateau, flare and post-flare) inside each individual flare shown in Fig.1. For further study, we concentrate on
the plots with 6 hour bins. Two sub-structures have been observed 
during Flare-1, we label them as flare-1(A) and flare-1(B). Flare-2 shows five sub-structures defined as flare-2(A), 
2(B), 2(C), 2(D) and 2(E). No sub-structure was seen during Flare-3 and Flare-5 while three sub-structures were 
noticed during Flare-4 and defined as flare-4(A), 4(B) and 4(C). All the different phases of activity have been marked 
with vertical broken red lines (see Fig.2 to Fig.13).
The time intervals which have TS $<$ 9 are rejected from the light curve 
analysis.
\subsection{\textbf{ Light Curves of Flares}}
As seen from Fig.1 we can clearly make out five major flaring episodes of PKS 1510-089. We have studied the temporal evolution of each flare separately.
In order to show the temporal evolution we have fitted the peaks by a sum of exponentials 
which give the decay and rising time for the different peaks shown in the light curve plots. The quiescent state 
(designated by light gray line in the Figs) is also presented in the light curve plots with the peaks of the flaring states.
The functional form of the sum of exponentials is
\begin{equation} \label{4}
 F(t) = 2F_0 [exp(\frac{t_0-t}{T_r}) + exp(\frac{t-t_0}{T_d})]^{-1}
\end{equation}
(\citealt{Abdo et al.2010b}), where $F_0$ is the flux at time $t_0$ representing the approximate flare amplitude, and $T_r$ and $T_d$ are
the rise and decay time of the flare.

\subsubsection{\textbf{ Flare-1}}
Fig.2 and Fig.3 show the light curves of flare-1(A) and flare-1(B) in time bins of 6  hr corresponding to the flaring
activity during MJD 54890--54927 and MJD 54935--54965 respectively.

In Fig.2 there is no Fermi-LAT data available in the time range MJD 54901.2--54905.6 and before 54899.0 the source was in a
quiescent state. We define this quiescent state as the pre-flare epoch of the source.
The flaring activity in flare-1(A) can be further divided into two parts---flare(I) and flare(II). The flare(I) phase was 
observed during MJD 54899.0 to 54910.3 where it shows two peaks P1 and P2 around MJD 54906.4 and 54909.1 with flux F$_{GeV}$ =
2.34$\pm$0.40 and 2.92$\pm$0.45 respectively. 
After this the source resides in a state where the flux exceeds the constant value of 0.64$\pm$0.07 for 
almost 5 days (MJD 54910.3-54915.0). 
This particular state which is neither the quiescent state nor a fully-fledged flaring state is referred to as the ``plateau''. 
After spending a few days in the so called plateau state (average flux = 1.38$\pm$0.06) the flux rises again (flare(II)) and 
shows one major peak P3 at 54916.9 with a flux of F$_{GeV}$ = 5.73$\pm$0.50.
A post-flare phase was also observed during MJD 54921 to 54927 with a flux almost close to that of the quiescent state. 
The decay and rising time of the peaks are tabulated in Table-1.
\par
A pre-flare was also observed during flare-1(B) (see Fig.3) whose flux of F$_{GeV}$ = 
0.61$\pm$0.04 (during time period MJD 54935 to 54944) is in close proximity to that of the quiescent state.

The flaring phase started from MJD $\sim$ 54944 and persisted for $\sim$ 7 days reaching a maximum flux of 
F$_{GeV}$ = 4.49$\pm$0.52 around MJD 54947.9 (P2).
The peak P1 was observed at MJD 54947.4 with a flux of F$_{GeV}$ = 3.85$\pm$0.55.
The peak P2 is followed by two peaks P3 and P4 at MJD 54948.6 and 54949.6 with flux of F$_{GeV}$ = 3.25$\pm$0.39 and 3.31$\pm$0.40
respectively.
Two post-flares were also observed in the vicinity of the quiescent state. However we did not consider them within the
flare region as the the amplitude of the first one is much lower than the other peaks in the flaring state and the second one
is far away from the main flaring phase. However it should be noted that the $\chi^{2}$ of the fit improves 
significantly if these additional small flares are included in the fit. A few outliers were also observed 
during this epoch for a very short time period (6 hr). The decay and rising time for the peaks are mentioned in Table-1.

\subsubsection{\textbf{ Flare-2}}
Similar to Flare-1, we carried out the 6 hr binning of Flare-2 (MJD 55732--56015) which shows the various 
sub-structures (flare-2(A), flare-2(B),
flare-2(C), flare-2(D), flare-2(E)) in various periods which are presented in Fig.4, Fig.5, Fig.6, Fig.7, and Fig.8 respectively. 
\par
Five activity phases have been observed in flare-2(A). A pre-flare was observed with a flux close to the quiescent 
state during the time period MJD 55732.0 to 55737.5. During MJD 55737.5--55741.0 (denoted by flare(I) as was done in the case 
of Flare-1), the flux starts rising from MJD 55737.9 and goes above 2.0 which is denoted as peak P1 (F$_{GeV}$ = 2.26$\pm$0.52) at MJD 
55738.9. After spending 3 days in the flare(I) phase the source comes back to its quiescent state.
However this duration of quiescence is quite short-lived and the flux starts rising slowly again. This rising part is 
considered as a plateau which has a time duration of MJD 55741.0 to 55743.5 with a average flux of F$_{GeV}$ = 0.79$\pm$0.10. 
The observation period of flare(II) MJD 55743.5 to 55751.0 shows three distinctive peaks P2, P3 and P4 at MJD 55743.9, 55744.9,
and 55746.4 with fluxes of F$_{GeV}$ = 2.37$\pm$0.55, 3.67$\pm$1.02 and 5.40$\pm$0.60 respectively. The modeling parameters have been
provided in Table-2.

\par
flare-2(B) shows the three phase pattern (pre-flare, flare and post-flare).  
A pre-flare phase has been observed with flux close to the quiescent state (and it also shows one outlier) and disconnected 
with the main flare during time period of MJD 55758 to 55765. A flaring activity happened from MJD 55765 to 55771 
during which the flux rises upto 
$\sim$4.0, denoted by peak P1 (F$_{GeV}$ = 3.81$\pm$0.46) at MJD 55767.4, and after spending around 5 days in the flaring 
state it returns to the quiescent state, where 
the flux is almost similar to that of the pre-flare epoch. The source resides in this quiescent state for a long time and we
consider this state as a post-flare from MJD 55771 to 55777. Details of the parameters in these phases are described in Table-2.

\par
The source exhibits a similar three phase pattern during flare-2(C) as well with a vaiation in the flux in the pre-flare 
region and the subsequent parameters for modeling this 
flaring episode is presented in Table-2. Incidentally, one of the brightest flare in the history of PKS 1510-089 
(\citealt{Foschini et al.(2013)}) was recorded during this period.
A major peak P1 (F$_{GeV}$ = 17.56$\pm$1.15) was observed at MJD 55853.9 accompanied by a pre-flare 
and post-flare observed during MJD 55846--55851 and MJD 55855--55860 respectively.

\par
In keeping with the earlier sub-structures flare-2(D) exhibits the typical phase of pre-flare, flare \& post-flare.  
The pre-flare and post-flare were observed during MJD 55860--55866 \& MJD 55878--55890. 
Three major peaks P1, P2, and P3 were observed during the flaring episode where the fluxes were F$_{GeV}$ = 6.38$\pm$0.63, 
7.62$\pm$0.73 and 8.88$\pm$0.77 at MJD 55867.9, 55868.4 and 55872.9 respectively. Peak P3 claims the distinction of becoming 
the 2$\rm ^{nd}$ highest peak in the history of PKS 1510-089. The modeling parameters have been described in Table-2.
\par
A four phase pattern (pre-flare, flare(I), flare(II) and post-flare) was observed during MJD 55965--56013 which we refer 
to as flare-2(E). During the pre-flare part slight fluctuations are noticed in the flux around the value 1.0. flare(I) 
comprises of four distinct major peaks P1, P2, P3 \& P4 at MJD 55980.4, 55982.9, 55988.7 and 55990.6 with the fluxes of 
F$_{GeV}$ = 4.20$\pm$0.51, 4.37$\pm$0.51, 3.36$\pm$0.44 \& 4.19$\pm$0.51 respectively. 
After spending 4-5 days in an almost quiescent state the flux starts rising again from MJD 55998 and shows a 
clear and major peak P5 at MJD 56002.4 with a flux of F$_{GeV}$ = 2.90$\pm$0.57. We refer to this peak as flare(II).
During MJD 56005 to 56013 a post-flare was observed whose flux instead of attaining a fixed value keeps fluctuating in the 
vicinity of the quiescent state flux. The modeling parameters have been described in Table-2.
\subsubsection{\textbf{ Flare-3}}
This is the first time that detailed study is being done on the flaring episode of PKS 1510-089 during 10 Sep-13 Oct 
2013 referred to as Flare-3. The characteristic temporal evolution of the flux of PKS 1510-089 during Flare-3 can be identified 
by a four phase pattern (pre-flare, flare(I), flare(II) and post-flare). Fig.9 shows a 6 hr bin light curve encompassing all the 
four phases and the modeling parameters have been provided in Table-3. The pre-flare phase observed during MJD 56545 to 56552 
exhibited a fluctuation in the flux around F$_{GeV}$ = 0.54$\pm$0.02. flare(I) was observed during MJD 56552--56561, where the flux rises
upto 3.5. The three peaks P1, P2 and P3 were observed at MJD 56554.1, 56556.4 and 56557.9 and the corresponding fluxes were 
F$_{GeV}$ = 3.47$\pm$0.47, 2.72$\pm$0.43 and 1.99$\pm$0.49 respectively. After this flaring state the source spent around 2 days in its 
quiescent state where the flux was around F$_{GeV}$ = 0.54$\pm$0.02. The flux again starts rising from MJD 56562.9 and reaches 
upto F$_{GeV}$ = 2.71$\pm$0.45, which is shown by peak P4 at MJD 56563.9. After spending around 6 days in 2nd flaring state the source
returns to its quiescent state. A post-flare period started from MJD 56570 and continued till MJD 56578 with the flux remaining steadily below 1.0.      
\subsubsection{\textbf{ Flare-4}}
A 6 hr binning of Flare-4 (MJD 57082--57265) was also carried out by us which revealed the underlying sub-structures with the three
distinctive features as flare-4(A), flare-4(B), flare-4(C). Fig.10, Fig.11 and Fig.12 shows these sub-structures along with their
different phases.
\par
flare-4(A) displays the usual three phase pattern (pre-flare, flare, post-flare). The details of the phase pattern are 
described in Table-4. The pre-flare and post-flare were observed before and after the flaring state during MJD 57106--57113 
and MJD 57118--57128. Even though there are substantial variations in the flux in both the pre and post flare regions, they 
are disconnected from the main flare under consideration and are hence not included in the analysis. 
The flaring duration lasted from MJD 57113 to 57118, 
during which the flux rose upto a value of 4.5. Two peaks P1 and P2 are clearly seen at MJD 57114.4 
and 57115.9 with fluxes of F$_{GeV}$ = 3.84$\pm$0.46 and 4.47$\pm$0.44 respectively.
\par
flare-4(B) shows a four phase pattern (pre-flare, flare(I), flare(II), post-flare) with flux variations in the pre-flare 
region. The detailed study is provided in  
Table-4. 
flare(I) was observed during MJD 57155 to 57163 where the flux reached a maximum of F$_{GeV}$ = 3.28$\pm$0.41 (P3). 
The peaks P1, P2 and P3 at 
MJD 57156.4, 57158.4 and 57159.9 notch up the peak fluxes of F$_{GeV}$ = 2.10$\pm$0.34, 2.02$\pm$0.33 and 3.28$\pm$0.41 respectively.
After spending around 7 days in the flaring state the source descends to its quiescent state where the flux is comparable to 
the pre-flare value. The source remains in this state for a duration of two and a half days. Surprisingly the flux again starts 
rising from MJD 57163 and reaches a maximum flux of F$_{GeV}$ = 3.56$\pm$0.47 (P5). This flare referred to as flare(II) was observed from 
MJD 57163 to 57171, during which three major peaks P4, P5 and P6 were noticed at MJD 57165.1, 57167.4 and 57170.4. The 
corresponding fluxes for these peaks were found to be F$_{GeV}$ = 2.32$\pm$0.37, 3.56$\pm$0.47 and 3.10$\pm$0.47 respectively. 
The post-flare epoch lasted from MJD 57171 to 57177 with a flux of around 1.0.
\par

flare-4(C) was recorded as the 3$\rm ^{rd}$ brightest flare in the history of PKS 1510-089. The flaring episode lasted
from MJD 57242 to 57250 during which the flux rose upto $\sim$8.60. Two major peaks P1 and P2 were observed at MJD 57244.6 
and 57245.4 with a flux of F$_{GeV}$ = 8.58$\pm$1.03 and 6.09$\pm$0.58 respectively. A pre-flare (MJD 57235 to 57242) and post-flare 
(MJD 57250 to 57259) were also observed with similar characteristics. For both the pre-flare and post-flare states the flux 
remains below 1.0. The detail about the parameters have been provided in Table-4.  

\subsubsection{\textbf{Flare-5}}
Another flare was observed in Aug-Sep 2016 during MJD 57628--57646. The maximum flux reached F$_{GeV}$ = 3.15$\pm$0.47 with 
TS=236.23. Three phase pattern (pre-flare, flare, post-flare) was observed during MJD 57628--57646. A clear peak P1 was 
observed in the flare phase at MJD 57634.625 (Fig.13). Peak P1 was fitted with the function given in equation (4) 
and rising and decay time have been provided in Table-5.
\par
All the above peaks during the flaring episodes were fitted with the constant state (value) the details of which 
are provided in Table-6.



\subsection{\textbf{Spectral Energy Distributions of Flares}}
This section is dedicated to studying the SEDs of flares and also to report the spectral features that will help to recognize
different phases of the flares.
We have produced the SEDs of PKS 1510-089 during different phases of the flares by using three different models PL, LP
and PLEC and their functional forms are given in eqs. 1, 2, and 3 respectively. We note that in all spectral models the
choice of reference energy does not affect the spectral shape (\citealt{Abdo et al.2010a}). It is fixed at 100 MeV for 
PL, at 300 MeV for LP and at 200 Mev for PLEC. 
\par
Spectral models (PL, LP and PLEC) have been plotted with the spectral data points
in cyan, black and red color respectively.
Fig.15 and Fig.16 show the spectral analysis of flare-1(A) and flare-1(B), their corresponding fitted parameters are 
given in the Table-7 and Table-8. 
The log(Likelihood) and $\Delta$log(Likelihood)
were calculated for each and every 
phase pattern where $\Delta$log(Likelihood) is defined as $\Delta$log($\mathcal{L}$) = (log $\mathcal{L}$(log-Parabola / PLEC)
- log $\mathcal{L}$(PL)), where $\mathcal{L}$ = Likelihood. 
A progressive spectral hardening have been noticed in flare-1(A) and flare-1(B) with increasing flux from one phase to another.
\par
Hardening in the spectrum is noticed during the flare-2(A), 2(B), 2(C), 2(D) and 2(E) as the flux increases
from pre-flare to flare phase.
The flare-2(B), 2(C) and 2(D) have shown significant spectral hardening as we move from pre-flare to flare with the value of spectral
index $\Gamma$  changing from 2.38$\pm$0.08 
to 2.17$\pm$0.04, 2.44$\pm$0.06 to 2.13$\pm$0.03 and 2.65$\pm$0.10 to 2.24$\pm$0.02 when fitted with PL distribution.

 
The SEDs for all the sub-flares of Flare-2 are plotted in Fig.17 to Fig.21 and the parameters describing all these 
sub-flares are provided in Table-9 to Table-13.
\par
Flare-3 shows progressive spectral hardening with increasing flux, $\Gamma$=2.47$\pm$0.01 changes to 2.35$\pm$0.00 and 
2.32$\pm$0.01 (PL fit) which are plotted in Fig.22. The values of the fitted parameters are displayed in Table-14.
\par
A significant amount of spectral hardening is also seen in sub-flares of Flare-4. For flare-4(A) and 4(B) the progressive
spectral hardening with increasing flux is seen as $\Gamma$ decreases from 2.32$\pm$0.03 to 2.14$\pm$0.02 and 2.40$\pm$0.05 to 2.19$\pm$0.02  in PL fit.
flare-4(C) also shows significant spectral hardening with increasing flux from pre-flare to flare as $\Gamma$ decreases from 2.42$\pm$0.09 
to 1.96$\pm$0.02  in PL fit.
Their SEDs are shown in Fig.23 to Fig.25 and the values of the fitted parameters are provided in Table-15 to Table-17
respectively.
\par
A progressive spectral hardening with increasing flux during pre-flare ($\Gamma$=2.58$\pm$0.08) to flare ($\Gamma$=2.39$\pm$0.04)
is also noted in Flare-5. The SED is shown in Fig.26 and the values of the fitted parameters are provided in Table-18.
\par
In Fig.14 we have plotted the photon spectral index as a function of integrated flux (F$_0$) for a few 
sub-flares. Our plots clearly show spectral hardening with increasing flux. 
The spectral hardening with increasing flux has been seen  previously in many other sources like 3C 454.3 
(\citealt{Britto et al.(2016)}) and Mrk 501(\citealt{Albert et al.(2007)}).

\section{\textbf{Results and Discussions}}
Being one of the most variable blazars in the Third Source Fermi Catalog (3FGL)
the light curve of PKS 1510-089  comprises of five major flares and each flare comprises of several sub-flares. 
Almost all the sub-flares shows various phases (pre-flare, flare, plateau, post-flare) and the flaring phases consist 
of peaks of different 
heights. Decay and rising times have been calculated for the major and clear peaks (P1, P2, P3..etc). Most of the peaks have rising 
and decay times of few hours (less than a day).

The brightest flare was observed during  Oct 2011 at MJD 55853.813. For 3 hr binning the flux was F$_{GeV}$ = 25.50$\pm$2.34 
with TS = 1340. 
A new flare was found in Aug 2015 (Fig.12) which has a peak P1 at MJD 57244.56 with a flux F$_{GeV}$ = 8.92$\pm$1.25
(TS = 397.18).
More recently a flare was also observed during 28 Aug--15 Sep, 2016 with a flux of F$_{GeV}$ = 3.15$\pm$0.47 at MJD 57634.61.
\par
Our results show in detail the presence of sub-flares within the flares, which we have 
 scanned separately by using the following function,
\begin{equation} \label{5}
F(t_2) = F(t_1).2^{(t_2-t_1)/ t_{d}},
\end{equation}
to calculate the minimum time of doubling/halving of flux between the time instants $t_1$ and $t_2$, 
$F(t_1)$ and $F(t_2)$ are respectively the fluxes measured at $t_1$ and $t_2$ and $t_d$ represents the doubling/halving
timescale.
The results are shown in Table-20 for the 3 hr bin of the light curve.
While scanning the light curve the following criteria was used : 
the flux should be double/half between two consecutive time instants and for these instants of time the condition
TS $>$ 25 ($\thicksim$ 5$\sigma$ detection) must be satisfied.
From Table-20, we find that the shortest observed variability time for the rising part is $t_{rise}$ = 1.43$\pm$0.22 hr between 
MJD 54945.438 and 54945.563 (flare-1(B)) and for  the decaying part  $t_{decay}$ = 1.30$\pm$0.18 hr between MJD 55852.063 and 55852.188 (flare-2(C)).
There were also some other time intervals  within which the flux changed by a factor of 2 but they did not satisfy the 
requirement of TS $>$ 25. 
Such time intervals are ignored in our analysis to find the fastest variability time scale.
The hour scale variability time has also been found earlier by \citet{Brown(2013)} and \citet{Saito et al.(2013)} for PKS 1510-089.
The variability time (The fastest halving/doubling time $t_d$ is the fastest variability time $t_{var}$) gives an idea about the size of the emission region, if 
we know the Doppler factor $\delta$ for the source. 
Variability time $t_{var}$, size of the emission region R and
Doppler factor $\delta$ are related by
\begin{equation} \label{6}
R \leq c t_{var} \delta (1+z)^{-1}
\end{equation}
where $z$ is the redshift of the source. The redshift corrected variability time ($\Delta t_{var}$ = $t_{var}(1+z)^{-1}$) is used to calculate 
the size of the emission region while modeling the SEDs of blazars. 
The apparent speed in the ultrarelativistic jet of PKS 1510-089 has been observed to be upto 46c (\citealt{Jorstad et al.(2005)}) which 
suggests that the Doppler factor could be very high for this source. From equation (6)
for $t_{var}$ = 1.30 hr, $\delta$ = 47
(\citealt{Kadota et al.(2012)}), we get an emission region of radius R $\thicksim$ $4.85 \times 10^{15}$ cm. 
A less extreme Doppler factor of 10 would imply an emission region of radius R $\thicksim$ $1.03 \times 10^{15}$ cm.
This is comparable to the estimates by \citet{Brown(2013)} and \citet{Saito et al.(2013)}, which are
$\thicksim$ $9.3 \times 10^{15}$ cm and $\thicksim$ $1.5 \times 10^{15} $cm, respectively. Such small emission 
regions are rather difficult to accomodate in the standard framework where the emission takes place from a large
distance from the central engine (see \citet{Tavecchio et al.(2010)} and references therein for a more detailed discussion).  

A multi-wavelength study of Flare-1 (Mar-Apr 2009) has been done by \citet{Abdo et al.2010a}. They found that the $\gamma$-ray flux 
had no correlation with the X-ray flux but it showed significant correlation with the optical flux. They also found that 
the optical flux was lagging 13 days behind the $\gamma$-ray flux. Moreover they estimated the isotropic luminosity
above 100 MeV during flare (II) of flare-1(A) to be more than $2 \times 10^{48}$ erg/s. 
The same flare has also been observed by HESS (\citealt{Abramowski et al.(2013)}) in very high energy gamma rays. According to their estimate  
 the integral flux in the very high energy (0.15 - 1.0 TeV) band is 
1.0$\pm$ $0.2_{stat}$ $\pm$ $0.2_{sys}) \times 10^{-11} cm^{-2} s^{-1}$ , which is $\approx$3$\%$ of the integral flux from
Crab nebula. It also shows the steepening in the photon spectrum with  spectral index 
5.4$\pm$ $0.7_{stat}$ $\pm$ $0.3_{sys}$ for PL distribution. \citet{Foschini et al.(2013)} have studied the outburst of Oct-Nov 2011. 
They estimated the shortest variability time ever detected in MeV-GeV energy regime as $\thicksim 20$ minutes at MJD 55852,
by using the GTI time binning. They have also mentioned about the hour scale variability (see Table-1 of \citet{Foschini et al.(2013)})
by using the 3 hr time binning which is consistent with the result of \citet{Brown(2013)} and \citet{Saito et al.(2013)}.
We note that our result shows that the shortest variability time is $\thicksim$1.30 hr 
(by using 3 hr binning) between MJD 55852.063 and 55852.188.

A multi-wavelength study of flare-2(E) has also been done previously by the MAGIC collaboration (\citealt{Aleksic et al.(2014)}). 
They used the data from Fermi-LAT observations during January 1 to April 7 in 2012 (MJD 55927--56024).
Within the time interval MJD 55974 to 55994 they estimated the shortest variability time scale as 
$t_{var}$ = 1.5$\pm$0.6 hr, which is very close to the value estimated by us $t_{var}$ = 1.84$\pm$0.28 hr (for almost the same time 
interval) given in Table-20 (flare-2E).
\par
Flare-3 has never been studied in the past.  The maximum flux of this flare was found to be around F$_{GeV}$ =
3.47$\pm$0.47 at MJD 56554.1 in our study. The fastest variability time for this flare was estimated as $t_{var}$ = 1.98$\pm$0.38 hr
(Table-20, flare-3) which is comparable to the fastest variability time found for other flares.
\par
We have also presented a detailed study of flare-4(A) (MJD 57100--57128) and flare-4(C) (MJD 57235--57259) for the first time
where flare-4(C) was identified as the 3$\rm ^{rd}$ brightest flare in the history of PKS 1510-089.
MAGIC collaboration (\citealt{Ahnen et al.(2016)}) has previously performed a multi-wavelength study of flare-4(B) observed in 
May 2015 (MJD 57143--57177).
\par
Flare-5 was found to be a very recent flare of PKS 1510-089. The shortest variability time was calculated as $t_{var}$ = 2.00$\pm$0.33 hr (Table-20).
\par

Figure~27 shows the histogram of the peak fluxes, rise and decay times of the peak fluxes as also enumerated in 
Table~1-5. The rise and decay times for the 
different peaks of the flares are distributed around a mean of 6.04$\pm$0.22 hr and 3.88$\pm$0.16 hr with a standard deviation of 2.40 hr
and 2.20 hr respectively, while the peak fluxes are distributed around a mean of 3.54$\pm$0.08 with standard deviation of 1.69.
Histogram of the constant fluxes are plotted in Fig.28. They are distributed with a mean of 0.51$\pm$0.01 and 
standard deviation of 0.20, which implies that the quiescent state of the source is pretty stable. A frequency 
distribution of all the flux data points are also plotted in the right hand panel of Figure~28. 
The plot shows a peaked distribution with slow rising part upto the peak and a fast decaying 
part beyond the peak. The peak value signifies the flux where the source spends most of the time. Above the 
peak, the flux values fall rapidly along with a few outliers which can be associated with large flux variations in the source. 
\citet{Tavecchio et al.(2010)} have studied flux variations and duty cycles with 1.5 years of data in two of the most variable 
sources, PKS~1510-089 and 3C 454.3. Our findings with a much larger data set also show very similar behaviour as compared to their study.

In Figure~29, we have plotted the histogram of redshift corrected variability time $\Delta t_{var}$ (see Table 20). 
One can clearly see  
that the distributions for rise and decay are not gaussian but the data points are distributed with mean of 
1.75$\pm$0.02 hr and 1.76$\pm$0.02 hr and with 
standard deviations of 0.35 hr and 0.40 hr respectively.

Fig.14 shows the variation of photon index as a function of integral fluxes for a few sub-flares. 
These plots reveal that when the source gets 
brighter its photon spectrum gets harder, a feature which has been also seen in many other blazars. 
A similar result was also reported earlier (see \citealt{Foschini et al.(2013)}).
\par 
We obtained the SEDs for different phases (pre-flare, plateau, flare and post-flare) and fitted them with different 
functional forms ( differential photon spectrum following PL, LP, PLEC distributions). To get the best fit we calculated 
the $\Delta$log(Likelihood) and reduced $\chi^2$ for each phase. 
\par
We compared the reduced $\chi^2$ values for PL, LP and PLEC fits and the spectral cut-off energies in Table-19 for
different flares. In almost all the cases, the best fit is found to be LP during flaring episodes.  
We also note that in the case of PLEC fit, the spectral cut-off energy varies from one flare to another. 
It is interesting to note that in a few cases where the reduced chi-square values for PLEC are comparable to the values obtained from 
LP fits, the cut-off energy is well constrained.
 This has strong physics implications regarding the location of the emission region.  
If the emission region is close to the core of the source, pair production optical depth would prevent the escape of very high 
energy gamma rays. As a result the highest energy gamma rays are expected from zones outside the BLR region, in the optically 
thin outer jet region (see \citealt{Aleksic et al.(2014)}, MAGIC Collaboration).
The variations in spectral fittings and spectral cut-off energies of the flares indicate that different flares might have originated 
from different zones along the length of the jet of PKS 1510-089. Earlier studies on blazar flares also indicated the 
possibility of multiple zones of emission during flares (Brown 2013, Dotson et al. (2012),(2015)). Detailed broadband 
spectral modeling with photon data ranging from radio to TeV
energy would be more useful in exploring the complex nature of flares of this highly variable source.
\section{\textbf{Concluding Remarks}}
We have studied the long term light curve of PKS 1510-089 with the data collected by Femi-LAT between Aug 2008 to Dec 2016.
The data have been binned in 7 days, 1 day and 6 hr to explore various features of the light curve. Five major flares 
along with many substructures have been 
detected in the weekly binning of the data which have been further studied in detail.
From a detailed study on variability, the shortest variability time has been found to be close to 1 hour. This 
puts a strong constraint on the size of
the emission region which has been estimated to be $\sim 10^{15}$ cm for reasonable values of the Doppler factor.
The spectral energy distributions have been fitted with three different functional forms PL, LP and PLEC. We find that 
in majority of the flares LP gives the best fit and in some cases PLEC can reasonably describe the data. 
Moreover, when PLEC gives the best fit the cut-off energies are found to vary from one 
flare to another. Our results indicate that the emission regions vary from one flare to another which is consistent 
with earlier results. 

\section{\textbf{Acknowledgment}}
We thank the referee for insightful comments which improved our work significantly.
It is also our pleasure to thank R. J. Britto and V. S. Paliya for many fruitful and helpful discussions on the topic.
This work has made use of public Fermi data obtained from the Fermi Science Support Center (FSSC), provided by NASA Goddard 
Space Flight Center.
\begin{table*}[htbp]
\caption{Results of temporal fitting with sum of exponentials (equation 4 in the text) for differents peaks of the flares (here Flare-1).
Column 2 represents the time (in MJD) at which the peaks are observed and the peak fluxes are given in column 3. The fitted rise ($T_r$)
and decay ($T_d$) times are mentioned in columns 4 \& 5}
\centering
\begin{tabular}{c c c c c}
\hline
&& flare-1(A)&& \\

Peak  & $t_0$  &$F_0$&  $T_r$ & $T_d$  \\[0.5ex]
     &[MJD] & [$10^{-6}$ ph cm$^{-2}$ s$^{-1}$] & [hr] & [hr]  \\
\hline
P1  & 54906.4 & 2.34$\pm$0.40 & 13.76$\pm$7.30 & 2.06$\pm$1.27  \\
P2  & 54909.1 & 2.92$\pm$0.45 & 10.97$\pm$2.32 & 7.26$\pm$2.25  \\
P3  & 54916.9 & 5.73$\pm$0.50 & 10.56$\pm$1.58 & 7.75$\pm$0.98  \\
\hline
&& flare-1(B) && \\

 \hline
P1  & 54947.4 & 3.85$\pm$0.55 & 6.43$\pm$2.66 & 4.04$\pm$2.52  \\
P2  & 54947.9 & 4.49$\pm$0.52 & 5.71$\pm$2.73 & 2.99$\pm$1.23  \\
P3  & 54948.6 & 3.25$\pm$0.39 & 1.93$\pm$1.98 & 4.83$\pm$2.08  \\
P4  & 54949.6 & 3.31$\pm$0.40 & 7.86$\pm$2.55 & 7.85$\pm$1.64  \\
\hline     
\end{tabular}
\label{Table:1}

\caption{All the columns represent the same parameters as mentioned in Table-1, here results are shown for Flare-2}
\centering
\begin{tabular}{c c c c c}
\hline
&& flare-2(A)&& \\
Peak  & $t_0$  &$F_0$&  $T_r$ & $T_d$  \\[0.5ex]
     &[MJD] & [$10^{-6}$ ph cm$^{-2}$ s$^{-1}$] & [hr] & [hr]  \\
\hline 
P1  & 55738.9 & 2.26$\pm$0.52 & 1.90$\pm$1.00 & 2.67$\pm$1.99  \\
P2  & 55743.9 & 2.37$\pm$0.55 & 4.70$\pm$1.42 & 3.72$\pm$1.44  \\
P3  & 55744.9 & 3.67$\pm$1.02 & 4.25$\pm$0.93 & 4.11$\pm$0.92  \\
P4  & 55746.4 & 5.40$\pm$0.60 & 7.86$\pm$0.96 & 3.98$\pm$0.61  \\
\hline
&& flare-2(B)&& \\
\hline 
P1  & 55767.4 & 3.81$\pm$0.46 & 7.38$\pm$0.73 & 5.10$\pm$0.72  \\
\hline
&& flare-2(C)&& \\
\hline 
P1  & 55853.9 & 17.56$\pm$1.15 & 2.92$\pm$0.89 & 2.50$\pm$0.27  \\
\hline
&& flare-2(D)&& \\
\hline
P1  & 55867.9 & 6.38$\pm$0.63 & 6.07$\pm$1.16 & 4.74$\pm$2.67  \\
P2  & 55868.4 & 7.62$\pm$0.73 & 7.08$\pm$2.50 & 3.81$\pm$1.43  \\
P3  & 55872.9 & 8.88$\pm$0.77 & 5.49$\pm$0.75 & 5.62$\pm$0.68  \\
\hline
&& flare-2(E)&& \\
\hline
P1  & 55980.4 & 4.20$\pm$0.51 & 8.41$\pm$1.36 & 8.78$\pm$1.42  \\
P2  & 55982.9 & 4.37$\pm$0.51 & 6.91$\pm$1.32 & 2.02$\pm$0.65  \\
P3  & 55988.7 & 3.36$\pm$0.44 & 7.06$\pm$2.86 & 9.39$\pm$1.96  \\
P3  & 55990.6 & 4.19$\pm$0.51 & 8.64$\pm$1.42 & 4.46$\pm$1.03 \\
P4  & 56002.4 & 2.90$\pm$0.57 & 15.07$\pm$2.72 & 9.50$\pm$2.29 \\
\hline
\end{tabular}
\label{Table:2}
\caption{All the columns represent the same parameters as mentioned in Table-1, here results are shown for Flare-3}
\centering
\begin{tabular}{c c c c c}
\hline
&& Flare-3&& \\
Peak  & $t_0$  &$F_0$&  $T_r$ & $T_d$  \\[0.5ex]
     &[MJD] & [$10^{-6}$ ph cm$^{-2}$ s$^{-1}$] & [hr] & [hr]  \\
\hline
P1  & 56554.1 & 3.47$\pm$0.47 &3.88$\pm$0.89 & 5.16$\pm$0.97  \\
P2  & 56556.4 & 2.72$\pm$0.43 &3.94$\pm$0.96 & 7.02$\pm$1.28  \\
P3  & 56557.9 & 1.99$\pm$0.49 &3.12$\pm$1.21 & 1.31$\pm$0.94  \\
P4  & 56563.9 & 2.71$\pm$0.45 &4.76$\pm$0.92 & 4.88$\pm$0.96  \\
\hline
\end{tabular}
\label{Table:3}

\caption{All the columns represent the same parameters as mentioned in Table-1, here results are shown for Flare-4}
\centering
\begin{tabular}{c c c c c}
\hline
&& flare-4(A)&& \\

Peak  & $t_0$  &$F_0$&  $T_r$ & $T_d$  \\[0.5ex]
     &[MJD] & [$10^{-6}$ ph cm$^{-2}$ s$^{-1}$] & [hr] & [hr]  \\
\hline
P1  & 57114.4 & 3.84$\pm$0.46 & 8.29$\pm$1.69 & 4.93$\pm$2.40  \\
P2  & 57115.9 & 4.47$\pm$0.44 & 8.27$\pm$2.98& 18.10$\pm$2.27  \\
\hline
& &flare-4(B)&& \\
\hline
P1  & 57156.4 & 2.10$\pm$0.34 & 6.83$\pm$2.12 & 9.50$\pm$3.99   \\
P2  & 57158.4 & 2.02$\pm$0.33 & 11.77$\pm$4.48 & 9.49$\pm$3.84   \\
P3  & 57159.9 & 3.28$\pm$0.41 & 8.00$\pm$2.27 & 5.35$\pm$2.92   \\
P4  & 57165.1 & 2.32$\pm$0.37 & 9.87$\pm$1.83 & 3.99$\pm$1.56   \\
P5  & 57167.4 & 3.56$\pm$0.47 & 6.35$\pm$2.09 & 11.82$\pm$1.61   \\
P6  & 57170.4 & 3.10$\pm$0.47 & 8.53$\pm$1.60 & 2.67$\pm$0.87   \\
\hline
& &flare-4(C)&& \\
\hline
P1  & 57244.6 & 8.58$\pm$1.03 & 7.59$\pm$0.85 & 2.66$\pm$0.98  \\
P2  & 57245.4 & 6.09$\pm$0.58 & 7.11$\pm$1.68 & 2.86$\pm$0.80  \\
\hline
\end{tabular}
\label{Table:4}

\caption{All the columns represent the same parameters as mentioned in Table-1, results are shown here for Flare-5}
\centering
\begin{tabular}{c c c c c}
\hline
&& Flare-5&& \\

Peak  & $t_0$  &$F_0$&  $T_r$ & $T_d$  \\[0.5ex]
     &[MJD] & [$10^{-6}$ ph cm$^{-2}$ s$^{-1}$] & [hr] & [hr]  \\
\hline
P1  & 57634.6 & 3.15$\pm$0.47 & 8.96$\pm$1.06 & 6.28$\pm$0.89  \\
\hline 
\end{tabular}
\label{Table:5}
\end{table*}

\begin{table*}[htbp]
\caption{The values of constant flux which are also fitted with the above peaks in the light curve. A histogram of the 
constant fluxes in different periods is shown in left panel of Figure~28}
\centering
\begin{tabular}{c c}
\hline
Flares/Sub-flares& Constant flux \\ 
& Flux F$_{0.1-300}$ $_{\rm{GeV}}$  \\
& [$10^{-6}$ ph cm$^{-2}$ s$^{-1}$]  \\
\hline
flare-1(A)  & 0.64$\pm$0.07 \\
flare-1(B)  & 0.61$\pm$0.04 \\
\hline
flare-2(A)  & 0.15$\pm$0.03 \\
flare-2(B)  & 0.35$\pm$0.03 \\
flare-2(C)  & 0.74$\pm$0.09 \\
flare-2(D)  & 0.53$\pm$0.05 \\
flare-2(E)  & 0.88$\pm$0.04 \\
\hline
Flare-3     & 0.54$\pm$0.02 \\
\hline 
flare-4(A)  & 0.69$\pm$0.05 \\
flare-4(B)  & 0.50$\pm$0.04 \\
flare-4(C)  & 0.41$\pm$0.04 \\
\hline
Flare-5     & 0.74$\pm$0.04 \\
\hline 
\end{tabular}
\label{Table:1}

\caption{Results of SEDs fitted with different spectral types like PL, LP, PLEC. Different periods of activity of the flares 
(here flare-1(A)) are mentioned in the 1st column. The fitted fluxes and the spectral indices are shown in the 
columns 2 \& 3. The goodness of unbinned fits
by log(Likelihood) is given in column 5 and the $\Delta$log(Likelihood) is calculated with respect to the log(Likelihood) of the 
PL fit (see text for more details).}
\centering
\begin{tabular}{c c c c c c}
\hline
&&PowerLaw (PL)&&& \\
Activity  &  F$_{0.1-300}$ $_{\rm{GeV}} $ & $\Gamma$ & &-log(Likelihood) &  \\[0.5ex]
 & ($10^{-6}$ ph cm$^{-2}$ s$^{-1}$)&&&&  \\
\hline
pre-flare    & 0.45$\pm$0.06 & 2.41$\pm$0.11 & - & 24496.5 & - \\
flare(I)    & 3.73$\pm$0.15 & 2.30$\pm$0.04 & - & 20608.5 & -  \\
plateau     & 3.26$\pm$0.14 & 2.29$\pm$0.04 & - & 20170.3 & - \\
flare(II)   & 4.57$\pm$0.12 & 2.24$\pm$0.02 & - & 38286.7 & - \\
post-flare   & 2.27$\pm$0.10 & 2.52$\pm$0.05 & - & 24715.6 & -  \\
\hline
 &&LogParabola (LP)&&&  \\
Activity  &  F$_{0.1-300}$ $_{\rm{GeV}} $ & $\alpha$ & $\beta$ & -log(Likelihood)& $\Delta$log(Likelihood) \\[0.5ex]
 & ($10^{-6}$ ph cm$^{-2}$ s$^{-1}$) &&&&  \\
\hline
pre-flare    & 0.44$\pm$0.06 & 2.40$\pm$0.15 & 0.00$\pm$0.00 & 24496.5 & 0.0  \\
flare(I)    & 3.72$\pm$0.15 & 2.28$\pm$0.05 & 0.00$\pm$0.00 & 20608.5 & 0.0  \\
plateau     & 3.18$\pm$0.14 & 2.18$\pm$0.06 & 0.08$\pm$0.03 & 20166.4 & -3.9 \\
flare(II)   & 4.45$\pm$0.12 & 2.14$\pm$0.04 & 0.07$\pm$0.02 & 38279.2 & -7.5  \\
post-flare   & 2.24$\pm$0.10 & 2.46$\pm$0.06 & 0.06$\pm$0.04 & 24714.2 & -1.4  \\
\hline
 &&PLExpCutoff (PLEC)&&&  \\
Activity  &  F$_{0.1-300}$ $_{\rm{GeV}} $ &$\Gamma_{PLEC}$ & E$_{cutoff}$ & -log(Likelihood) & $\Delta$log(Likelihood)  \\
 &($10^{-6}$ ph cm$^{-2}$ s$^{-1}$)&& [GeV] &&  \\
\hline
pre-flare    & 0.44$\pm$0.06 & 2.32$\pm$0.15 & 9.359$\pm$7.506 & 24496.3  & -0.2 \\
flare(I)    & 3.71$\pm$0.15 & 2.26$\pm$0.04 & 30.000$\pm$0.253 & 20610.0 & 1.5  \\
plateau     & 3.18$\pm$0.14 & 2.11$\pm$0.08 & 5.185 $\pm$2.394 & 20164.8 & -5.5 \\
flare(II)   & 4.50$\pm$0.12 & 2.16$\pm$0.04 & 15.980$\pm$6.358 & 38281.0 & -5.7  \\
post-flare   & 2.24$\pm$0.10 & 2.40$\pm$0.08 & 6.081$\pm$3.856 & 24713.4  & -2.2  \\
\hline
\end{tabular}
\label{Table:1}

\caption{All the columns represent the same parameters as mentioned in Table-7, here results are shown for flare-1(B)}
\centering
\begin{tabular}{c c c c c c}
\hline
 &&PowerLaw (PL)&&&  \\
Activity  &  F$_{0.1-300}$ $_{\rm{GeV}} $ & $\Gamma$ & &-log(Likelihood) &  \\[0.5ex]
 & ($10^{-6}$ ph cm$^{-2}$ s$^{-1}$)&&&&  \\
\hline
pre-flare   & 1.12 $\pm$ 0.07 & 2.50 $\pm$ 0.06 & - & 33110.1 & -\\
flare      & 5.20 $\pm$ 0.15 & 2.41 $\pm$ 0.03 & - & 36271.2 & -\\
post-flare  & 2.56 $\pm$ 0.09 & 2.33 $\pm$ 0.03 & - & 49194.4 & -\\
\hline
 &&LogParabola (LP)&&&  \\
Activity  &  F$_{0.1-300}$ $_{\rm{GeV}} $ & $\alpha$ & $\beta$ & -log(Likelihood)& $\Delta$log(Likelihood)\\[0.5ex]
 & ($10^{-6}$ ph cm$^{-2}$ s$^{-1}$) &&&&  \\
\hline
pre-flare   & 1.10 $\pm$ 0.07 & 2.42 $\pm$ 0.08 & 0.08 $\pm$0.05 & 33108.8 &  -1.3 \\
flare      & 5.05 $\pm$ 0.15 & 2.30 $\pm$ 0.04 & 0.11 $\pm$0.03 & 36260.6 & -10.6 \\
post-flare  & 2.48 $\pm$ 0.09 & 2.21 $\pm$ 0.05 & 0.09 $\pm$0.03 & 49187.9 & -6.5  \\
\hline
 &&PLExpCutoff (PLEC)&&&  \\
Activity  &  F$_{0.1-300}$ $_{\rm{GeV}} $ &$\Gamma_{PLEC}$ & E$_{cutoff}$ & -log(Likelihood) & $\Delta$log(Likelihood) \\
 &($10^{-6}$ ph cm$^{-2}$ s$^{-1}$) && [GeV] &&  \\
\hline
pre-flare   & 1.10 $\pm$ 0.07 & 2.37 $\pm$ 0.10 & 5.948 $\pm$ 4.510 & 33108.5 &  -1.6 \\
flare      & 5.10 $\pm$ 0.15 & 2.27 $\pm$ 0.05 & 5.740 $\pm$ 1.830 & 36262.9 &  -8.3 \\
post-flare  & 2.52 $\pm$ 0.09 & 2.24 $\pm$ 0.05 & 11.670$\pm$ 5.692  & 49191.0 & -3.4  \\
\hline
\end{tabular}
\label{table:1}
\end{table*}

\begin{table*}[htbp]
\caption{All the columns represent the same parameters as mentioned in Table-7, here results are shown for flare-2(A)}
\centering
\begin{tabular}{c c c c c c}
\hline
 &&PowerLaw (PL)&&&  \\
Activity  &  F$_{0.1-300}$ $_{\rm{GeV}} $ & $\Gamma$ & &-log(Likelihood) &  \\[0.5ex]
 & ($10^{-6}$ ph cm$^{-2}$ s$^{-1}$)&&&&  \\
\hline
pre-flare   & 0.40 $\pm$0.07 & 2.22$\pm$0.13 & - & 14835.3 & -  \\
flare(I)   & 2.80 $\pm$0.20 & 2.19$\pm$0.06 & - & 9462.8  & -  \\
plateau    & 2.12 $\pm$0.21 & 2.32$\pm$0.09 & - & 6705.2  & -  \\
flare(II)  & 2.89 $\pm$0.14 & 2.21$\pm$0.04 & - & 21374.7 & -  \\
post-flare  & 0.56 $\pm$0.08 & 2.23$\pm$0.10 & - & 15417.3 & -  \\
\hline
 &&LogParabola (LP)&&&  \\
Activity  &  F$_{0.1-300}$ $_{\rm{GeV}} $ & $\alpha$ & $\beta$ & log(Likelihood)& $\Delta$log(Likelihood) \\[0.5ex]
 & ($10^{-6}$ ph cm$^{-2}$ s$^{-1}$) &&&&  \\
\hline
pre-flare   & 0.39 $\pm$0.07 & 2.13$\pm$0.21 & 0.04$\pm$0.08 & 14835.1 & -0.2 \\
flare(I)   & 2.68 $\pm$0.21 & 2.06$\pm$0.10 & 0.07$\pm$0.05 & 9461.2  & -1.6 \\
plateau    & 1.79 $\pm$0.21 & 2.07$\pm$0.15 & 0.10$\pm$0.07 & 6697.8  & -7.4 \\
flare(II)  & 2.65 $\pm$0.14 & 1.95$\pm$0.07 & 0.17$\pm$0.04 & 21362.3 & -12.4 \\
post-flare  & 0.50 $\pm$0.08 & 1.97$\pm$0.21 & 0.14$\pm$0.10 & 15416.0 & -1.3 \\
\hline
 &&PLExpCutoff (PLEC)&&&  \\
Activity  &  F$_{0.1-300}$ $_{\rm{GeV}} $ &$\Gamma_{PLEC}$ & E$_{cutoff}$ & -log(Likelihood) & $\Delta$log(Likelihood)  \\
 &($10^{-6}$ ph cm$^{-2}$ s$^{-1}$) && [GeV] &&  \\
\hline
pre-flare   & 0.38 $\pm$0.07 & 2.08$\pm$0.22 & 9.546$\pm$12.560 & 14834.6 & -0.7 \\
flare(I)   & 2.71 $\pm$0.20 & 2.07$\pm$0.10 & 11.270$\pm$8.127 & 9461.0  & -1.8 \\
plateau    & 1.80 $\pm$0.21 & 2.03$\pm$0.16 & 5.316$\pm$4.204 & 6697.3 & -7.9 \\
flare(II)  & 2.69 $\pm$0.14 & 1.86$\pm$0.09 & 2.699$\pm$0.733 & 21359.9 & -14.8 \\
post-flare  & 0.51 $\pm$0.08 & 1.94$\pm$0.21 & 4.121$\pm$3.184 & 15415.7 & -1.6 \\
\hline
\end{tabular}
\label{Table:2(A)}

\caption{All the columns represent the same parameters as mentioned in Table-7, here results are shown for flare-2(B)}
\centering
\begin{tabular}{c c c c c c}
\hline
&&PowerLaw (PL)&&&  \\
Activity  &  F$_{0.1-300}$ $_{\rm{GeV}} $ & $\Gamma$ & &-log(Likelihood) &  \\[0.5ex]
 & ($10^{-6}$ ph cm$^{-2}$ s$^{-1}$)&&&&  \\
\hline
pre-flare   & 1.07 $\pm$0.10 & 2.38$\pm$0.08 & - & 17894.3  & -   \\
flare      & 2.15 $\pm$0.11 & 2.17$\pm$0.04 & - & 21202.8  & -   \\
post-flare  & 0.71 $\pm$0.07 & 2.57$\pm$0.10 & - & 20681.9  & -   \\
\hline
&&LogParabola (LP)&&&  \\
Activity  &  F$_{0.1-300}$ $_{\rm{GeV}} $ & $\alpha$ & $\beta$ & -log(Likelihood)& $\Delta$log(Likelihood) \\[0.5ex]
 & ($10^{-6}$ ph cm$^{-2}$ s$^{-1}$) &&&&  \\
\hline
pre-flare   & 1.02 $\pm$0.11 & 2.26$\pm$0.13 & 0.09$\pm$0.07 & 17893.1 & -1.2 \\
flare      & 2.03 $\pm$0.11 & 1.97$\pm$0.07 & 0.13$\pm$0.04 & 21196.0 & -6.8 \\
post-flare  & 0.70 $\pm$0.07 & 2.48$\pm$0.13 & 0.11$\pm$0.10 & 20681.2 & -0.7 \\
\hline
&&PLExpCutoff (PLEC)&&&  \\
Activity  &  F$_{0.1-300}$ $_{\rm{GeV}} $ &$\Gamma_{PLEC}$ & E$_{cutoff}$ & -log(Likelihood) & $\Delta$log(Likelihood)  \\
 &($10^{-6}$ ph cm$^{-2}$ s$^{-1}$)&& [GeV] &&  \\
\hline
pre-flare   & 1.03 $\pm$0.10 & 2.23$\pm$0.15 & 6.094$\pm$5.426 & 17893.1 & -1.2 \\
flare      & 2.06 $\pm$0.11 & 1.98$\pm$0.08 & 5.818$\pm$2.364 & 21196.6 & -6.2\\
post-flare  & 0.70 $\pm$0.07 & 2.46$\pm$0.17 & 6.339$\pm$8.688 & 20681.5 & -0.4 \\
\hline
\end{tabular}
\label{Table:2(B)}

\caption{All the columns represent the same parameters as mentioned in Table-7, here results are shown for flare-2(C)}
\centering
\begin{tabular}{c c c c c c}
\hline
&&PowerLaw (PL)&&&  \\
Activity  &  F$_{0.1-300}$ $_{\rm{GeV}} $ & $\Gamma$ & &-log(Likelihood) &  \\[0.5ex]
 & ($10^{-6}$ ph cm$^{-2}$ s$^{-1}$)&&&&  \\
\hline
pre-flare   & 2.55 $\pm$0.17 & 2.44$\pm$0.06 & - & 13635.9 & -  \\
flare      & 9.16  $\pm$0.30 & 2.13$\pm$0.03 & - & 17028.5 & -  \\
post-flare  & 2.25  $\pm$0.17 & 2.30$\pm$0.07 & - & 11397.9 & -  \\
\hline
&&LogParabola (LP)&&&  \\
Activity  &  F$_{0.1-300}$ $_{\rm{GeV}} $ & $\alpha$ & $\beta$ & -log(Likelihood)& $\Delta$log(Likelihood) \\[0.5ex]
 & ($10^{-6}$ ph cm$^{-2}$ s$^{-1}$) &&&&  \\
\hline
pre-flare   & 2.70 $\pm$0.17 & 2.45$\pm$0.08 & 0.051$\pm$0.050 & 13642.8 & 6.9 \\
flare     & 8.92  $\pm$0.30 & 2.03$\pm$0.04 & 0.06$\pm$0.02 & 17023.4 & -5.1 \\
post-flare & 2.25  $\pm$0.17 & 2.30$\pm$0.07 & 0.00$\pm$0.00 & 11397.9 &  0.0\\
\hline
&&PLExpCutoff (PLEC)&&&  \\
Activity  &  F$_{0.1-300}$ $_{\rm{GeV}} $ &$\Gamma_{PLEC}$ & E$_{cutoff}$ & -log(Likelihood) & $\Delta$log(Likelihood)  \\
 &($10^{-6}$ ph cm$^{-2}$ s$^{-1}$) && [GeV] &&  \\
\hline
pre-flare   & 2.50 $\pm$0.17 & 2.34$\pm$0.10 & 9.067$\pm$8.024 & 13634.8 & -1.1 \\
flare     & 9.00 $\pm$0.31 & 2.05$\pm$0.04 & 18.030$\pm$7.530& 17023.2 &  -5.3\\
post-flare & 2.22 $\pm$0.17 & 2.26$\pm$0.07 & 30.000$\pm$0.080& 11398.6 & 0.7 \\
\hline
\end{tabular}
\label{Table:2(C)}
\end{table*}

\begin{table*}[htbp]
\caption{All the columns represent the same parameters as mentioned in Table-7, here results are shown for flare-2(D)}
\centering
\begin{tabular}{c c c c c c}
\hline
&&PowerLaw (PL)&&&  \\
Activity  &  F$_{0.1-300}$ $_{\rm{GeV}} $ & $\Gamma$ & &-log(Likelihood) &  \\[0.5ex]
 & ($10^{-6}$ ph cm$^{-2}$ s$^{-1}$)&&&&  \\
\hline
pre-flare  & 1.45$\pm$0.12 & 2.65$\pm$0.10 & - & 14564.0 & -  \\
flare     & 5.55$\pm$0.13 & 2.24$\pm$0.02 & - & 59688.8 & -  \\
post-flare & 1.65$\pm$0.07 & 2.48$\pm$0.04 & - & 47148.3 & -  \\
\hline
&&LogParabola (LP)&&&  \\
Activity  &  F$_{0.1-300}$ $_{\rm{GeV}} $ & $\alpha$ & $\beta$ & -log(Likelihood)& $\Delta$log(Likelihood) \\[0.5ex]
 & ($10^{-6}$ ph cm$^{-2}$ s$^{-1}$) &&&&  \\
\hline
pre-flare  & 1.41$\pm$0.13 & 2.55$\pm$0.12 & 0.14$\pm$0.10 & 14562.7 & -1.3 \\
flare     & 5.35$\pm$0.13 & 2.11$\pm$0.03 & 0.09$\pm$0.02 & 59672.5 & -16.3 \\
post-flare & 1.63$\pm$0.07 & 2.42$\pm$0.05 & 0.05$\pm$0.03 & 47147.0 & -1.3  \\
\hline
&&PLExpCutoff (PLEC)&&&  \\
Activity  &  F$_{0.1-300}$ $_{\rm{GeV}} $ &$\Gamma_{PLEC}$ & E$_{cutoff}$ & -log(Likelihood) & $\Delta$log(Likelihood)  \\
 &($10^{-6}$ ph cm$^{-2}$ s$^{-1}$) && [GeV] &&  \\
\hline
pre-flare  & 1.42$\pm$0.13 & 2.44$\pm$0.18 & 3.140$\pm$2.743  &  14562.9  &  -1.9 \\
flare     & 5.44$\pm$0.13 & 2.14$\pm$0.03 & 12.310$\pm$3.515 &  59677.4  &  -11.4 \\
post-flare & 1.63$\pm$0.07 & 2.38$\pm$0.07 & 8.401$\pm$5.696  &  47146.2  & -2.1  \\
\hline
\end{tabular}
\label{Table:2(D)}


\caption{All the columns represent the same parameters as mentioned in Table-7, here results are shown for flare-2(E)}
\centering
\begin{tabular}{c c c c c c}
\hline
&&PowerLaw (PL)&&&  \\
Activity  &  F$_{0.1-300}$ $_{\rm{GeV}} $ & $\Gamma$ & &-log(Likelihood) &  \\[0.5ex]
 & ($10^{-6}$ ph cm$^{-2}$ s$^{-1}$)&&&&  \\
\hline
pre-flare    & 2.91$\pm$0.09 & 2.40$\pm$0.03 & - & 49012.7 & -  \\
flare(I)    & 4.94$\pm$0.09 & 2.29$\pm$0.02 & - & 70122.5 & -  \\
flare(II)   & 4.13$\pm$0.13 & 2.49$\pm$0.03 & - & 26676.5 & -  \\
post-flare   & 1.26$\pm$1.05 & 2.64$\pm$0.08 & - & 13838.9 & -  \\
\hline
&&LogParabola (LP)&&&  \\
Activity  &  F$_{0.1-300}$ $_{\rm{GeV}} $ & $\alpha$ & $\beta$ & -log(Likelihood)& $\Delta$log(Likelihood) \\[0.5ex]
 & ($10^{-6}$ ph cm$^{-2}$ s$^{-1}$) &&&&  \\
\hline
pre-flare    & 2.87$\pm$0.10 & 2.36$\pm$0.04 & 0.03$\pm$0.02 & 49011.6 & -1.1 \\
flare(I)    & 4.78$\pm$0.09 & 2.17$\pm$0.02 & 0.09$\pm$0.01 & 70096.3 & -26.2 \\
flare(II)   & 4.06$\pm$0.14 & 2.44$\pm$0.04 & 0.06$\pm$0.03 & 26674.2 & -2.3  \\
post-flare   & 1.24$\pm$0.11 & 2.58$\pm$0.11 & 0.07$\pm$0.06 & 13838.4 & -0.5 \\
\hline
&&PLExpCutoff (PLEC)&&&  \\
Activity  &  F$_{0.1-300}$ $_{\rm{GeV}} $ &$\Gamma_{PLEC}$ & E$_{cutoff}$ & -log(Likelihood) & $\Delta$log(Likelihood)  \\
 &($10^{-6}$ ph cm$^{-2}$ s$^{-1}$) && [GeV] &&  \\
\hline
pre-flare    & 2.89$\pm$0.09 & 2.37$\pm$0.03 & 29.970$\pm$3.615 & 49011.7 &  -1.0 \\
flare(I)    & 4.83$\pm$0.09 & 2.16$\pm$0.03 & 7.612$\pm$1.533  & 70099.0 & -23.5 \\
flare(II)   & 4.08$\pm$0.14 & 2.41$\pm$0.05 & 9.709$\pm$5.118  & 26673.7 &  -2.8 \\
post-flare   & 1.24$\pm$0.11 & 2.54$\pm$0.14 & 7.244$\pm$8.785  & 13838.4 &  -0.5 \\
\hline
\end{tabular}
\label{Table:2(E)}

\caption{All the columns represent the same parameters as mentioned in Table-7, here results are shown for Flare-3}
\centering
\begin{tabular}{c c c c c c}
\hline
&&PowerLaw (PL)&&&  \\
Activity  &  F$_{0.1-300}$ $_{\rm{GeV}} $ & $\Gamma$ & &-log(Likelihood) &  \\[0.5ex]
& ($10^{-6}$ ph cm$^{-2}$ s$^{-1}$)&&&&  \\
\hline
pre-flare   & 1.28$\pm$0.02 & 2.47$\pm$0.01 & - & 15158.0 & -  \\
flare(I)   & 2.88$\pm$0.06 & 2.32$\pm$0.01 & - & 24129.8 & -  \\
flare(II)  & 2.22$\pm$0.01 & 2.35$\pm$0.02 & - & 29884.4 & -  \\
post-flare  & 1.78$\pm$0.02 & 2.40$\pm$0.01 & - & 30937.3 & -  \\
\hline
&&LogParabola (LP)&&&  \\
Activity  &  F$_{0.1-300}$ $_{\rm{GeV}} $ & $\alpha$ & $\beta$ & -log(Likelihood)& $\Delta$log(Likelihood) \\[0.5ex]
& ($10^{-6}$ ph cm$^{-2}$ s$^{-1}$) &&&&  \\
\hline
pre-flare   & 1.29$\pm$0.11 & 2.45$\pm$0.09 & 0.02$\pm$0.00 & 15158.0 & 0.0 \\
flare(I)   & 2.74$\pm$0.08 & 2.19$\pm$0.05 & 0.10$\pm$0.04 & 24123.9 & -5.9 \\
flare(II)  & 2.17$\pm$0.06 & 2.27$\pm$0.03 & 0.06$\pm$0.03 & 29883.2 & -1.2 \\
post-flare  & 1.76$\pm$0.08 & 2.36$\pm$0.04 & 0.04$\pm$0.01 & 30936.7 &  -0.6\\
\hline
&&PLExpCutoff (PLEC)&&&  \\
Activity  &  F$_{0.1-300}$ $_{\rm{GeV}} $ &$\Gamma_{PLEC}$ & E$_{cutoff}$ & log(Likelihood) & $\Delta$log(Likelihood)  \\
 &($10^{-6}$ ph cm$^{-2}$ s$^{-1}$) && [GeV] &&  \\
\hline
pre-flare   & 1.27$\pm$0.10 & 2.41$\pm$0.08 & 16.580$\pm$2.897 & 15157.1 & -0.7 \\
flare(I)   & 2.78$\pm$0.10 & 2.15$\pm$0.05 & 5.022$\pm$0.180  & 24123.7 & -6.1 \\
flare(II)   & 2.17$\pm$0.05 & 2.24$\pm$0.03 & 9.043$\pm$0.223  & 29881.6 & -2.8 \\
post-flare  & 1.76$\pm$0.06 & 2.35$\pm$0.05 & 18.030$\pm$1.724 & 30936.5 &  -0.8 \\
\hline
\end{tabular}
\label{Table:3}

\end{table*}

\maxdeadcycles=1000

\begin{table*}[htbp]
\caption{All the columns represent the same parameters as mentioned in Table-7, here results are shown for flare-4(A)}
\centering
\begin{tabular}{c c c c c c}
\hline
&&PowerLaw (PL)&&&  \\
Activity  &  F$_{0.1-300}$ $_{\rm{GeV}} $ & $\Gamma$ & &-log(Likelihood) &  \\[0.5ex]
& ($10^{-6}$ ph cm$^{-2}$ s$^{-1}$)&&&&  \\
\hline
pre-flare   & 2.21$\pm$0.07 & 2.32$\pm$0.03 & - & 50195.9 & -  \\
flare      & 6.41$\pm$0.17 & 2.14$\pm$0.04 & - & 30084.0 & -  \\
post-flare  & 2.92$\pm$0.12 & 2.42$\pm$0.04 & - & 34033.0 & -  \\
\hline
&&LogParabola (LP)&&&  \\
Activity  &  F$_{0.1-300}$ $_{\rm{GeV}} $ & $\alpha$ & $\beta$ & -log(Likelihood)& $\Delta$log(Likelihood) \\[0.5ex]
& ($10^{-6}$ ph cm$^{-2}$ s$^{-1}$) &&&&  \\
\hline
pre-flare   & 2.14$\pm$0.07 & 2.20$\pm$0.04 & 0.09$\pm$0.03 & 50188.6 & -7.3 \\
flare      & 6.03$\pm$0.18 & 2.03$\pm$0.06 & 0.06$\pm$0.03 & 30067.9 & -16.1 \\
post-flare  & 2.83$\pm$0.12 & 2.33$\pm$0.05 & 0.08$\pm$0.03 & 34029.4 & -3.6 \\
\hline
&&PLExpCutoff (PLEC)&&&  \\
Activity  &  F$_{0.1-300}$ $_{\rm{GeV}} $ &$\Gamma_{PLEC}$ & E$_{cutoff}$ & -log(Likelihood) & $\Delta$log(Likelihood)  \\
 &($10^{-6}$ ph cm$^{-2}$ s$^{-1}$) && [GeV] &&  \\
\hline
pre-flare   & 2.17$\pm$0.07 & 2.21$\pm$0.05 & 9.657$\pm$3.964  & 50191.3 & -4.6 \\
flare      & 6.35$\pm$0.17 & 2.04$\pm$0.07 & 12.785$\pm$8.115 & 30080.7 & -3.3 \\
post-flare  & 2.86$\pm$0.12 & 2.32$\pm$0.06 & 9.118$\pm$5.157  & 34030.2 & -2.8 \\
\hline
\end{tabular}
\label{Table:4(A)}

\caption{All the column represents the same parameters as mentioned in Table-7, here results are shown for flare-4(B)}
\centering
\begin{tabular}{c c c c c c}
\hline
&&PowerLaw (PL)&&&  \\
Activity  &  F$_{0.1-300}$ $_{\rm{GeV}} $ & $\Gamma$ & &-log(Likelihood) &  \\[0.5ex]
& ($10^{-6}$ ph cm$^{-2}$ s$^{-1}$)&&&&  \\
\hline
pre-flare   & 1.64$\pm$0.08 & 2.40$\pm$0.05 & - & 38092.3 & -  \\
flare(I)   & 4.22$\pm$0.11 & 2.19$\pm$0.04 & - & 39757.9 & -  \\
flare(II)  & 3.75$\pm$0.11 & 2.20$\pm$0.04 & - & 36387.9 & -  \\
post-flare  & 1.59$\pm$0.12 & 2.39$\pm$0.07 & - & 16506.6 & -  \\
\hline
&&LogParabola (LP)&&&  \\
Activity  &  F$_{0.1-300}$ $_{\rm{GeV}} $ & $\alpha$ & $\beta$ & -log(Likelihood)& $\Delta$log(Likelihood) \\[0.5ex]
& ($10^{-6}$ ph cm$^{-2}$ s$^{-1}$) &&&&  \\
\hline
pre-flare   & 1.62$\pm$0.08 & 2.37$\pm$0.06 & 0.02$\pm$0.03 & 38092.0 & -0.3 \\
flare(I)   & 4.12$\pm$0.11 & 2.09$\pm$0.05 & 0.07$\pm$0.00 & 39749.7 & -8.2 \\
flare(II)  & 3.61$\pm$0.11 & 2.06$\pm$0.04 & 0.10$\pm$0.02 & 36375.6 & -12.3 \\
post-flare  & 1.49$\pm$0.12 & 2.21$\pm$0.11 & 0.15$\pm$0.07 & 16503.3 &  -3.3\\
\hline
&&PLExpCutoff (PLEC)&&&  \\
Activity  &  F$_{0.1-300}$ $_{\rm{GeV}} $ &$\Gamma_{PLEC}$ & E$_{cutoff}$ & -log(Likelihood) & $\Delta$log(Likelihood)  \\
 &($10^{-6}$ ph cm$^{-2}$ s$^{-1}$) && [GeV] &&  \\
\hline
pre-flare    & 1.62$\pm$0.08 & 2.36$\pm$0.05 & 30.000$\pm$0.050 & 38093.6 & 1.3 \\
flare(I)    & 4.14$\pm$0.11 & 2.06$\pm$0.05 & 9.073$\pm$0.308  & 39747.0 & -10.9 \\
flare(II)   & 3.67$\pm$0.11 & 2.08$\pm$0.04 & 9.743$\pm$3.159  & 36378.8 & -9.1 \\
post-flare   & 1.51$\pm$0.12 & 2.12$\pm$0.14 & 3.060$\pm$1.610  & 16503.0 & -3.6 \\
\hline
\end{tabular}
\label{Table:4(B)}

\caption{All the columns represent the same parameters as mentioned in Table-7, here results are shown for flare-4(C)}
\centering
\begin{tabular}{c c c c c c}
\hline
&&PowerLaw (PL)&&&  \\
Activity  &  F$_{0.1-300}$ $_{\rm{GeV}} $ & $\Gamma$ & &-log(Likelihood) &  \\[0.5ex]
& ($10^{-6}$ ph cm$^{-2}$ s$^{-1}$)&&&&  \\
\hline
pre-flare    & 1.15$\pm$0.11 & 2.42$\pm$0.09 & - & 15800.4 & -  \\
flare	    & 4.89$\pm$0.15 & 1.96$\pm$0.02 & - & 30002.9 & -  \\
post-flare   & 1.22$\pm$0.08 & 2.42$\pm$0.06 & - & 27063.2 & -  \\
\hline
&&LogParabola (LP)&&&  \\
Activity  &  F$_{0.1-300}$ $_{\rm{GeV}} $ & $\alpha$ & $\beta$ & -log(Likelihood)& $\Delta$log(Likelihood) \\[0.5ex]
& ($10^{-6}$ ph cm$^{-2}$ s$^{-1}$) &&&&  \\
\hline
pre-flare    & 1.35$\pm$0.11 & 2.55$\pm$0.09 & 0.00$\pm$0.00 & 15817.6 & 17.2 \\
flare       & 4.64$\pm$0.16 & 1.81$\pm$0.04 & 0.06$\pm$0.01 & 29993.2 & -9.7 \\
post-flare   & 1.16$\pm$0.09 & 2.29$\pm$0.10 & 0.11$\pm$0.06 & 27061.0 & -2.2 \\
\hline
&&PLExpCutoff (PLEC)&&&  \\
Activity  &  F$_{0.1-300}$ $_{\rm{GeV}} $ &$\Gamma_{PLEC}$ & E$_{cutoff}$ & -log(Likelihood) & $\Delta$log(Likelihood)  \\
 &($10^{-6}$ ph cm$^{-2}$ s$^{-1}$) && [GeV] &&  \\
\hline
pre-flare    & 1.12$\pm$0.11 & 2.31$\pm$0.13 & 8.931$\pm$9.809 & 15799.8 & -0.6 \\
flare	    & 4.75$\pm$0.16 & 1.88$\pm$0.03 & 29.710$\pm$8.166& 29994.5 & -8.4 \\
post-flare   & 1.19$\pm$0.09 & 2.29$\pm$0.11 & 7.354$\pm$5.857& 27061.7 & -1.5 \\
\hline
\end{tabular}
\label{Table:4(C)}
\end{table*}

\begin{table*}[htbp]
\caption{All the columns represent the same parameters as mentioned in Table-7, here results are shown for Flare-5}
\centering
\begin{tabular}{c c c c c c}
\hline
&&PowerLaw (PL)&&&  \\
Activity  &  F$_{0.1-300}$ $_{\rm{GeV}} $ & $\Gamma$ & &-log(Likelihood) &  \\[0.5ex]
& ($10^{-6}$ ph cm$^{-2}$ s$^{-1}$)&&&&  \\
\hline
pre-flare  &  1.64$\pm$0.12  & 2.58$\pm$0.08 & - & 14412.7  &  - \\
flare     &  3.01$\pm$0.11  & 2.39$\pm$0.04 & - & 28013.0  &  - \\
post-flare &  1.85$\pm$0.09  & 2.38$\pm$0.04 & - & 29145.4  &  - \\
\hline
&&LogParabola (LP)&&&  \\
Activity  &  F$_{0.1-300}$ $_{\rm{GeV}} $ & $\alpha$ & $\beta$ & -log(Likelihood)& $\Delta$log(Likelihood) \\[0.5ex]
& ($10^{-6}$ ph cm$^{-2}$ s$^{-1}$)&&&&  \\
\hline
pre-flare  & 1.63$\pm$0.12  & 2.55$\pm$0.10 & 0.04$\pm$0.06 & 14412.5 &  -0.2 \\
flare     & 2.93$\pm$0.11  & 2.28$\pm$0.05 & 0.10$\pm$0.03 & 28007.3 & -5.7  \\
post-flare & 1.82$\pm$0.09  & 2.31$\pm$0.06 & 0.06$\pm$0.04 & 29143.9 & -1.5  \\
\hline 
&&PLExpCutoff (PLEC)&&&  \\
Activity  &  F$_{0.1-300}$ $_{\rm{GeV}} $ &$\Gamma_{PLEC}$ & E$_{cutoff}$ & -log(Likelihood) & $\Delta$log(Likelihood)  \\
 &($10^{-6}$ ph cm$^{-2}$ s$^{-1}$) && [GeV] &&  \\
\hline
pre-flare  & 1.62$\pm$0.12 & 2.44$\pm$0.14 & 4.640$\pm$4.235 & 14411.7 & -1.0  \\
flare    & 2.95$\pm$0.11 & 2.23$\pm$0.07 & 5.016$\pm$2.001  & 28007.2 & -5.8  \\
post-flare & 1.83$\pm$0.09 & 2.33$\pm$0.06 & 19.040$\pm$17.200& 29144.5 & -0.9  \\
\hline 
\end{tabular}
\label{Table:5}
\end{table*}

\begin{table*}[htbp]
\caption{The reduced-$\chi^2$ for SEDs fitted by PowerLaw (PL), LogParabola (LP) and PowerLaw ExpCutoff (PLEC) for the flaring 
episodes are displayed below. In most cases LP and in a few cases PLEC provide the best fit to the data. Cutoff energies found with
PLEC vary from one flare to another, which could be due to different emission regions of these flares}
\centering
\begin{tabular} {c |c c c| c}
\hline

Activity  & \multicolumn{3}{|c|}{Reduced-$\chi^2$} & $E_{cutoff}$ for PLEC (GeV)  \\
\hline
 flare-1(A) & PL & LP & PLEC & \\
\hline 
flare(I) & 2.28 & 2.31 & 1.98 &  30.00$\pm$0.25   \\
flare(II)& 2.90 & 0.12 & 1.09 &  15.98$\pm$6.36    \\
\hline
flare-1(B) &&&& \\
\hline 
   flare & 5.06 & 0.58 & 1.03 & 5.74$\pm$1.83  \\  
\hline    
flare-2(A) &&&& \\
\hline 
flare(I) & 3.66 & 1.91 & 2.40 &  11.27$\pm$8.13  \\
flare(II)& 2.84 & 0.92 & 0.48 &  2.70$\pm$0.73   \\
\hline
flare-2(B) &&&& \\
\hline 
flare & 2.15 & 0.23 & 0.43 & 5.82$\pm$2.36  \\
\hline
flare-2(C) &&&&   \\
\hline 
flare & 1.73 & 0.41 & 0.83 & 18.03$\pm$7.53  \\
\hline
 flare-2(D) &&& &  \\
\hline 
flare & 8.14 & 0.43 & 2.83 & 12.31$\pm$3.51  \\
\hline 
flare-2(E) &&&&  \\
\hline 
flare(I) & 10.23 & 1.63 & 2.41 & 7.61$\pm$1.53  \\
flare(II) & 0.43 & 0.15 & 0.06 & 9.71$\pm$5.12  \\
\hline 
Flare-3   &&&& \\
\hline 
flare(I) & 2.73 & 0.91 & 1.19 & 5.02$\pm$0.18  \\
flare(II)& 0.41 & 0.42 & 0.34 & 9.04$\pm$0.22 \\
\hline 
flare-4(A)   &&&& \\
\hline 
flare   & 11.93 & 3.25 & 5.82 & 12.78$\pm$8.11 \\
\hline 
flare-4(B)   &&&& \\
\hline 
flare(I) & 2.41 & 2.95 & 1.78 & 9.07$\pm$0.31\\
flare(II)& 8.60 & 0.50 & 3.30 & 9.74$\pm$3.16 \\
\hline 
flare-4(C)  &&&&  \\
\hline
flare & 4.41 & 1.00 & 1.84 & 29.71$\pm$8.16  \\
\hline
Flare-5   &&&&  \\
\hline 
flare & 1.55 & 0.43 & 0.50 & 5.01$\pm$2.00  \\

\hline
\end{tabular}
\label{Table:6}
\end{table*}

\begin{table*}[htbp]
\caption{Details of fastest variability time scale of PKS 1510-089 for the whole 8 years data has been presented here. Data which 
has a significance of at least 5$\sigma$ has been considered (see text for details). 
Here $t_{var}$ represents the observed characteristic time scale and 
$\Delta t_{var}$ = $t_{var}$ $(1+z)^{-1}$. R (rise) and D(decay) 
represent the behavior of the flux in a particular time interval}
\centering
\begin{tabular}{c c c c c c c }
\hline
\hline
$T_{start}(t_1)$ & $T_{stop}(t_2)$ & Flux start ($F_1$) & Flux stop ($F_2$) & $t_{var}$(hr) & $\Delta t_{var}$(hr) & Rise/Decay  \\
 (MJD) & (MJD) & ($10^{-6}$ ph cm$^{-2}$ s$^{-1}$) & ($10^{-6}$ ph cm$^{-2}$ s$^{-1}$)&&& \\
\hline
&& flare-1(A) &&&& \\
\hline 
54916.563 & 54916.688 & 1.95$\pm$0.45 & 4.25$\pm$0.59 &2.67$\pm$0.32 &1.96$\pm$0.17 & R   \\
54917.188 & 54917.313 & 4.55$\pm$0.68 & 1.75$\pm$0.43 &-2.18$\pm$0.22 &-1.60$\pm$0.12& D   \\
54917.938 & 54918.063 & 0.69$\pm$0.25 & 1.40$\pm$0.43 &2.96$\pm$0.23 &2.17$\pm$0.12 & R   \\
\hline
&& flare-1(B) &&&& \\
\hline
54945.438 & 54945.563 & 0.72$\pm$0.29 & 3.09$\pm$0.53 &1.43$\pm$0.22 &1.05$\pm$0.12 & R   \\
54948.938 & 54949.063 & 2.25$\pm$0.68 & 5.49$\pm$1.91 &2.34$\pm$0.12 &1.72$\pm$0.07 & R   \\
54949.688 & 54949.813 & 3.10$\pm$0.55 & 1.37$\pm$0.36 &-2.56$\pm$0.26 &-1.88$\pm$0.14 & D   \\
\hline 
&& flare-2(A)  &&&& \\
\hline 
55739.313 & 55739.438 & 0.89$\pm$0.36 & 2.31$\pm$0.52 &2.17$\pm$0.42 &1.59$\pm$0.22 & R   \\
55745.563 & 55745.688 & 3.03$\pm$0.63 & 0.95$\pm$0.32 &-1.79$\pm$0.20 &-1.32$\pm$0.11 & D   \\
55745.688 & 55745.813 & 0.95$\pm$0.32 & 1.87$\pm$0.70 &3.07$\pm$0.15 &2.26$\pm$0.08 & R   \\
55746.063 & 55746.188 & 6.10$\pm$1.50 & 2.95$\pm$0.66 &-2.88$\pm$0.09&-2.12$\pm$0.05 & D   \\
55746.438 & 55746.563 & 7.01$\pm$0.95 & 3.48$\pm$0.66 & -2.98$\pm$0.23&-2.19$\pm$0.12 & D   \\
55746.563 & 55746.688 & 3.48$\pm$0.66 & 1.19$\pm$0.43 & -1.94$\pm$0.31&-1.42$\pm$0.17 & D   \\ 
\hline
&& flare-2(B) &&&& \\
\hline 
55767.063 & 55767.188 & 1.11$\pm$0.45 & 2.98$\pm$0.62 &2.11$\pm$0.42 &1.55$\pm$0.23 & R  \\
55767.813 & 55767.938 & 4.35$\pm$1.15 & 1.94$\pm$0.60 &-2.59$\pm$0.14 &-1.90$\pm$0.08 & D  \\
\hline 
&& flare-2(C) &&&& \\
\hline
55852.063 & 55852.188 & 5.80$\pm$0.84 & 1.17$\pm$0.43 &-1.30$\pm$0.18 &-0.95$\pm$0.10 & D  \\
55852.313 & 55852.438 & 0.91$\pm$0.37 & 2.57$\pm$0.87 &2.00$\pm$0.13 &1.47$\pm$0.07 & R  \\
55852.438 & 55852.563 & 2.57$\pm$0.87 & 5.84$\pm$1.75 &2.53$\pm$0.13 &1.86$\pm$0.07 & R  \\
55853.063 & 55853.188 & 3.11$\pm$0.65 & 6.28$\pm$0.87 &2.97$\pm$0.30 &2.18$\pm$0.16 & R   \\
55853.188 & 55853.313 & 6.28$\pm$0.87 & 3.00$\pm$0.60 &-2.81$\pm$0.24&-2.07$\pm$0.13& D   \\
55853.563 & 55853.688 & 3.46$\pm$1.43 & 7.20$\pm$2.54 &2.84$\pm$0.24 &2.09$\pm$0.13 & R   \\
55853.688 & 55853.813 & 7.20$\pm$2.54 & 25.50$\pm$2.34&1.64$\pm$0.34 &1.21$\pm$0.18 & R   \\
55853.938 & 55854.063 & 13.35$\pm$1.27& 4.94$\pm$0.76 &-2.09$\pm$0.12&-1.54$\pm$0.07 & D   \\
\hline 
&& flare-2(D) &&&& \\
\hline
55867.313 & 55867.438 & 3.49$\pm$0.70 & 1.38$\pm$0.59 &-2.24$\pm$0.54 &-1.64$\pm$0.29 & D   \\
55868.438 & 55868.563 & 6.92$\pm$1.09 & 2.74$\pm$1.15 &-2.25$\pm$0.64 &-1.65$\pm$0.35 & D   \\
55868.688 & 55868.813 & 1.62$\pm$0.72 & 3.55$\pm$0.74 &2.66$\pm$0.81 &1.95$\pm$0.43 & R   \\
55869.063 & 55869.188 & 4.78$\pm$0.81 & 2.19$\pm$0.52 &-2.67$\pm$0.24&-1.96$\pm$0.13 & D   \\
55869.188 & 55869.313 & 2.19$\pm$0.52 & 4.50$\pm$0.77 &2.89$\pm$0.27&2.12$\pm$0.15 & R   \\
55870.313 & 55870.438 & 2.05$\pm$0.58 & 4.10$\pm$0.90 &3.00$\pm$0.28&2.20$\pm$0.15 & R   \\
55872.563 & 55872.688 & 2.66$\pm$0.87 & 6.11$\pm$0.86 &2.50$\pm$0.56 &1.84$\pm$0.30 & R   \\
\hline
&& flare-2(E) &&&& \\
\hline
55989.188 & 55989.313 & 3.54$\pm$0.59 & 1.15$\pm$0.39 &-1.84$\pm$0.28&-1.35$\pm$0.15 & D   \\
55989.313 & 55989.438 & 1.15$\pm$0.39 & 2.77$\pm$0.56 &2.36$\pm$0.36 &1.74$\pm$0.20 & R   \\
55990.063 & 55990.188 & 1.35$\pm$0.38 & 2.67$\pm$0.53 &3.06$\pm$0.36 &2.25$\pm$0.20 & R   \\
55990.438 & 55990.563 & 1.94$\pm$0.54 & 4.33$\pm$0.78 &2.59$\pm$0.32 &1.90$\pm$0.17 & R   \\
55991.313 & 55991.438 & 1.01$\pm$0.43 & 2.07$\pm$0.50 &2.89$\pm$0.73 &2.12$\pm$0.40 & R   \\
55991.813 & 55991.938 & 1.82$\pm$0.43 & 0.84$\pm$0.32 &-2.68$\pm$0.50&-1.97$\pm$0.27 & D   \\
55998.938 & 55999.063 & 0.78$\pm$0.30 & 1.80$\pm$0.54 &2.49$\pm$0.26 &1.83$\pm$0.14 & R   \\
56000.188 & 56000.313 & 1.25$\pm$0.41 & 0.63$\pm$0.27 &-3.04$\pm$0.44&-2.23$\pm$0.24 & D    \\
56000.688 & 56000.813 & 1.35$\pm$0.37 & 2.75$\pm$0.52 &2.94$\pm$0.35 &2.16$\pm$0.19 & R   \\
56001.063 & 56001.188 & 2.12$\pm$0.45 & 1.07$\pm$0.39 &-3.03$\pm$0.61 &-2.23$\pm$0.33 & D   \\
\hline 
&& Flare-3 &&&& \\
\hline
56556.188 & 56556.313 & 1.61$\pm$0.49 & 3.88$\pm$0.66 &2.37$\pm$0.36 &1.74$\pm$0.19 & R  \\
56563.313 & 56563.438 & 2.05$\pm$0.52 & 0.93$\pm$0.39 &-1.98$\pm$0.38&-1.45$\pm$0.21 & D  \\
56568.063 & 56568.188 & 2.10$\pm$0.49 & 0.99$\pm$0.35 &-2.76$\pm$0.53&-2.03$\pm$0.29 & D  \\
\hline 
&& flare-4(A) &&&&\\
\hline 
57113.188 & 57113.313 & 0.30$\pm$0.13 & 0.93$\pm$0.33 &1.84$\pm$0.15 &1.35$\pm$0.08 & R  \\
57116.938 & 57117.063 & 3.65$\pm$0.49 & 1.73$\pm$0.42 &-2.78$\pm$0.40&-2.04$\pm$0.22 & D  \\
\hline
&& flare-4(B) &&&& \\
\hline 
57164.063 & 57164.188 & 1.40$\pm$0.38 & 0.63$\pm$0.29 &-2.61$\pm$0.62 &-1.92$\pm$0.34 & D   \\
57165.688 & 57165.813 & 0.97$\pm$0.34 & 2.06$\pm$0.47 &2.78$\pm$0.44 &2.04$\pm$-0.24 & R   \\
57166.188 & 57166.313 & 1.30$\pm$0.46 & 2.85$\pm$0.61 &2.66$\pm$0.48 &1.95$\pm$0.26 & R   \\
57166.438 & 57166.563 & 2.43$\pm$0.51 & 0.98$\pm$0.36 &-2.29$\pm$0.40&-1.68$\pm$0.21 & D   \\
57166.688 & 57166.813 & 1.27$\pm$0.41 & 3.05$\pm$0.67 &2.38$\pm$0.29 &1.75$\pm$0.16 & R   \\
57169.688 & 57169.813 & 1.15$\pm$0.40 & 2.21$\pm$0.52 &3.19$\pm$0.58&2.34$\pm$0.31 & R   \\
57170.438 & 57170.563 & 3.61$\pm$0.69 & 0.81$\pm$0.34 &-1.39$\pm$0.21 &-1.02$\pm$0.11 & D   \\
\hline
&& flare-4(C) &&&& \\
\hline
57243.438 & 57243.563 & 0.57$\pm$0.25 & 1.98$\pm$0.60 &1.67$\pm$0.19 &1.23$\pm$0.10 & R  \\
57245.813 & 57245.938 & 4.60$\pm$1.55 & 2.37$\pm$0.94 &-3.14$\pm$0.29&-2.31$\pm$0.16 & D  \\
57249.563 & 57249.688 & 0.78$\pm$0.32 & 1.94$\pm$0.66 &2.27$\pm$0.18 &1.67$\pm$0.10 & R  \\
\hline
&& Flare-5 &&&&  \\
\hline
57632.563 & 57632.688 & 1.03$\pm$0.37 & 2.09$\pm$0.54  & 2.00$\pm$0.33&1.47$\pm$0.18 & R  \\
57634.938 & 57635.063 & 2.30$\pm$0.50 & 1.06$\pm$0.35  &-2.68$\pm$0.39&-1.97$\pm$0.21 & D  \\
57635.063 & 57635.188 & 1.06$\pm$0.35 & 2.13$\pm$0.49  &2.98$\pm$0.43&2.19$\pm$0.23 & R  \\
57635.188 & 57635.313 & 2.13$\pm$0.49 & 0.80$\pm$0.30  &-2.12$\pm$0.31&-1.56$\pm$0.17 & D  \\
\hline 
\end{tabular}

\end{table*}

\begin{figure*}
\begin{center}
\includegraphics[scale=0.5]{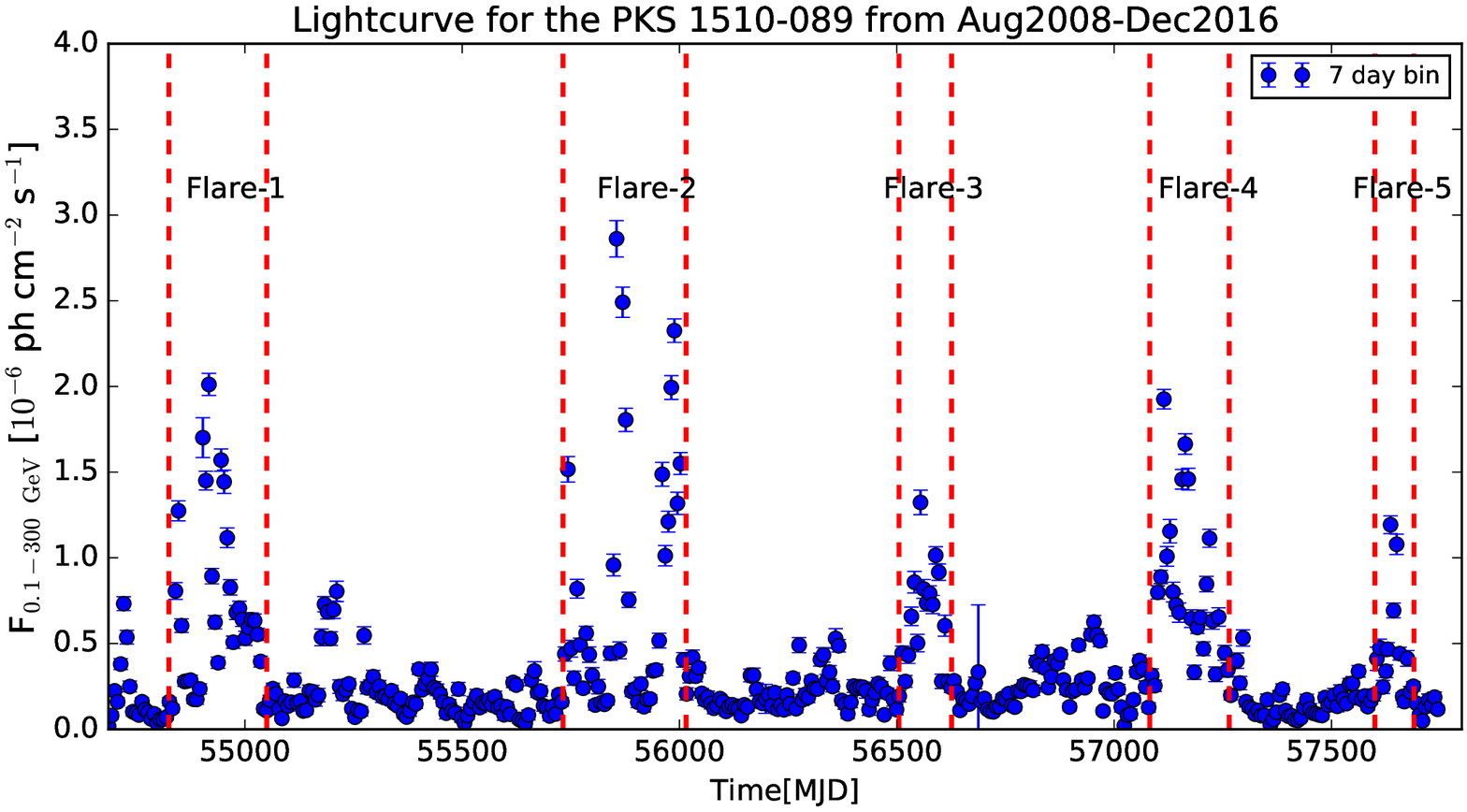}
\end{center}
\caption{Light curve history of the PKS 1510-089. Five flare episodes have been identified and further studied. Their time durations are
the following: MJD 54825--55050, MJD 55732--56015, MJD 56505--56626, MJD 57082--57265 and MJD 57657--57753, which are shown by broken
red lines.}
\label{fig:A}

\begin{center}
\includegraphics[scale=0.5]{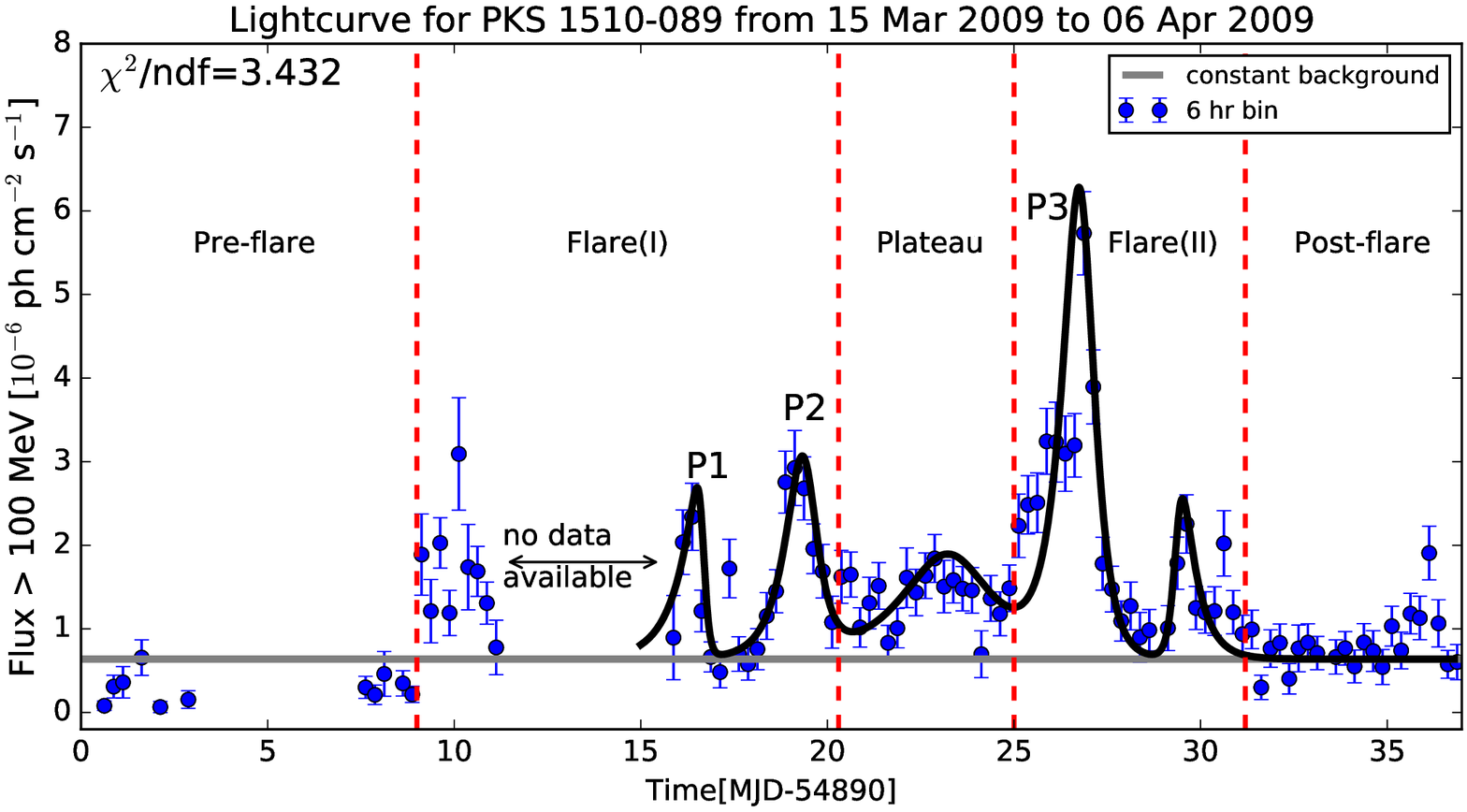}
\end{center}
\caption{Light curve for the flare-1(A) fitted by the sum of exponentials (see text for details). The fitted parameters 
are given in Table-1.
All the different periods of activity have been separated by broken red lines and the light grey line represents the 
constant state/flux.}
\label{fig:B1}

\begin{center}
\includegraphics[scale=0.5]{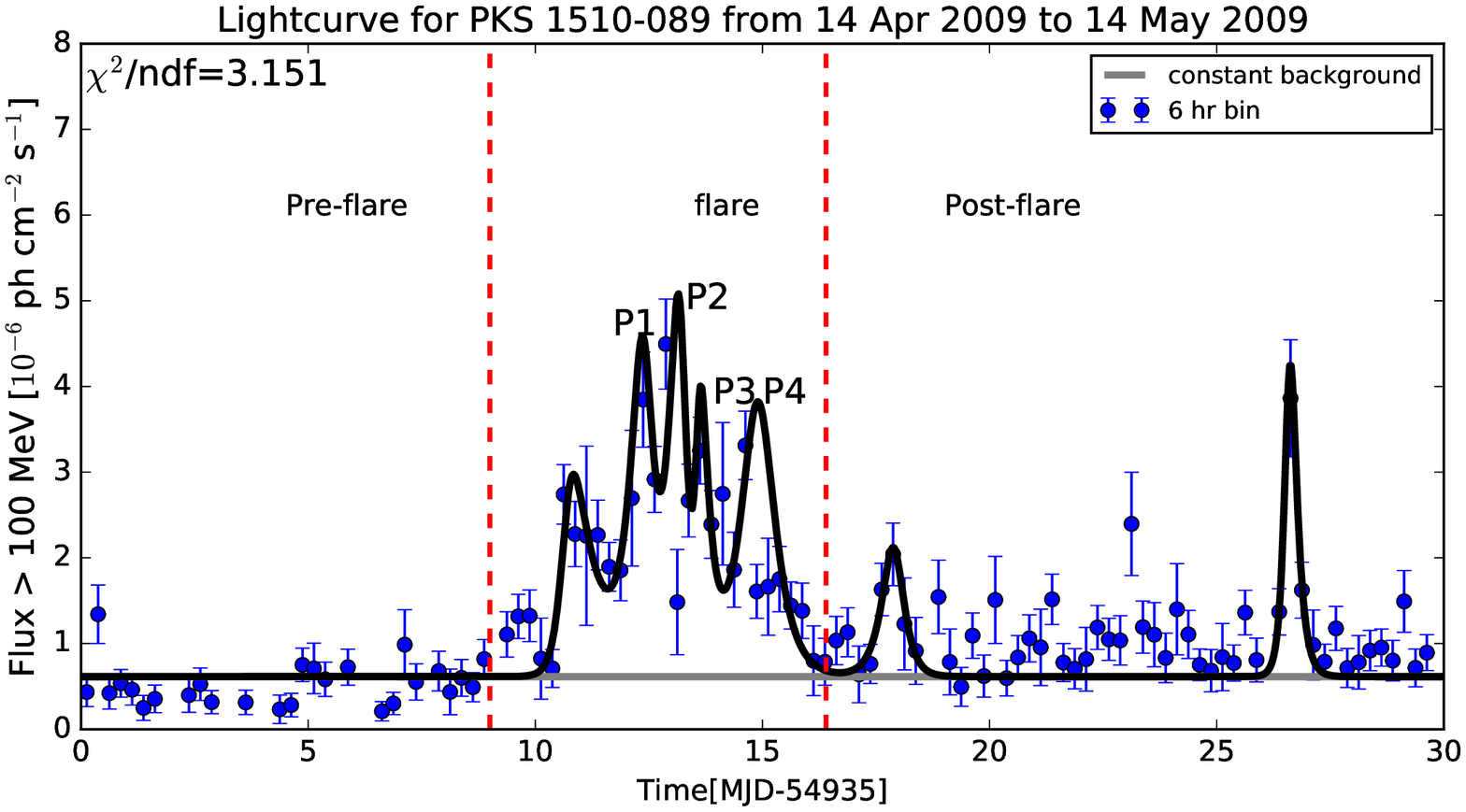}
\end{center}
\caption{ Light curve for the flare-1(B) fitted by the sum of exponentials (see text for details). The fitted parameters 
are given in Table-1.
All the different periods of activity have been separated by broken red lines and the light grey line represents the         
constant state/flux.}
\label{fig:B2}
\end{figure*}

\begin{figure*}
\begin{center}
\includegraphics[scale=0.5]{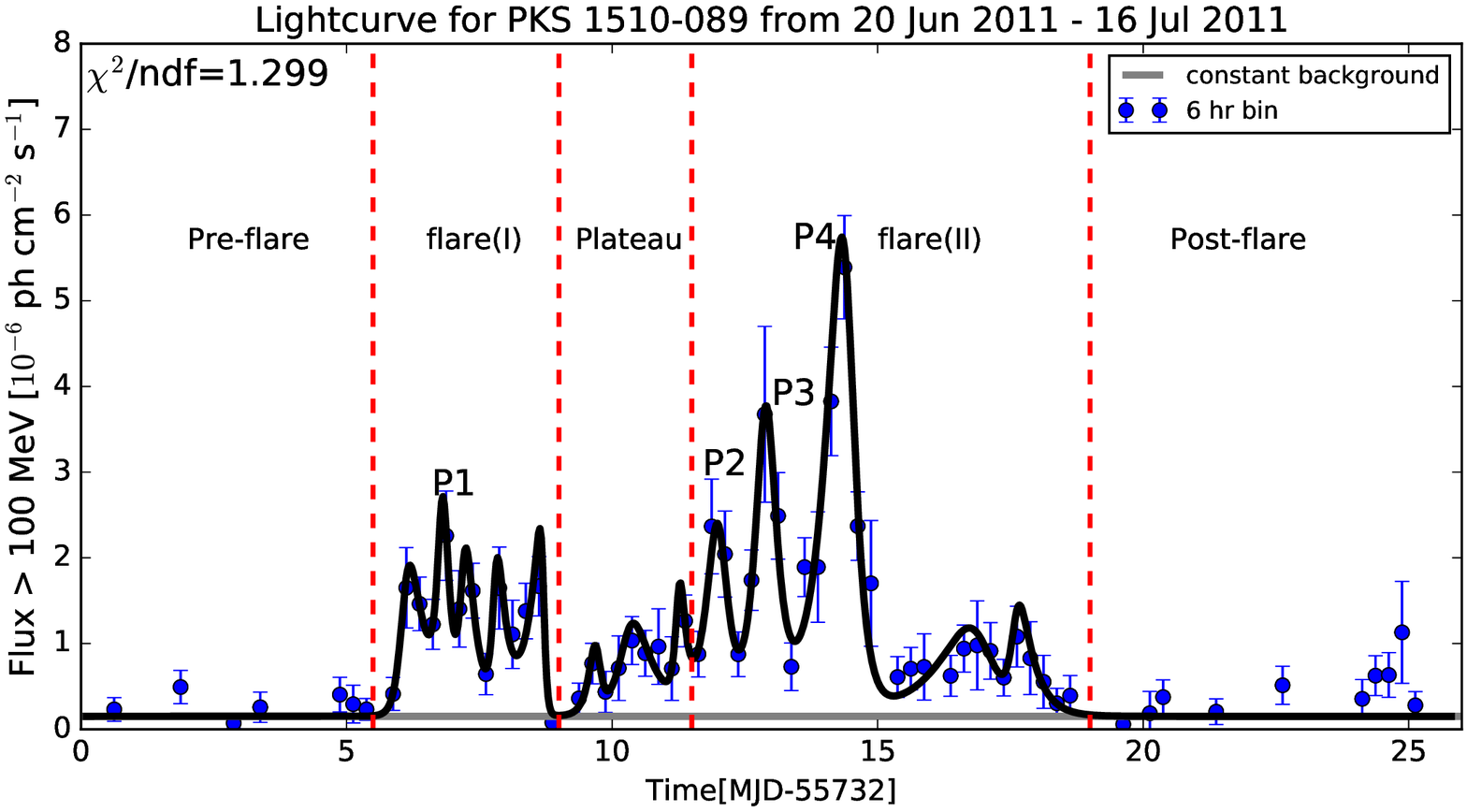}
\end{center}
\caption{Light curve for the flare-2(A) fitted by the sum of exponentials (see text for details). The fitted parameters
are given in Table-2.
All the different periods of activity have been separated by broken red lines and the light grey line represents the
constant state/flux.}
\label{fig:C1}
\begin{center}
\includegraphics[scale=0.5]{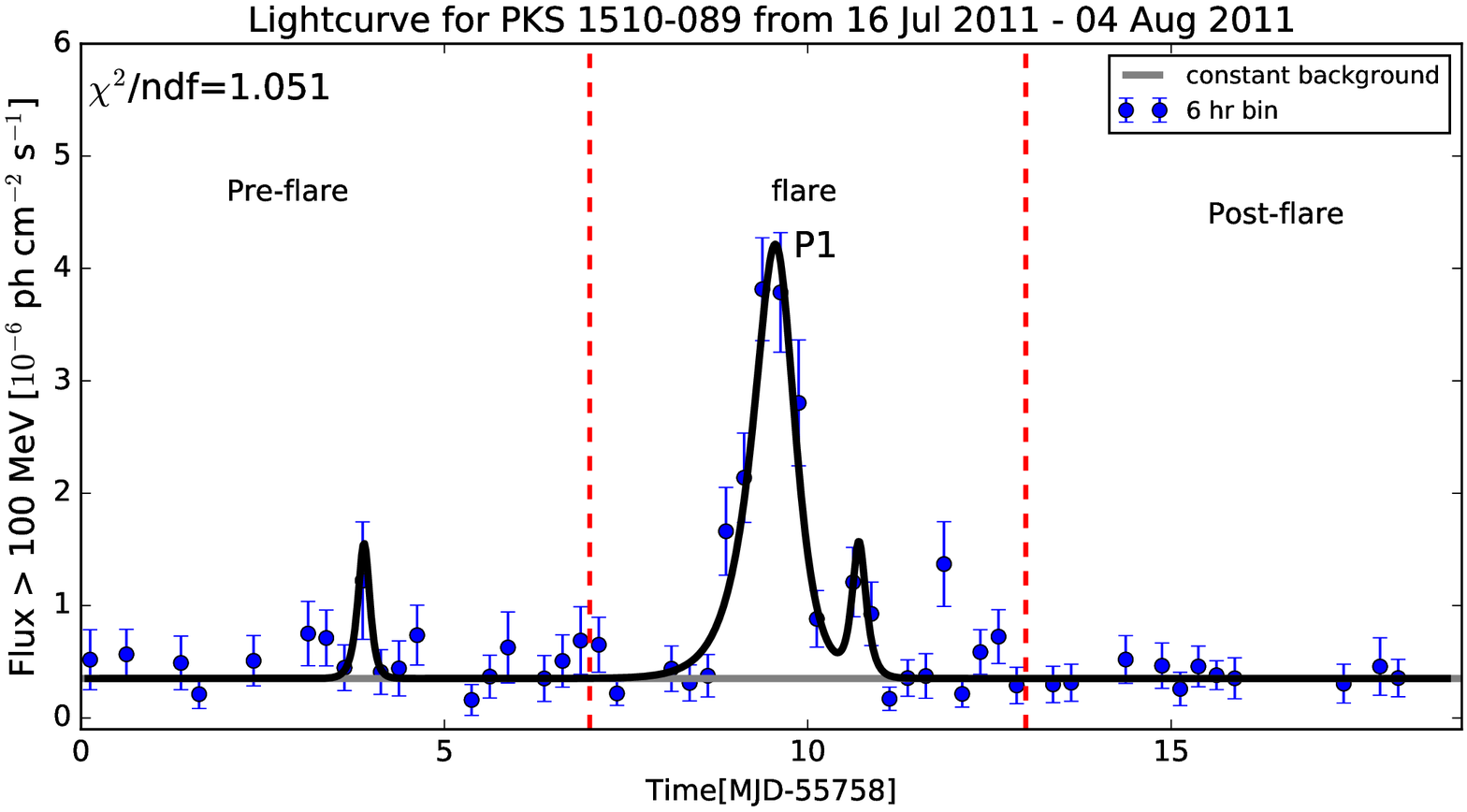}
\end{center}
\caption{Light curve for the flare-2(B) fitted by the sum of exponentials (see text for details). The fitted parameters
are given in Table-2.
All the different periods of activity have been separated by broken red lines and the light grey line represents the
constant state/flux.}
\label{fig:C2}
\begin{center}
\includegraphics[scale=0.5]{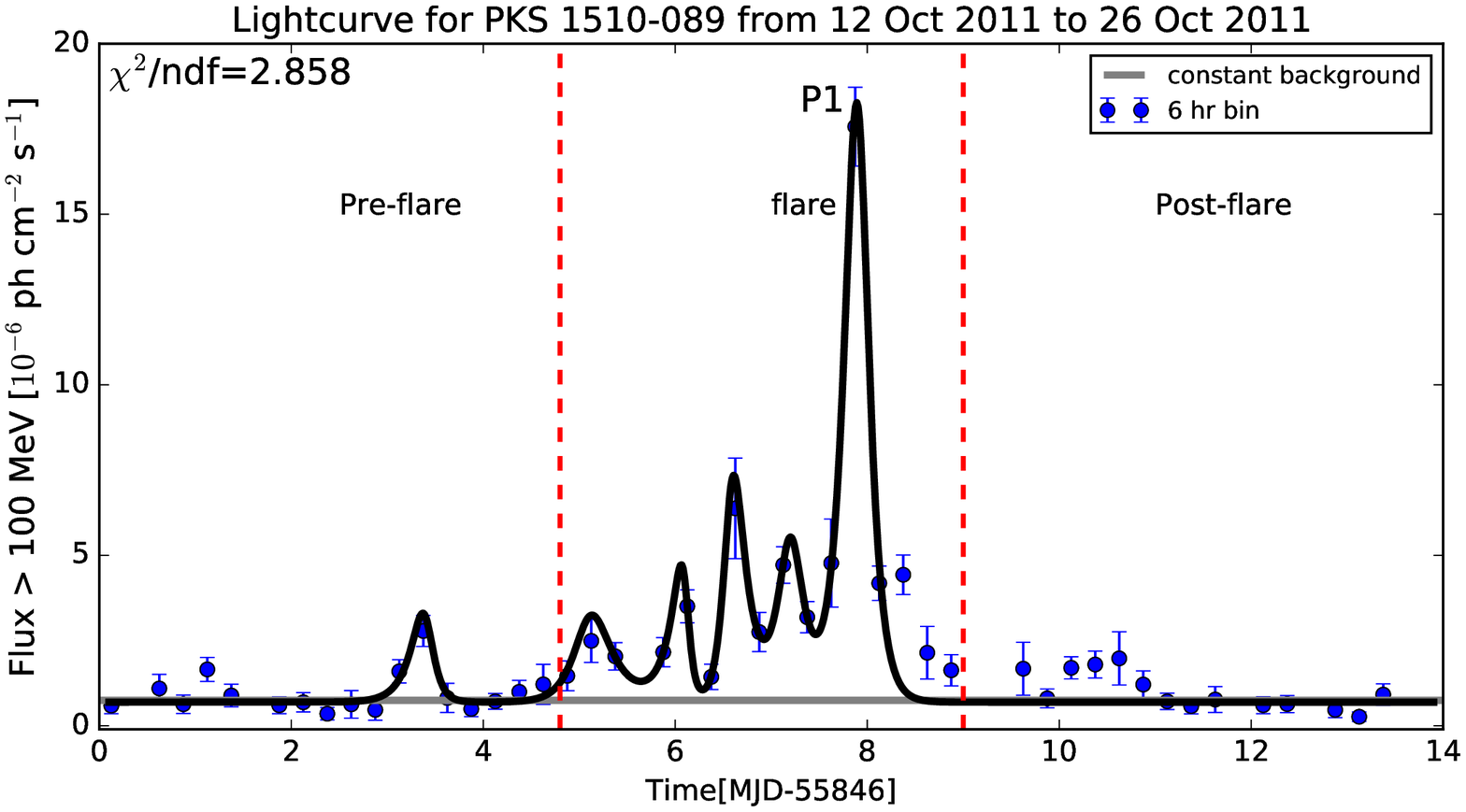}
\end{center}
\caption{Light curve for the flare-2(C) fitted by the sum of exponentials (see text for details). The fitted parameters
are given in Table-2.
All the different periods of activity have been separated by broken red lines and the light grey line represents the
constant state/flux.}
\label{fig:C3}
\end{figure*}

\begin{figure*}
\begin{center}
\includegraphics[scale=0.5]{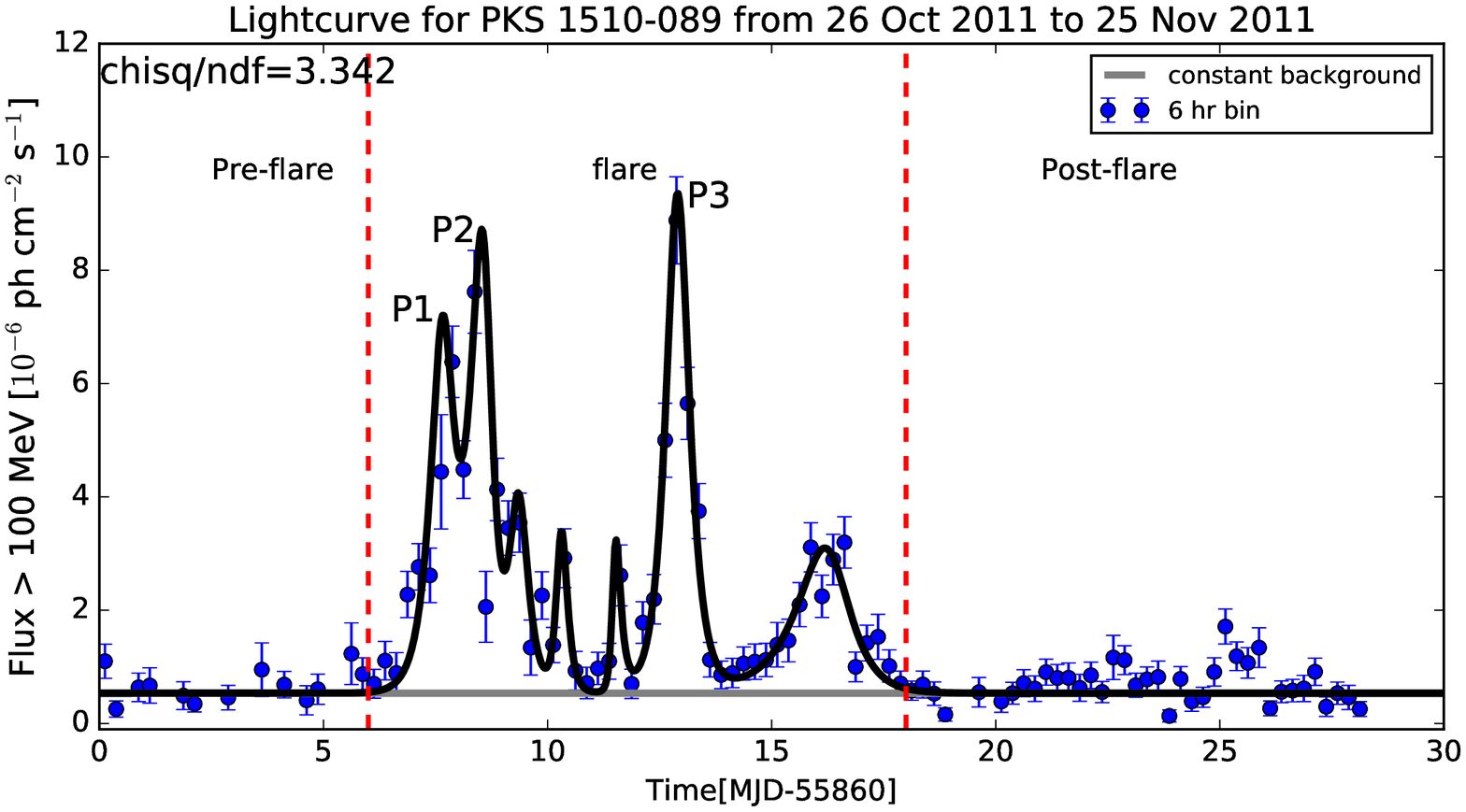}
\end{center}
\caption{Light curve for the flare-2(D) fitted by the sum of exponentials (see text for details). The fitted parameters
are given in Table-2.
All the different periods of activity have been separated by broken red lines and the light grey line represents the
constant state/flux.}
\label{fig:C4}
\begin{center}
\includegraphics[scale=0.5]{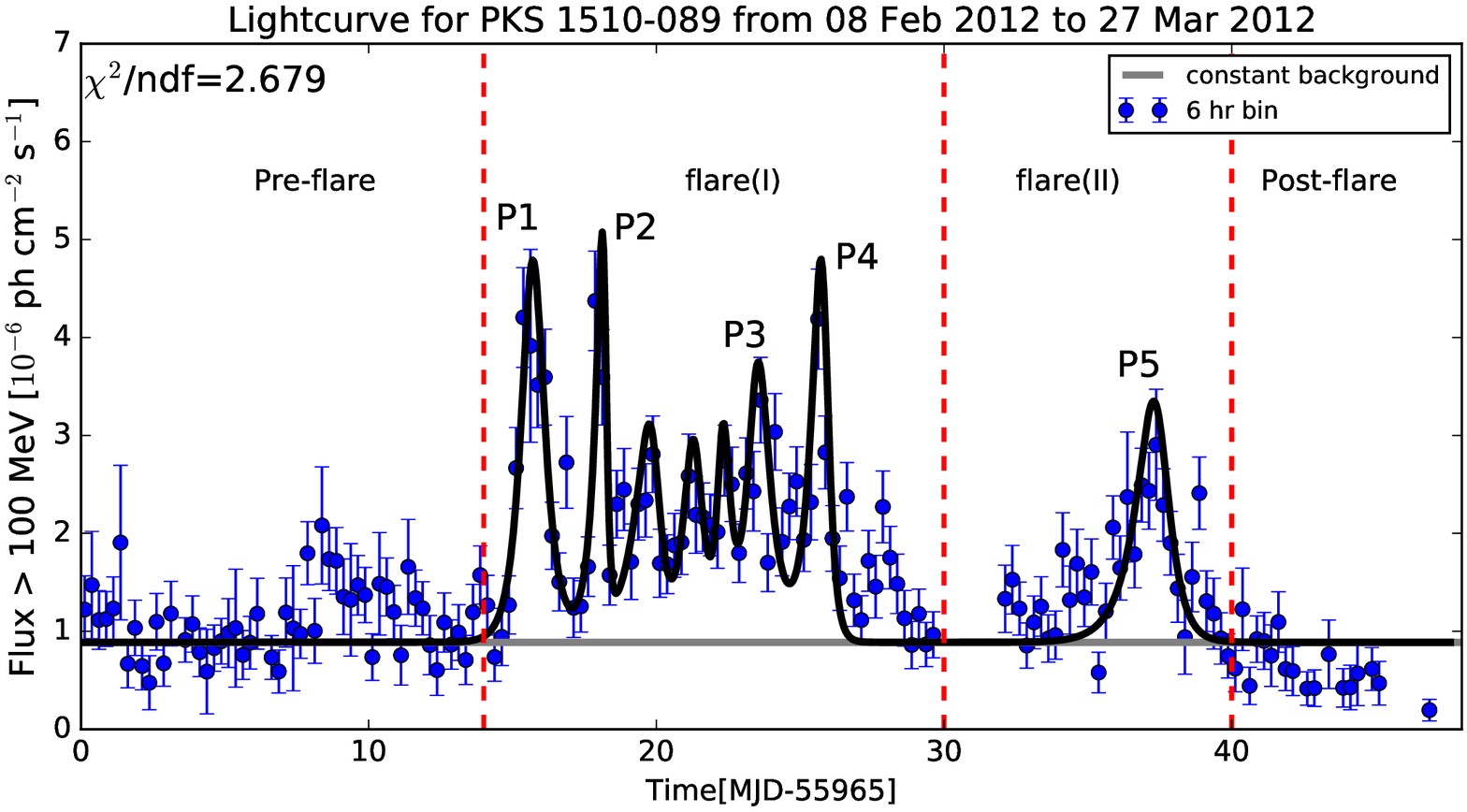}
\end{center}
\caption{Light curve for the flare-2(E) fitted by the sum of exponentials (see text for details). The fitted parameters
are given in Table-2.
All the different periods of activity have been separated by broken red lines and the light grey line represents the
constant state/flux.}
\label{fig:C5}
\begin{center}
\includegraphics[scale=0.5]{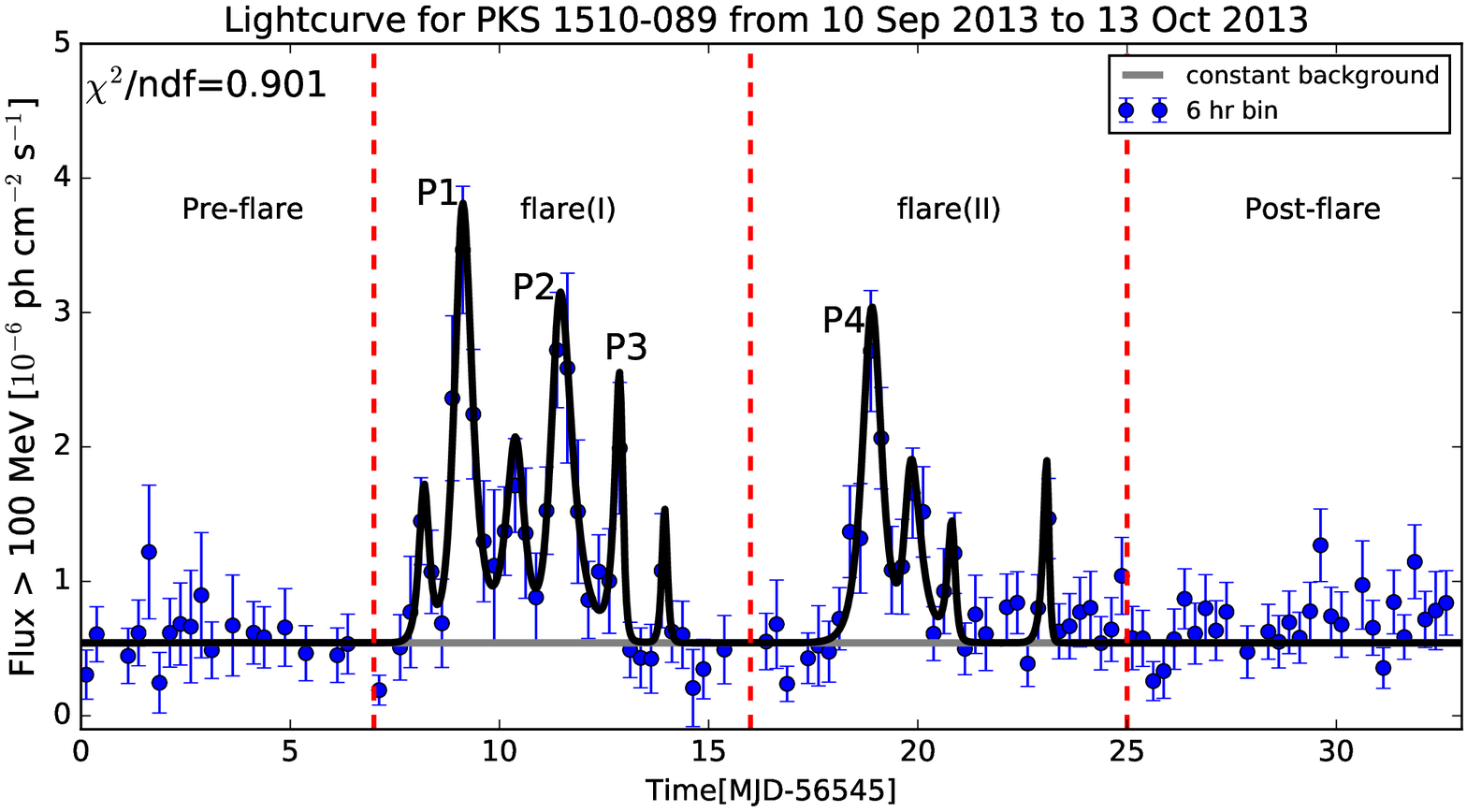}
\end{center}
\caption{Light curve for the flare-3 fitted by the sum of exponentials (see text for details). The fitted parameters
are given in Table-3.
All the different periods of activity have been separated by broken red lines and the light grey line represents the
constant state/flux.}
\label{fig:D}
\end{figure*}

\begin{figure*}
\begin{center}
\includegraphics[scale=0.5]{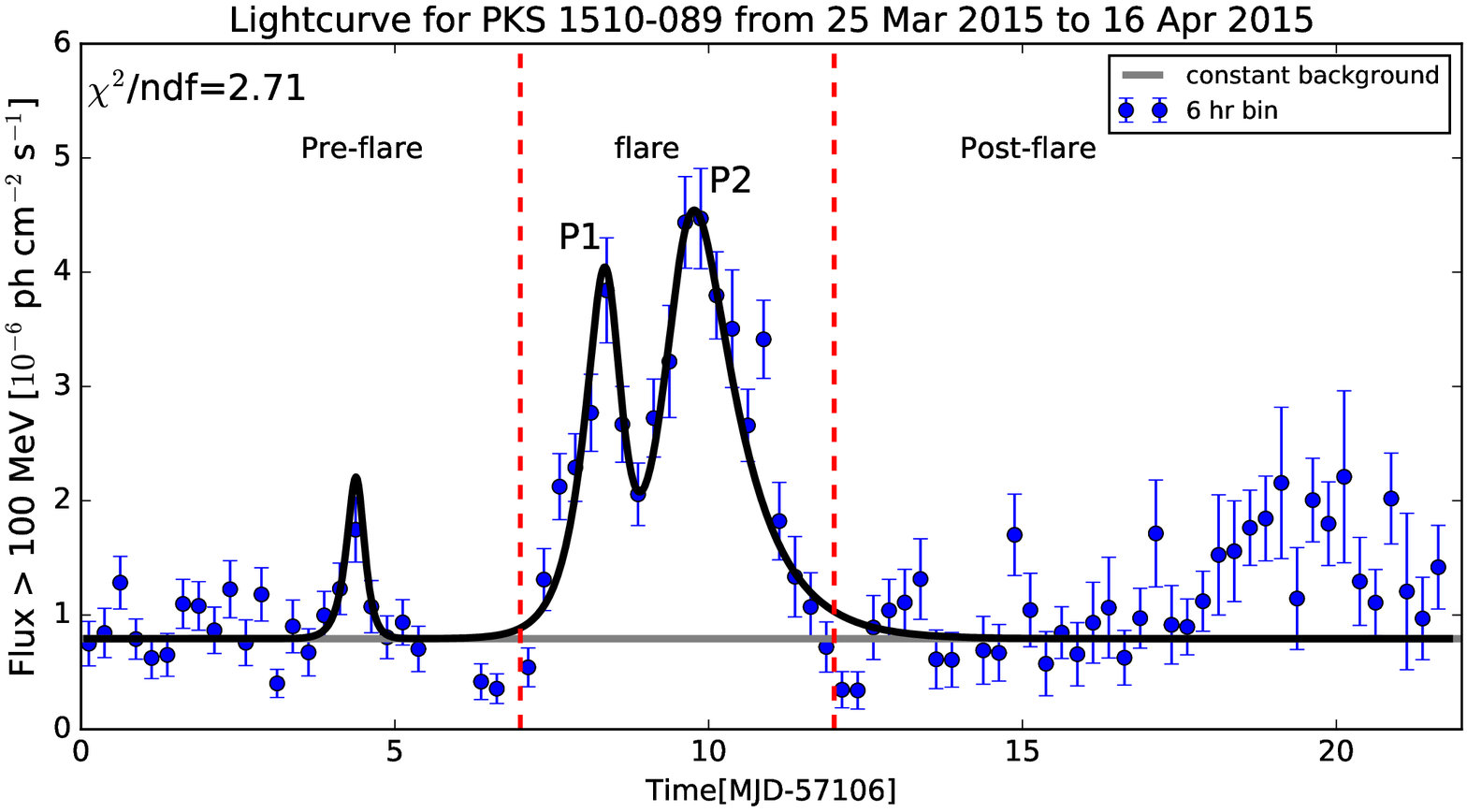}
\end{center}
\caption{Light curve for the flare-4(A) fitted by the sum of exponentials (see text for details). The fitted parameters
are given in Table-4.
All the different periods of activity have been separated by broken red lines and the light grey line represents the
constant state/flux.}
\label{fig:E1}
\begin{center}
\includegraphics[scale=0.5]{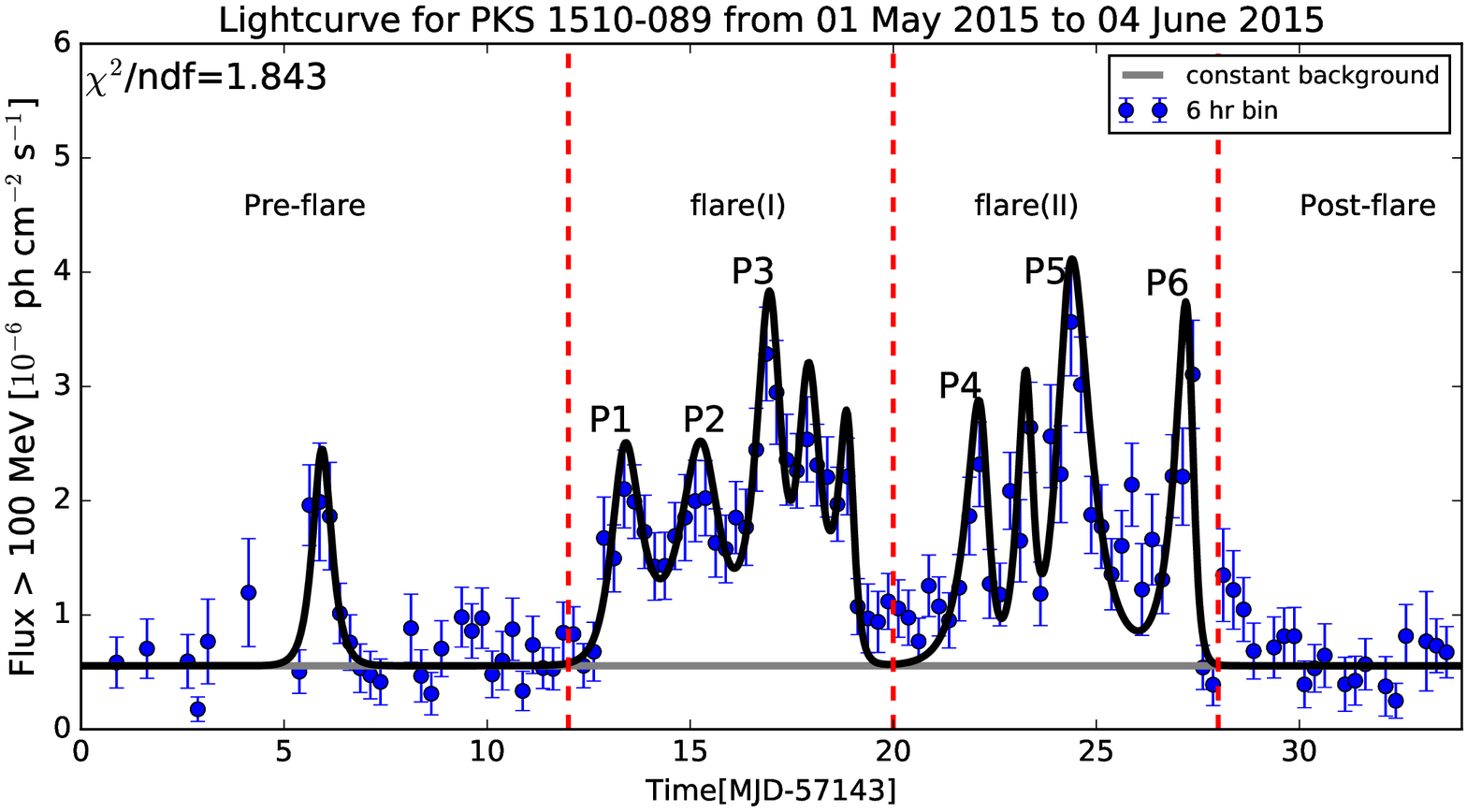}
\end{center}
\caption{Light curve for the flare-4(B) fitted by the sum of exponentials (see text for details). The fitted parameters
are given in Table-4.
All the different periods of activity have been separated by broken red lines and the light grey line represents the
constant state/flux.}
\label{fig:E2}
\begin{center}
\includegraphics[scale=0.5]{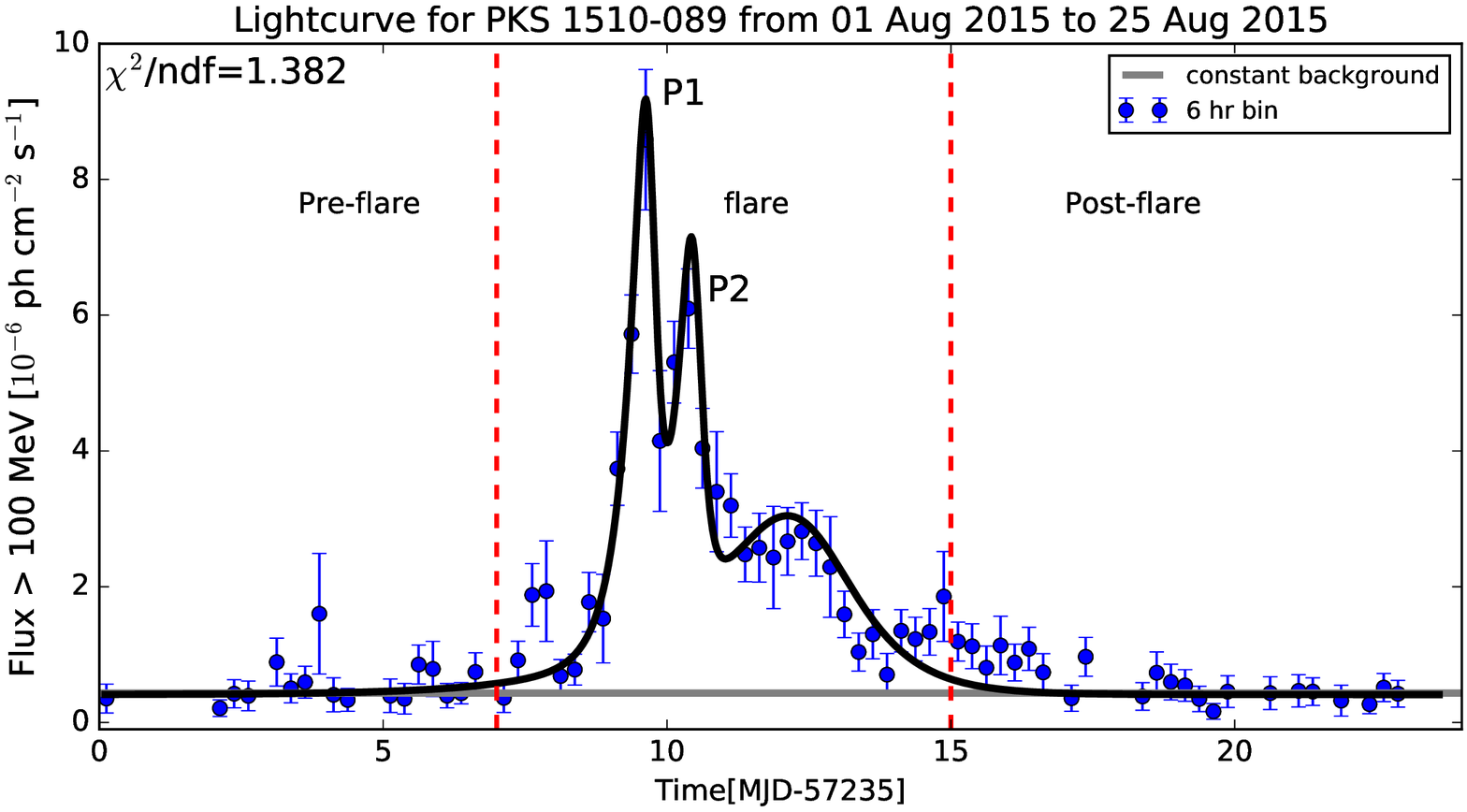}
\end{center}
\caption{Light curve for the flare-4(C) fitted by the sum of exponentials (see text for details). The fitted parameters
are given in Table-4.
All the different periods of activity have been separated by broken red lines and the light grey line represents the
constant state/flux.}
\label{fig:E3}
\end{figure*}

\begin{figure*}
\begin{center}
\includegraphics[scale=0.5]{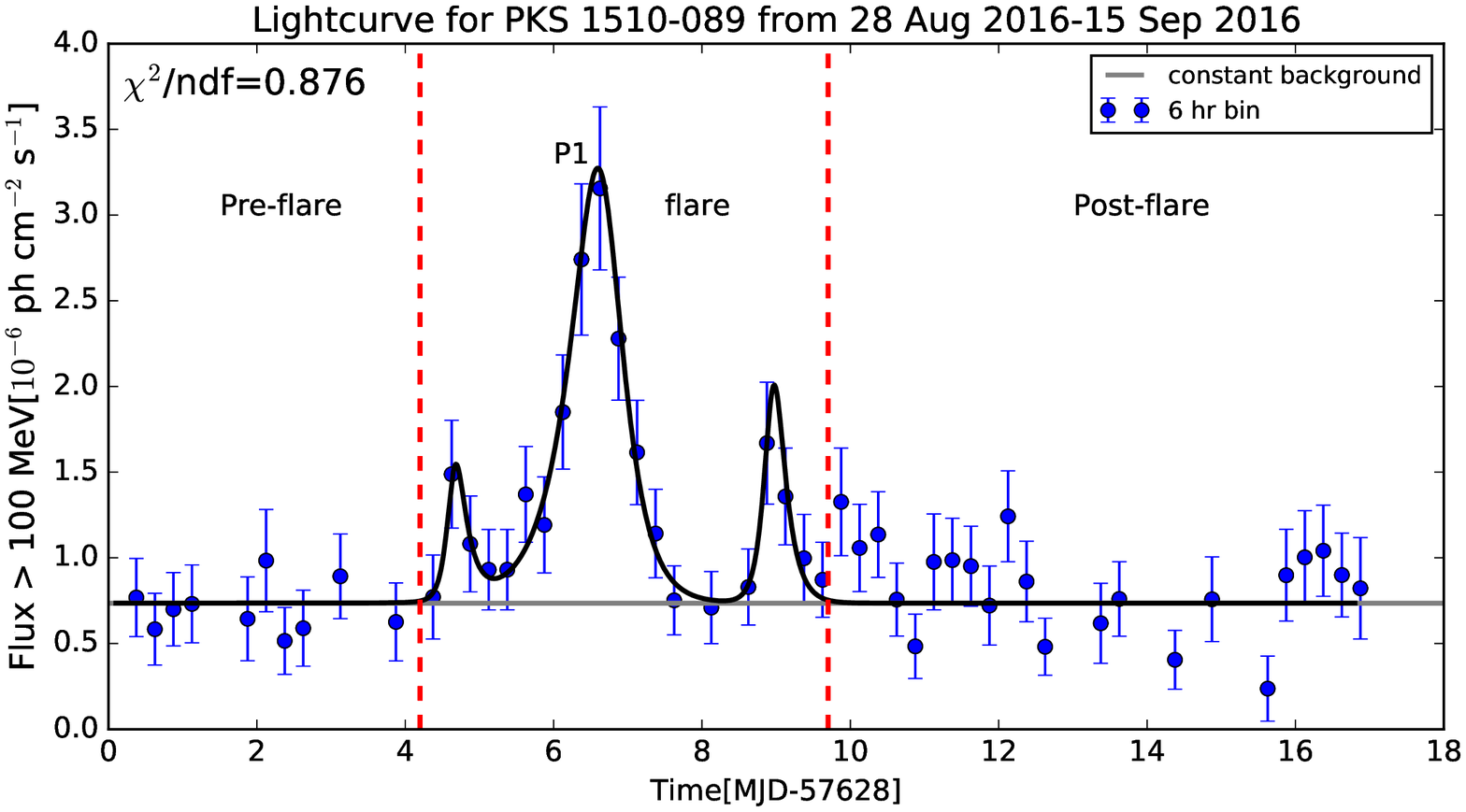}
\end{center}
\caption{Light curve for the flare-5 fitted by the sum of exponentials (see text for details). The fitted parameters
are given in Table-5.
All the different periods of activity have been separated by broken red lines and the light grey line represents the
constant state/flux.}
\label{fig:E4}

\begin{center}
 \includegraphics[scale=0.30]{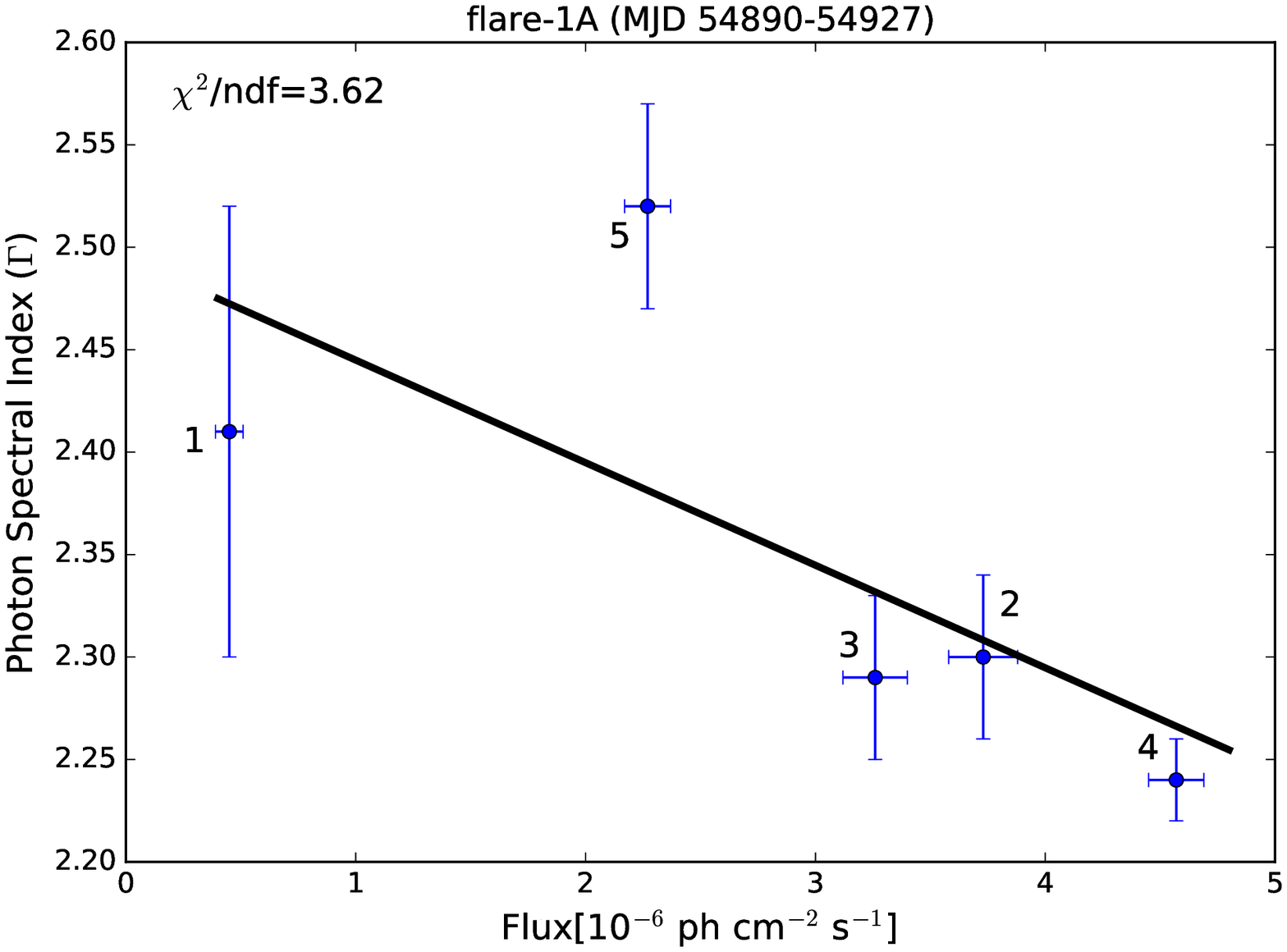}
 \includegraphics[scale=0.30]{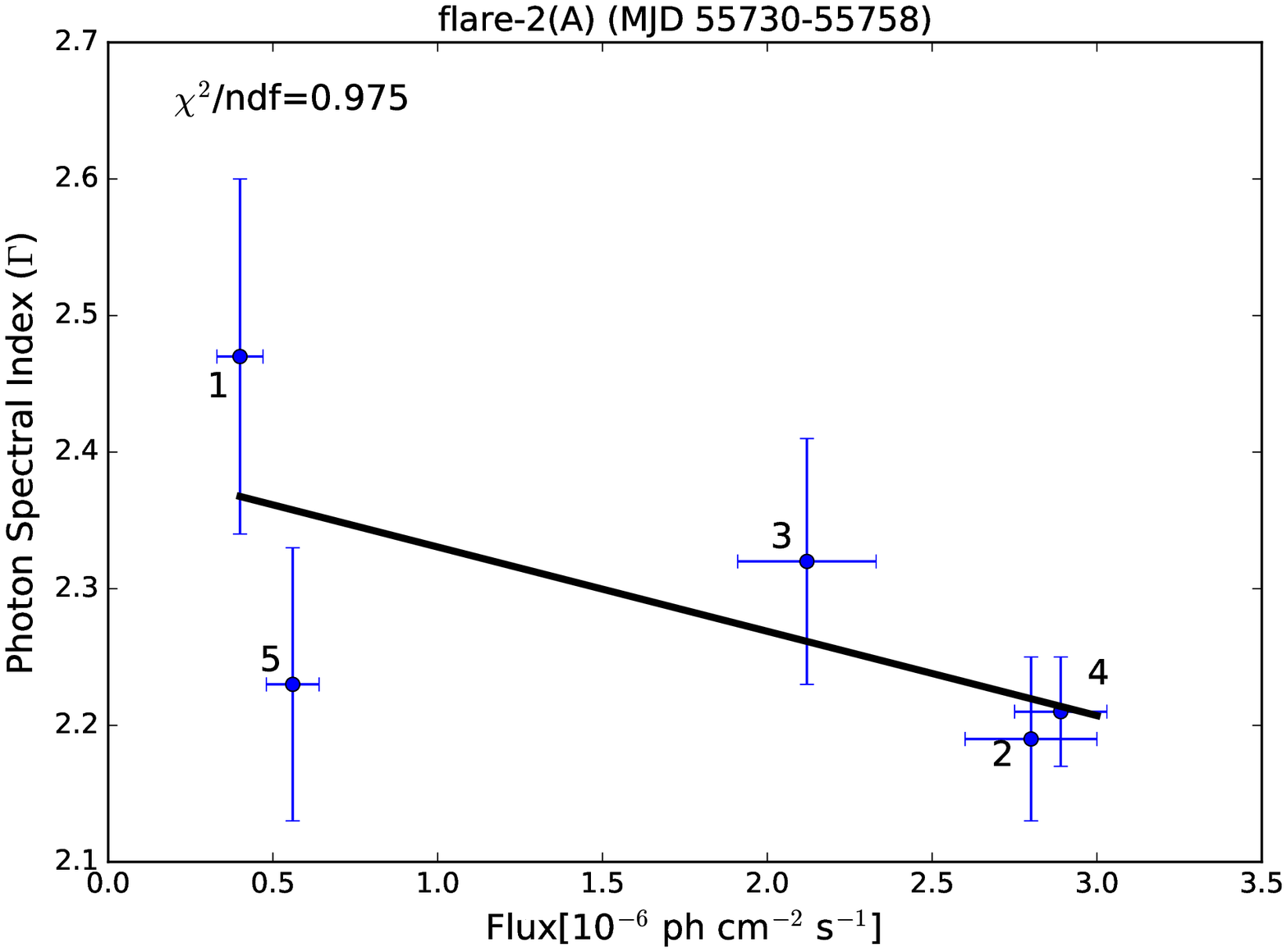}
\end{center}
\begin{center} 
 \includegraphics[scale=0.30]{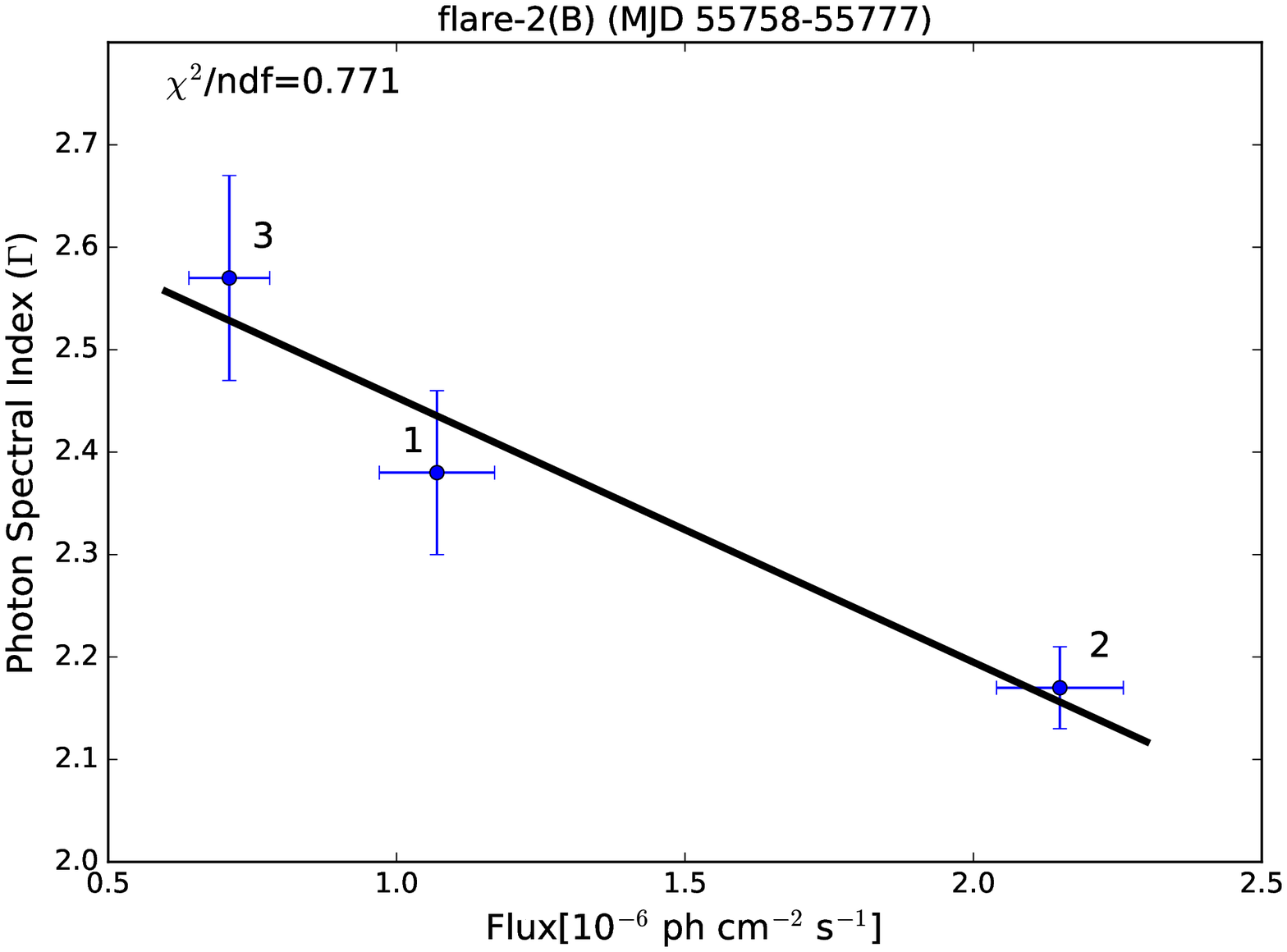}
 \includegraphics[scale=0.30]{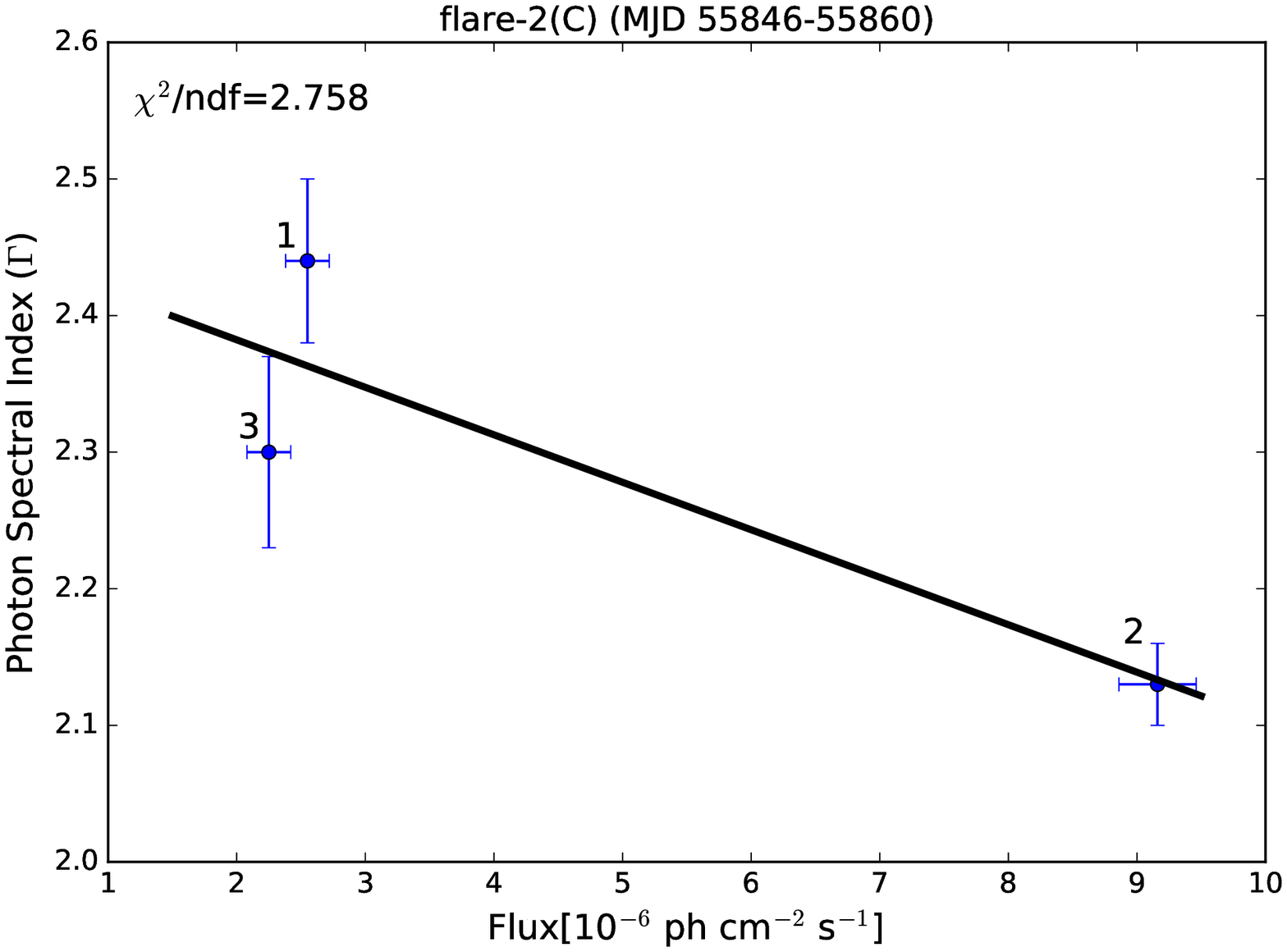}
\end{center}
\begin{center} 
 \includegraphics[scale=0.30]{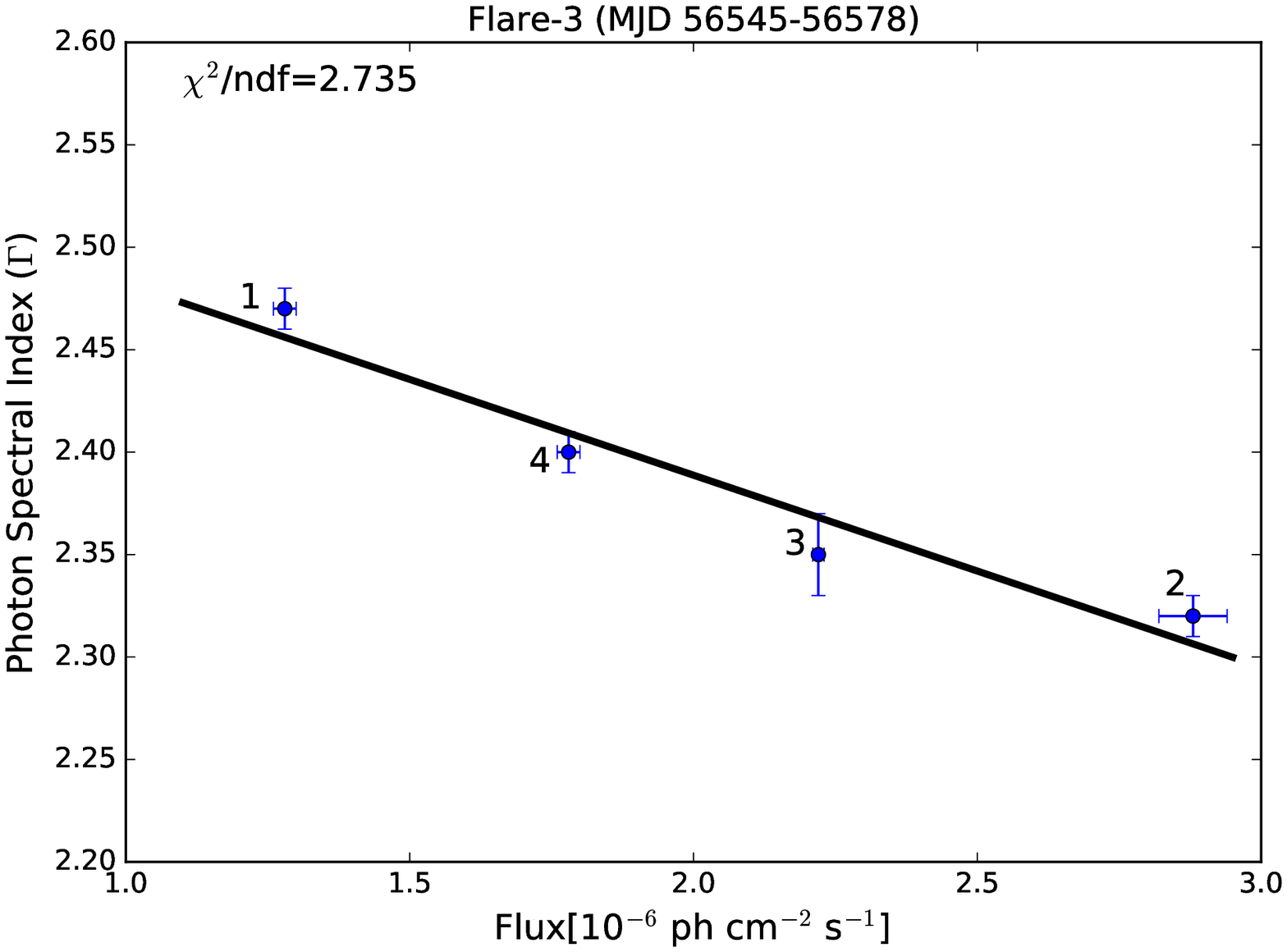}
 \includegraphics[scale=0.30]{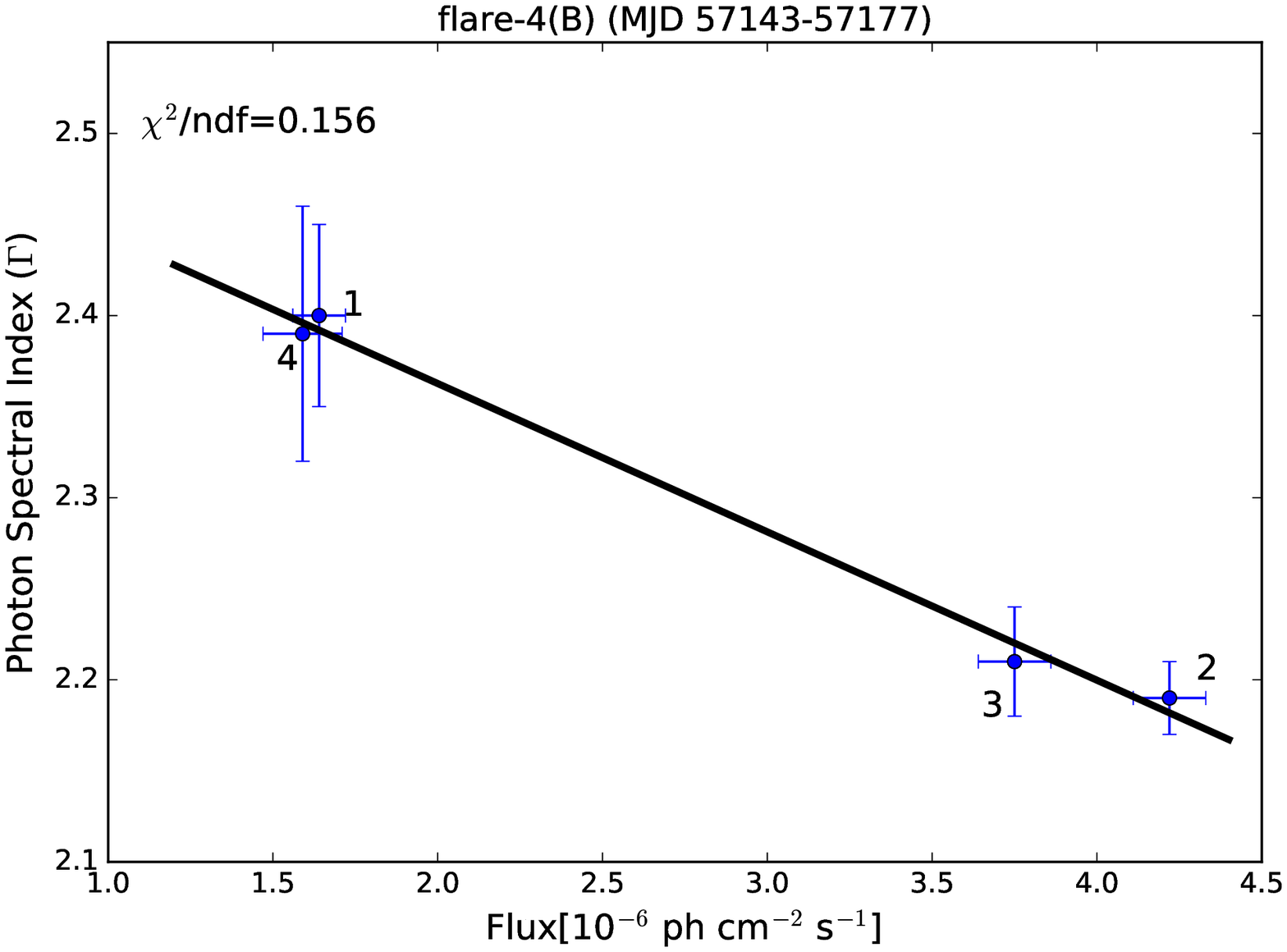}
 \end{center}
 \caption{Photon index vs flux are plotted for few sub-flares. And the numbers 1, 2, 3, 4, 5 represent the different time 
 periods. Top panel: 1, 2, 3, 4 and 5 represents the pre-flare, flare(I), plateau, flare(II) and post-flare states respectively. 
 Middle panel: 1, 2 and 3 represents the pre-flare, flare and post-flare respectively. Bottom panel: 1, 2, 3, and 4 represents 
 the pre-flare, flare(I), flare(II) and post-flare respectively. All the points have been fitted by the PL spectral type and the 
 corresponding reduced $\chi^2$ have been mention in the plots. Errors, associated with each data points, are 
 statistical only.} 
\label{fig:E4}
\end{figure*}

\begin{figure*}[htbp]
\begin{center}
\includegraphics[scale=0.35]{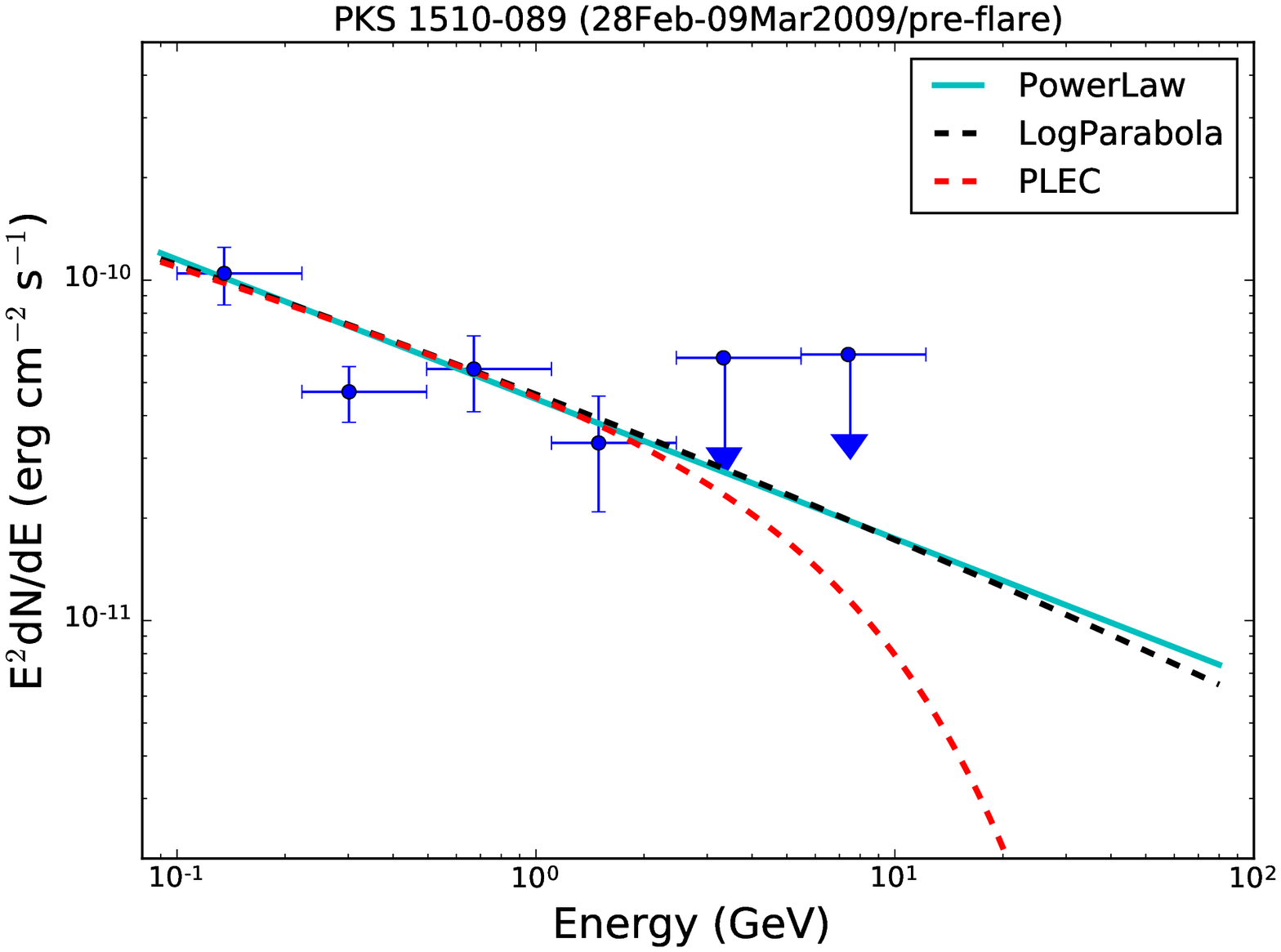}
\includegraphics[scale=0.35]{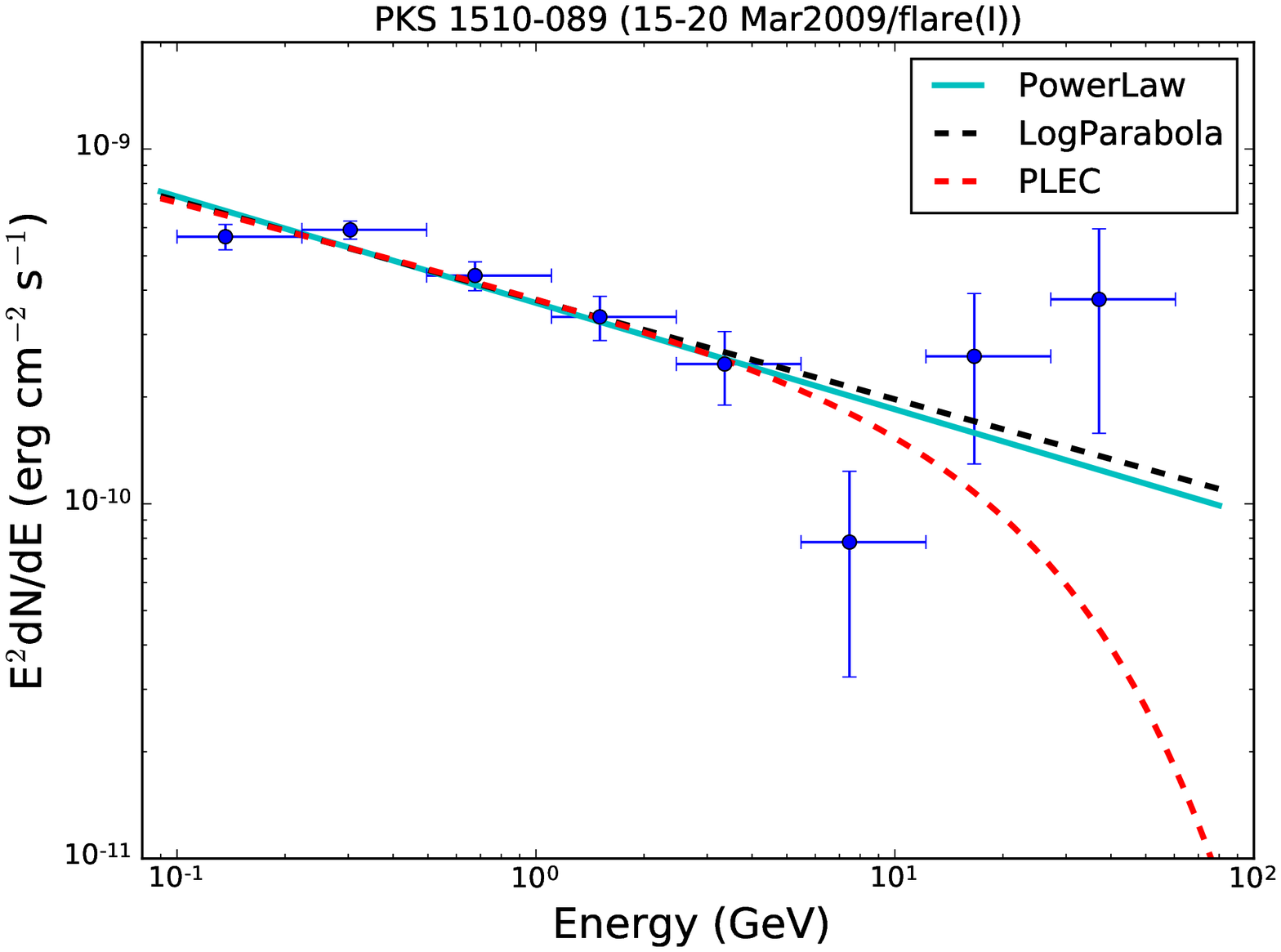}
\end{center}
\begin{center}
\includegraphics[scale=0.35]{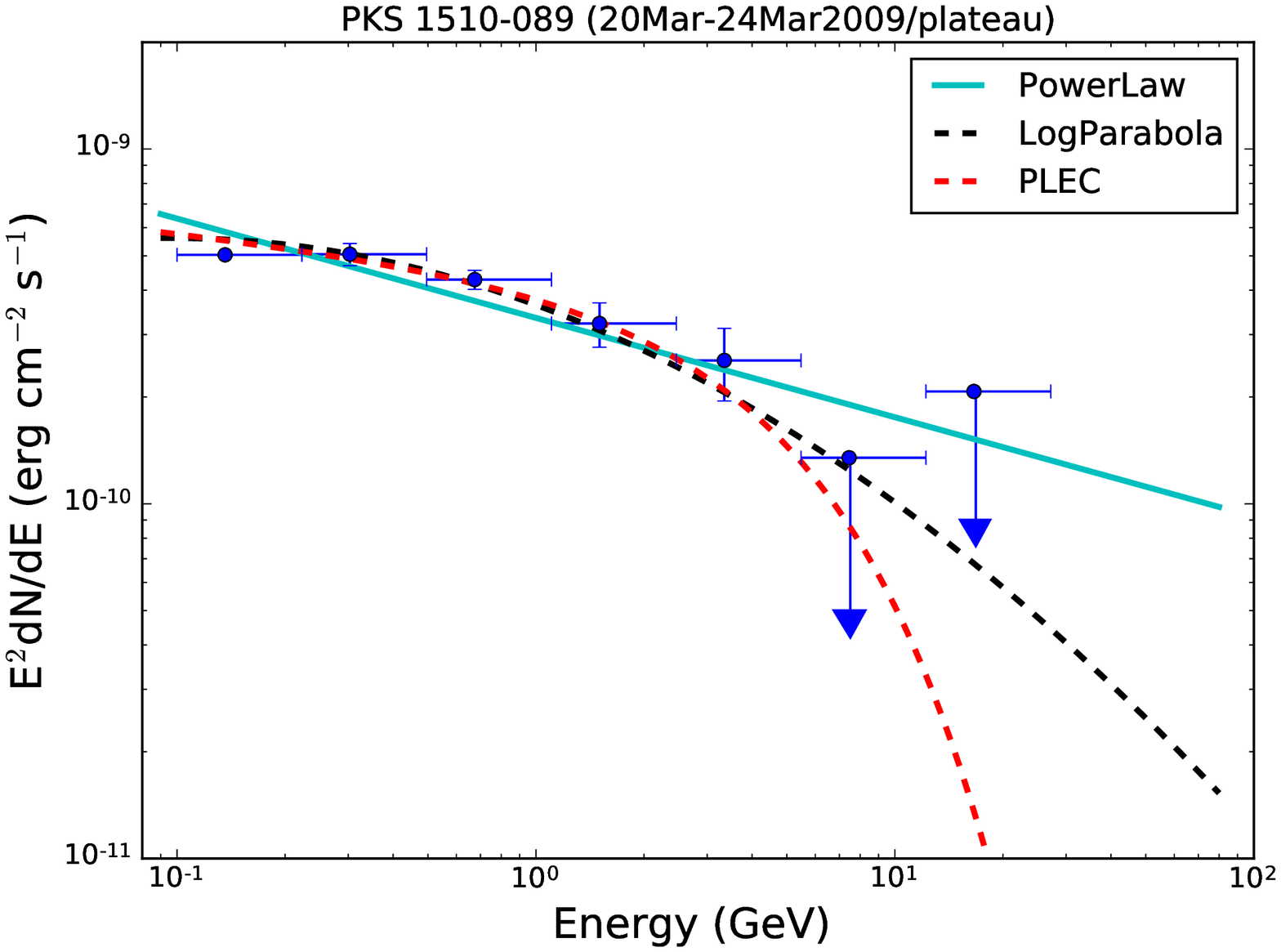}
\includegraphics[scale=0.35]{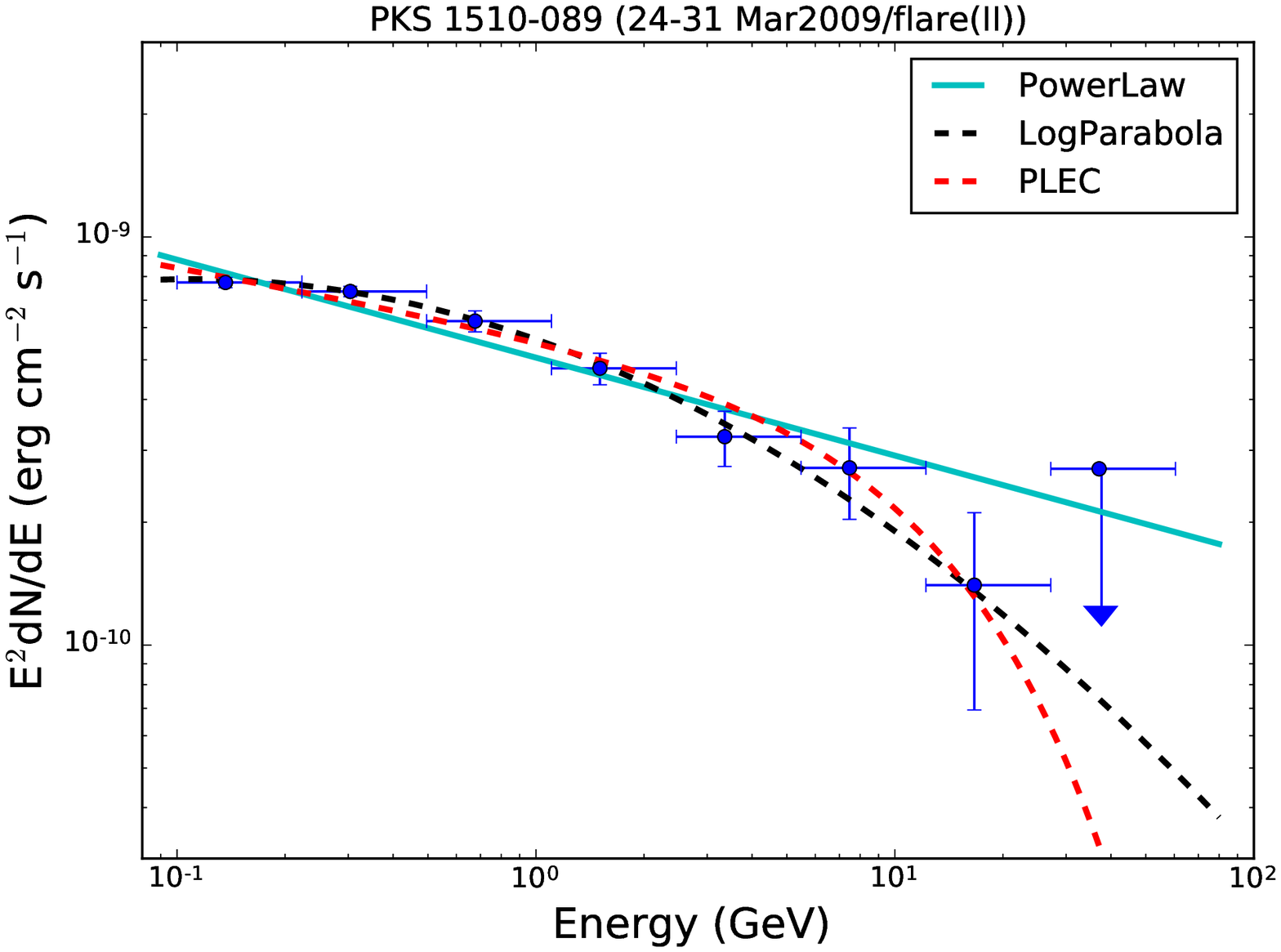}
\end{center}
\begin{center}
\includegraphics[scale=0.35]{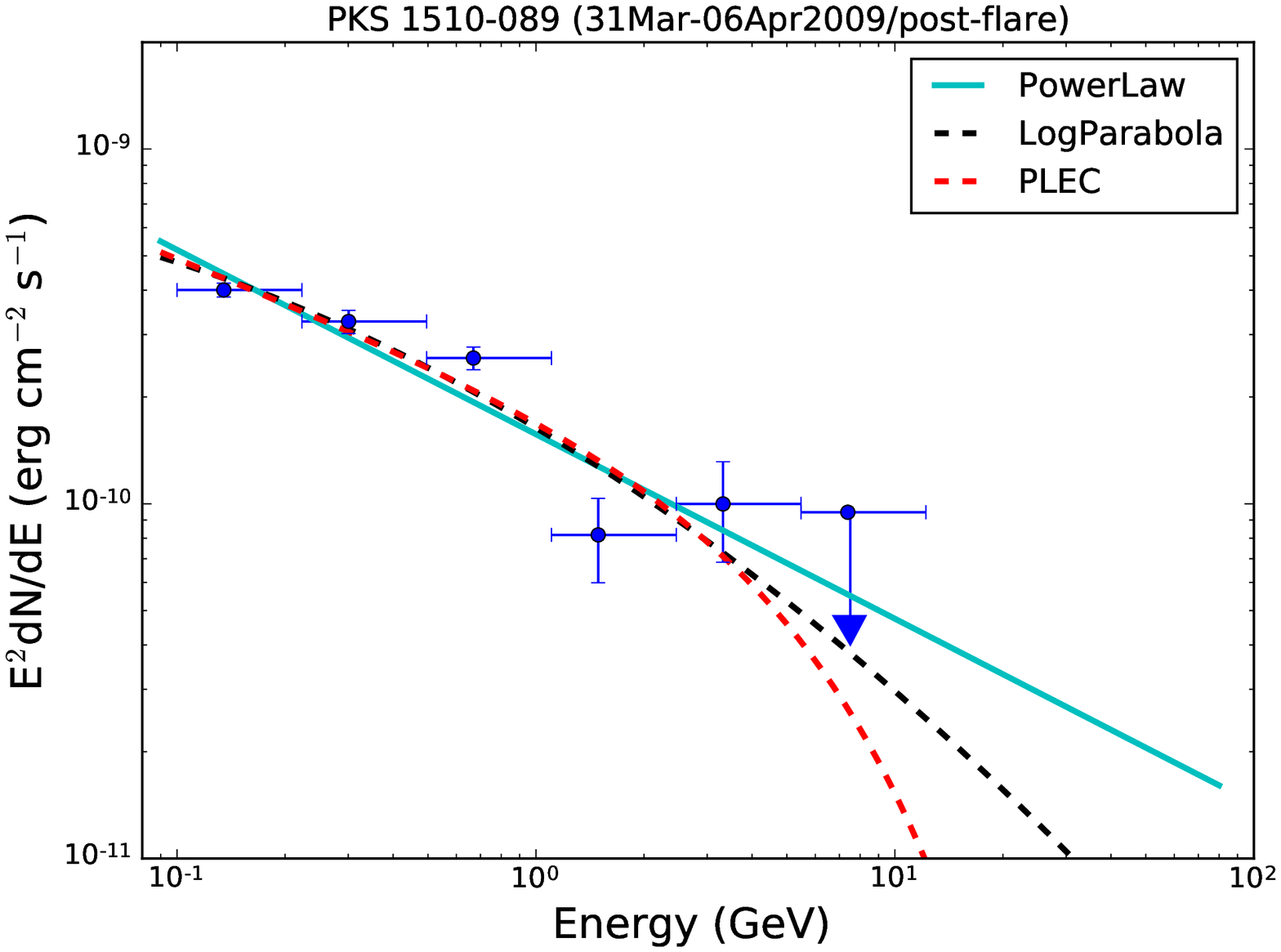}
\end{center}
\caption{Fermi-LAT SEDs during different activity states of flare-1(A) as defined in Fig.2 . PL, LP, PLEC models are 
shown in cyan, black and red color and there respective parameters are given in the Table-7. }
\label{fig:flare-1(A)}

\begin{center}
\includegraphics[scale=0.35]{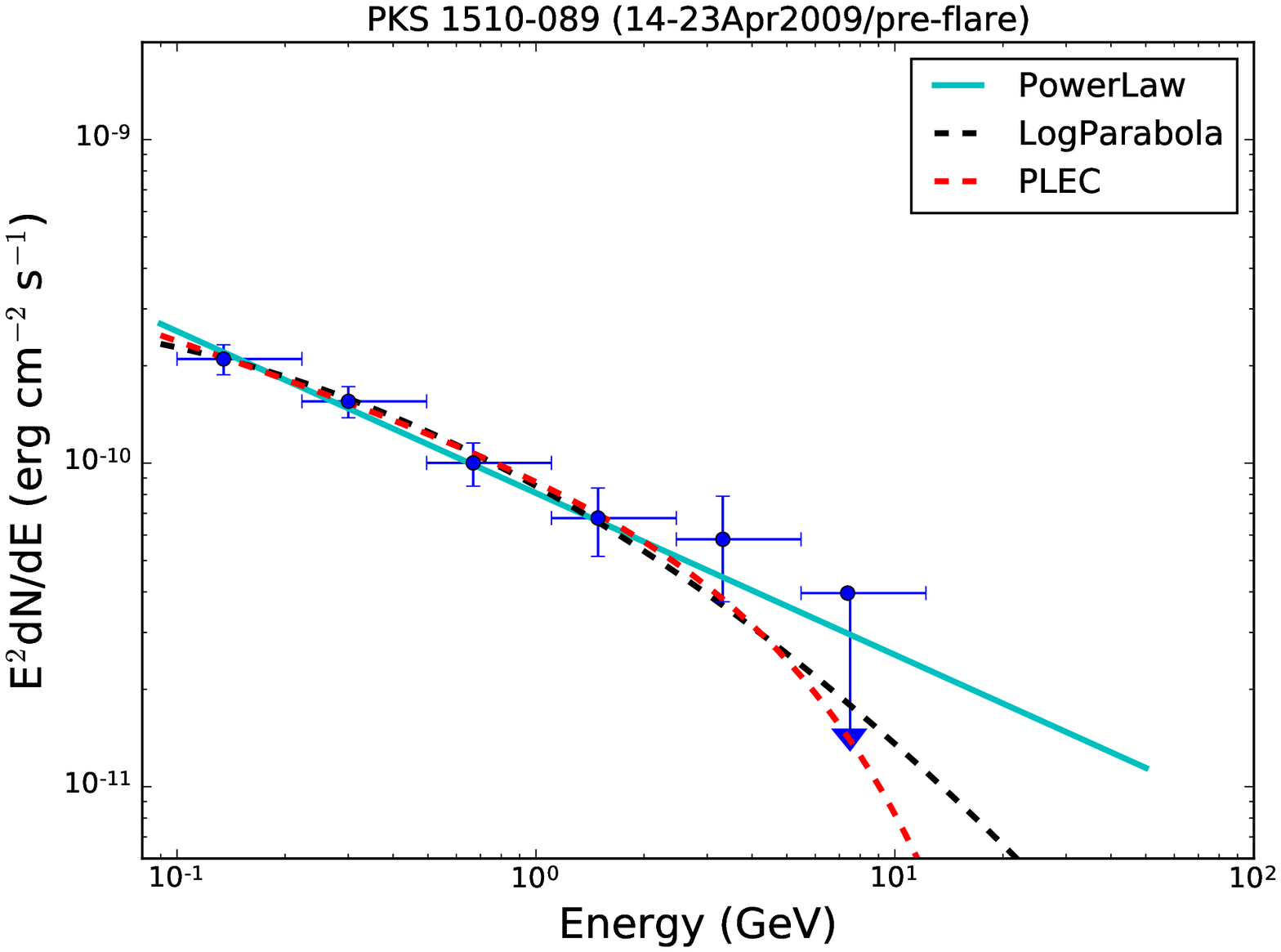}
\includegraphics[scale=0.35]{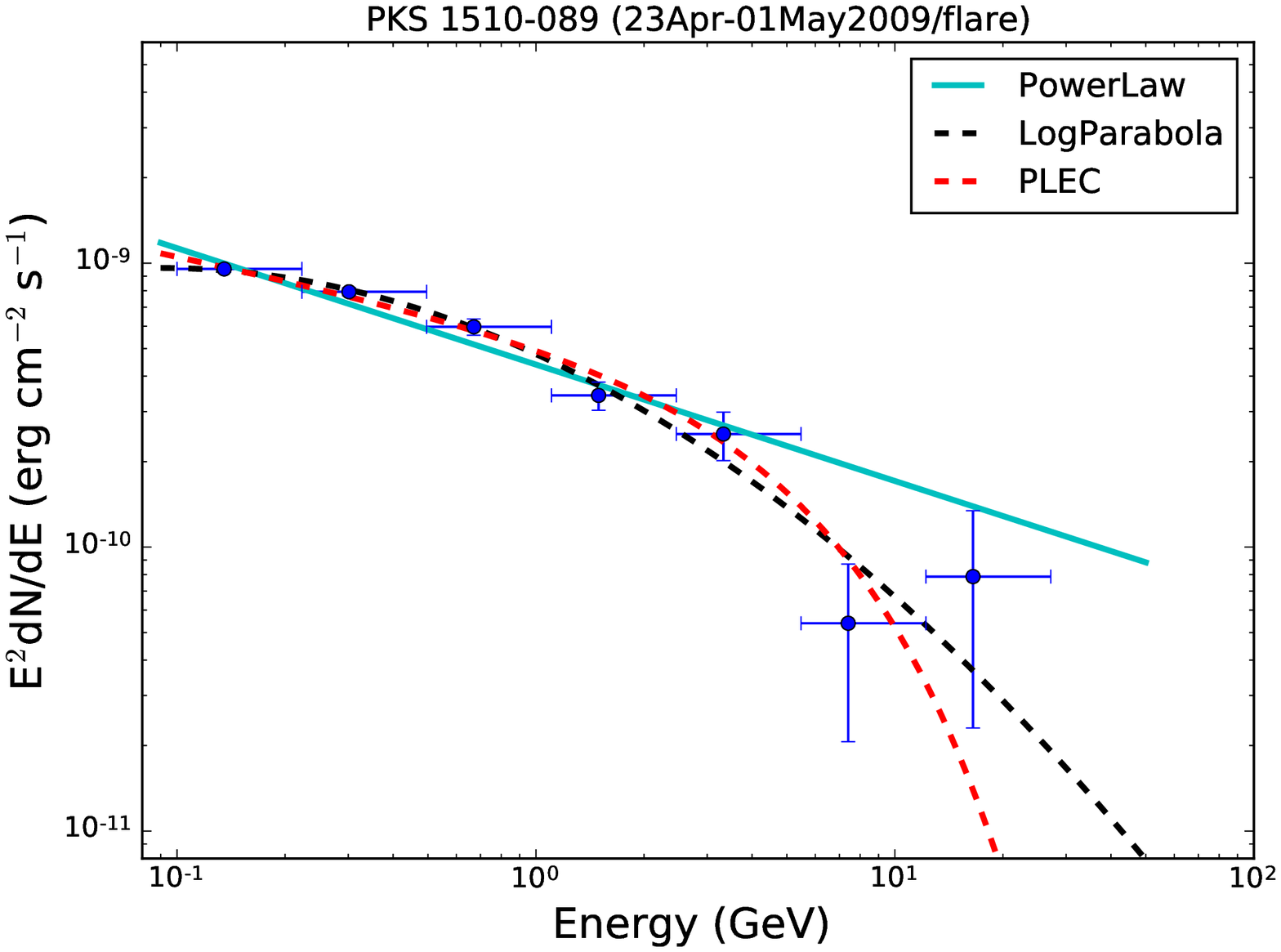}
\end{center} 
\end{figure*}

\begin{figure*}[htbp]
\begin{center}
\includegraphics[scale=0.35]{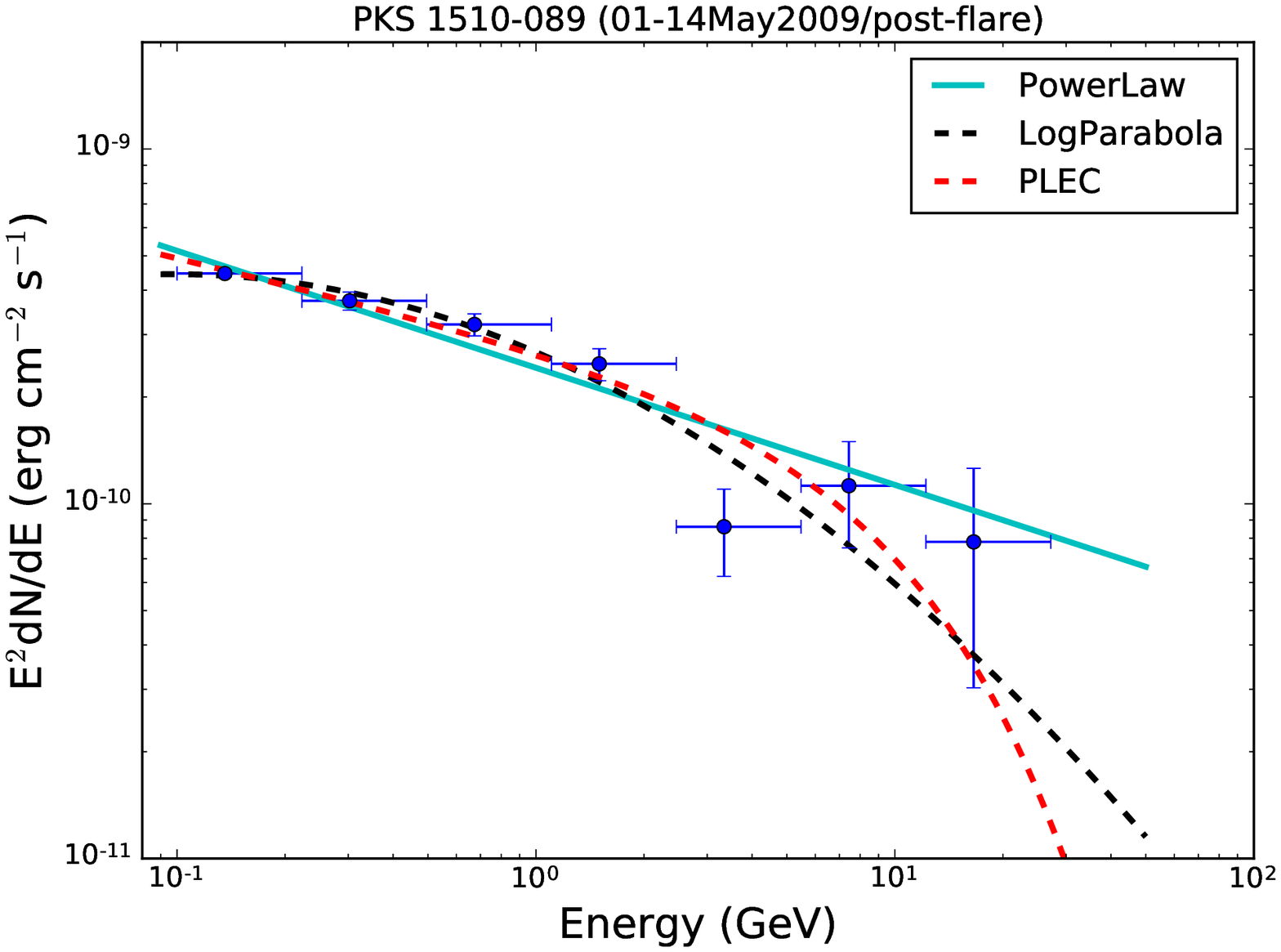}
\end{center}
\caption{Fermi-LAT SEDs during different activity states of flare-1(B) as defined in Fig.3 . PL, LP, PLEC models are 
shown in cyan, black and red color and there respective parameters are given in the Table-8. }
\label{fig:flare-1(B)}
\begin{center}
\includegraphics[scale=0.35]{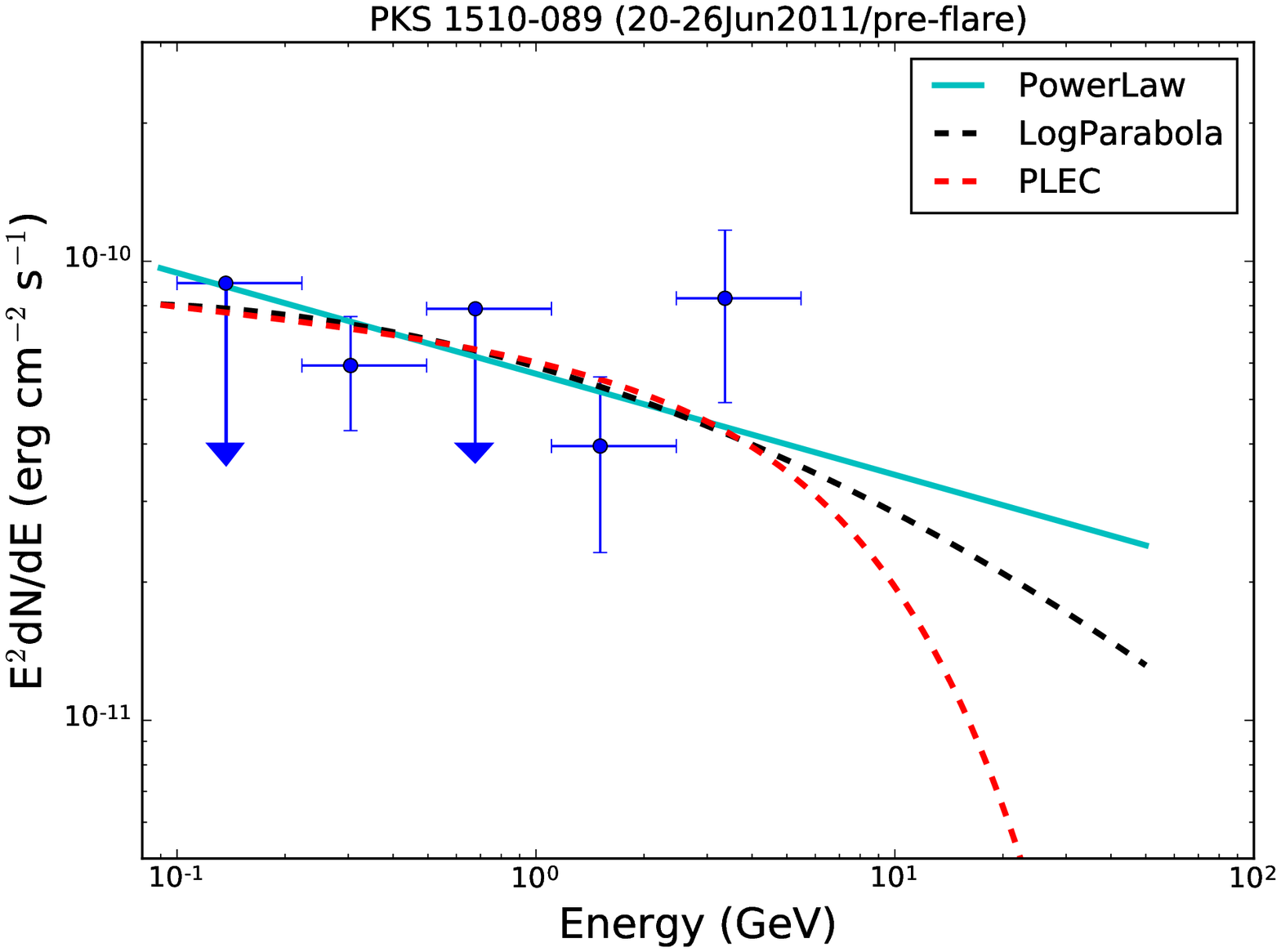}
\includegraphics[scale=0.35]{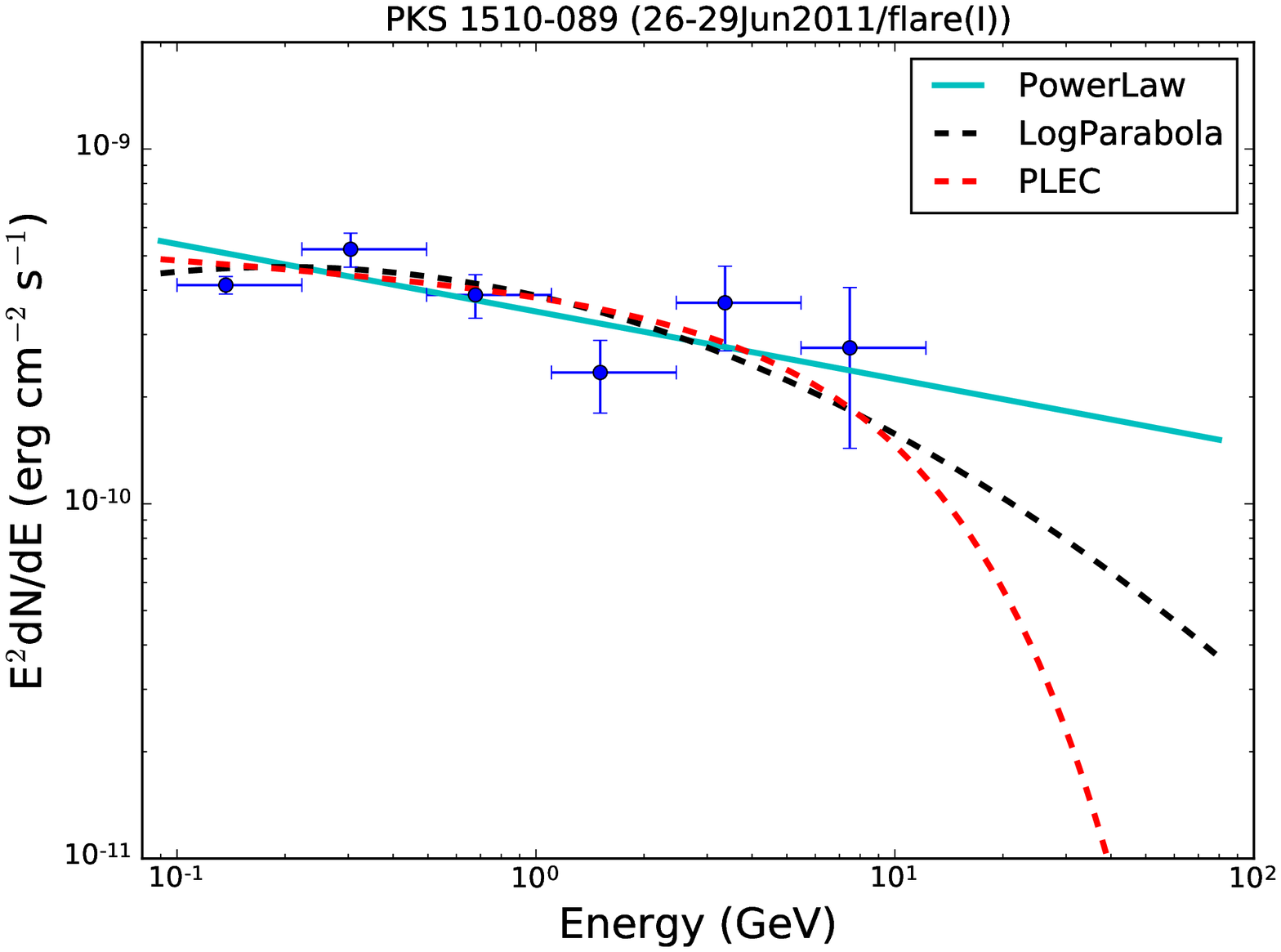}
\end{center}
\begin{center}
\includegraphics[scale=0.35]{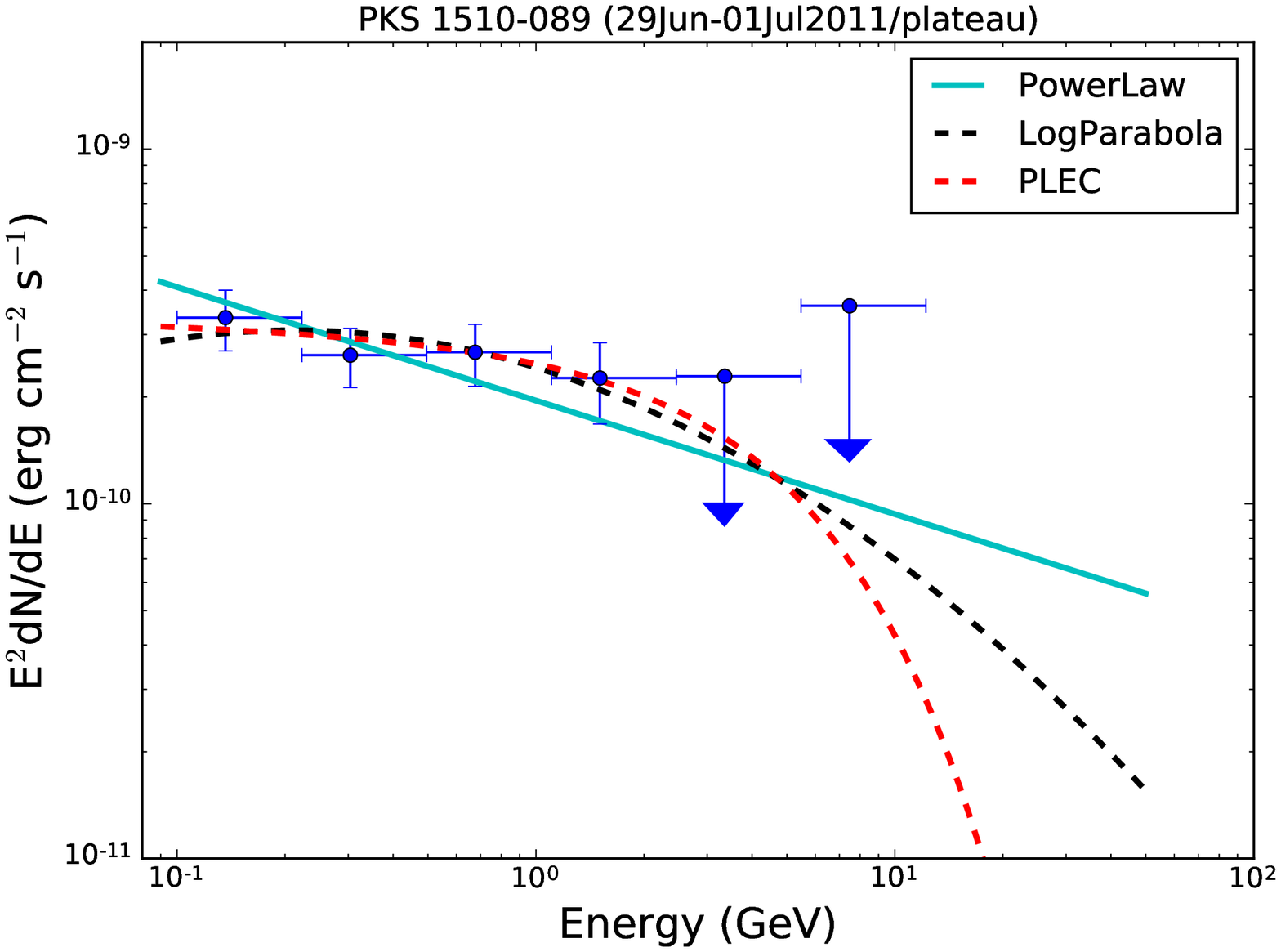}
\includegraphics[scale=0.35]{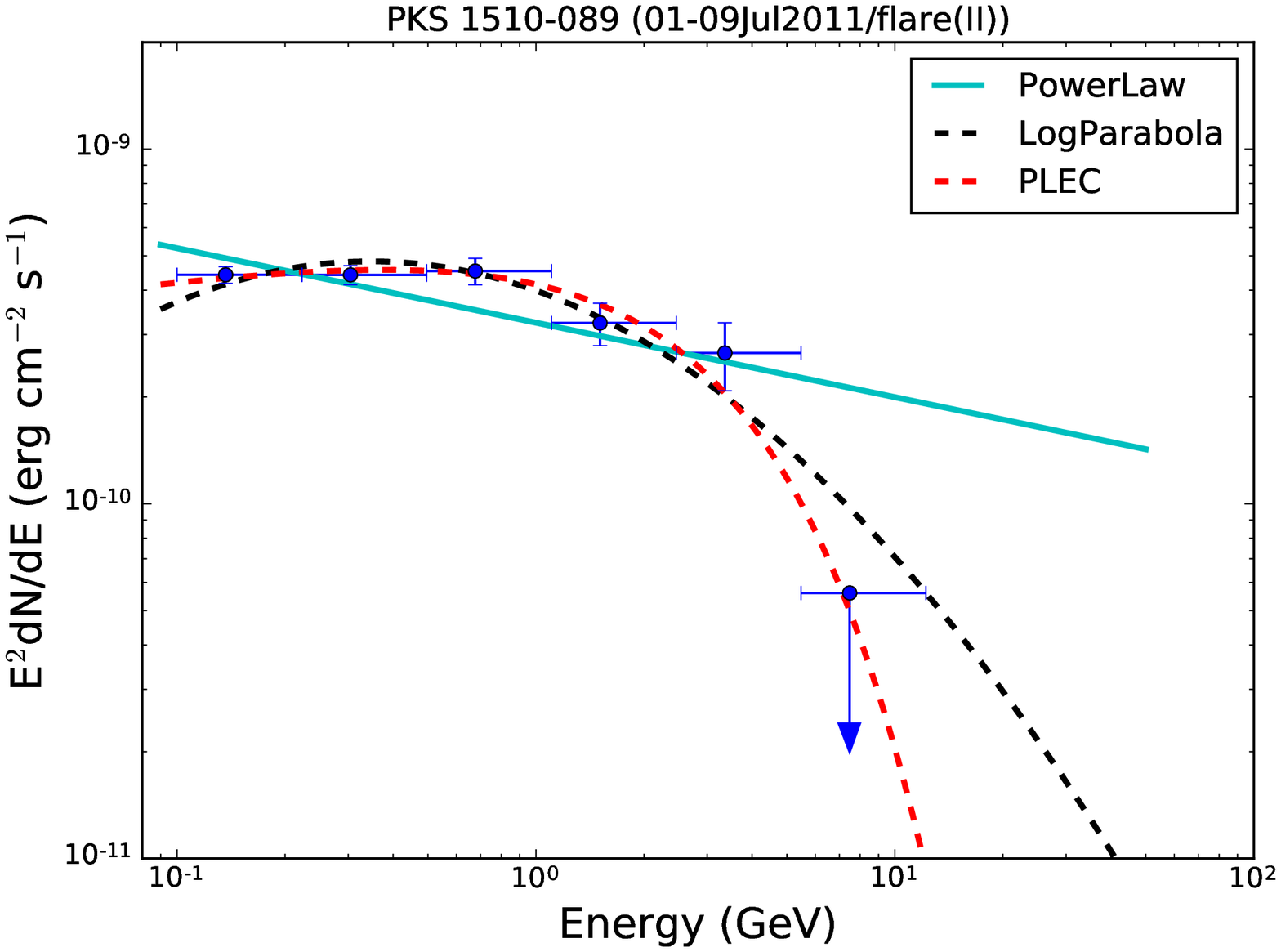}
\end{center}
\begin{center}
\includegraphics[scale=0.35]{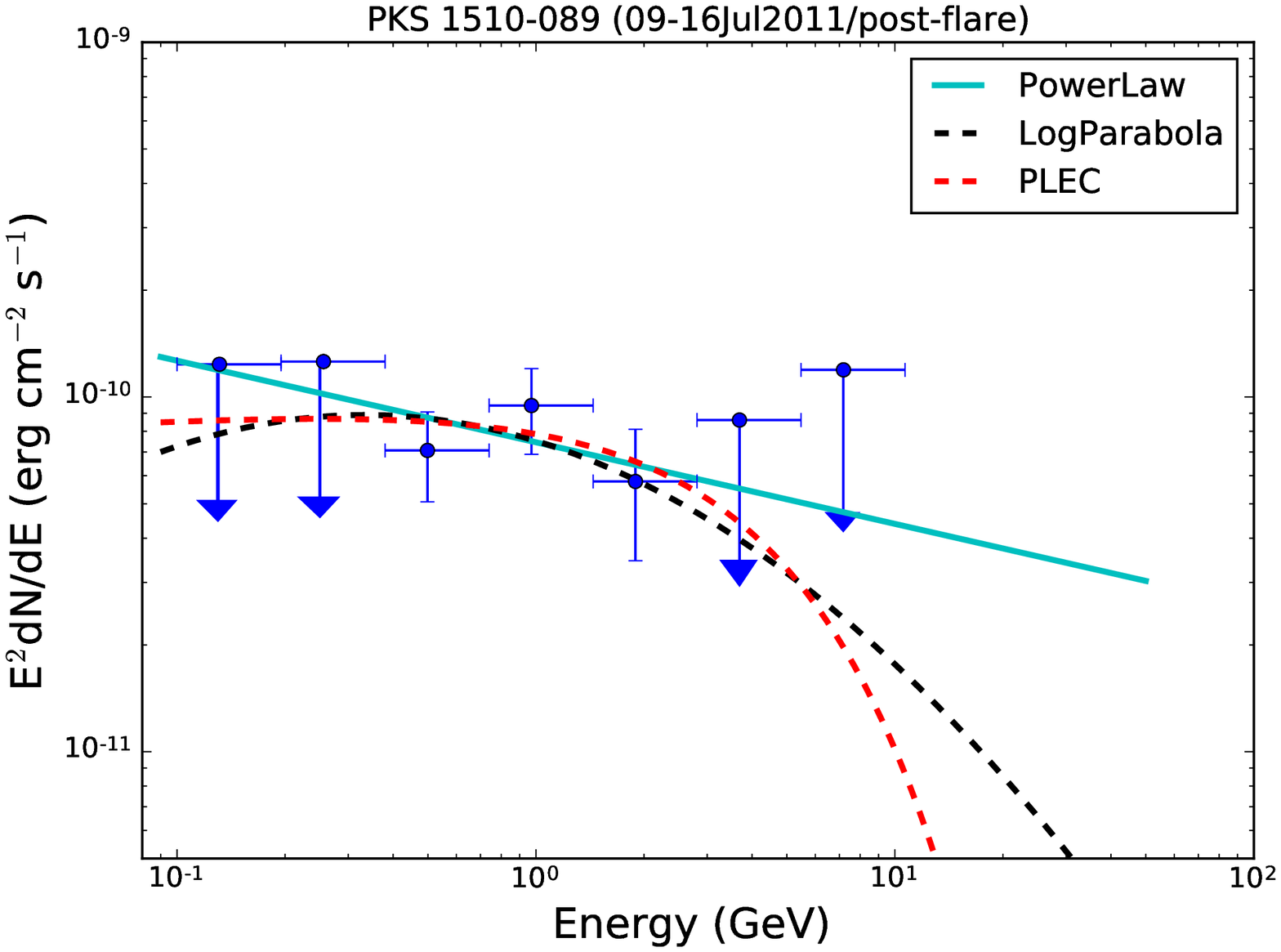} 
\end{center}

\caption{Fermi-LAT SEDs during different activity states of flare-2(A) as defined in Fig.4 . PL, LP, PLEC models are 
shown in cyan, black and red color and there respective parameters are given in the Table-9. }
\label{fig:flare-2(a)}
\end{figure*}

\begin{figure*}[htbp]
\begin{center}
\includegraphics[scale=0.35]{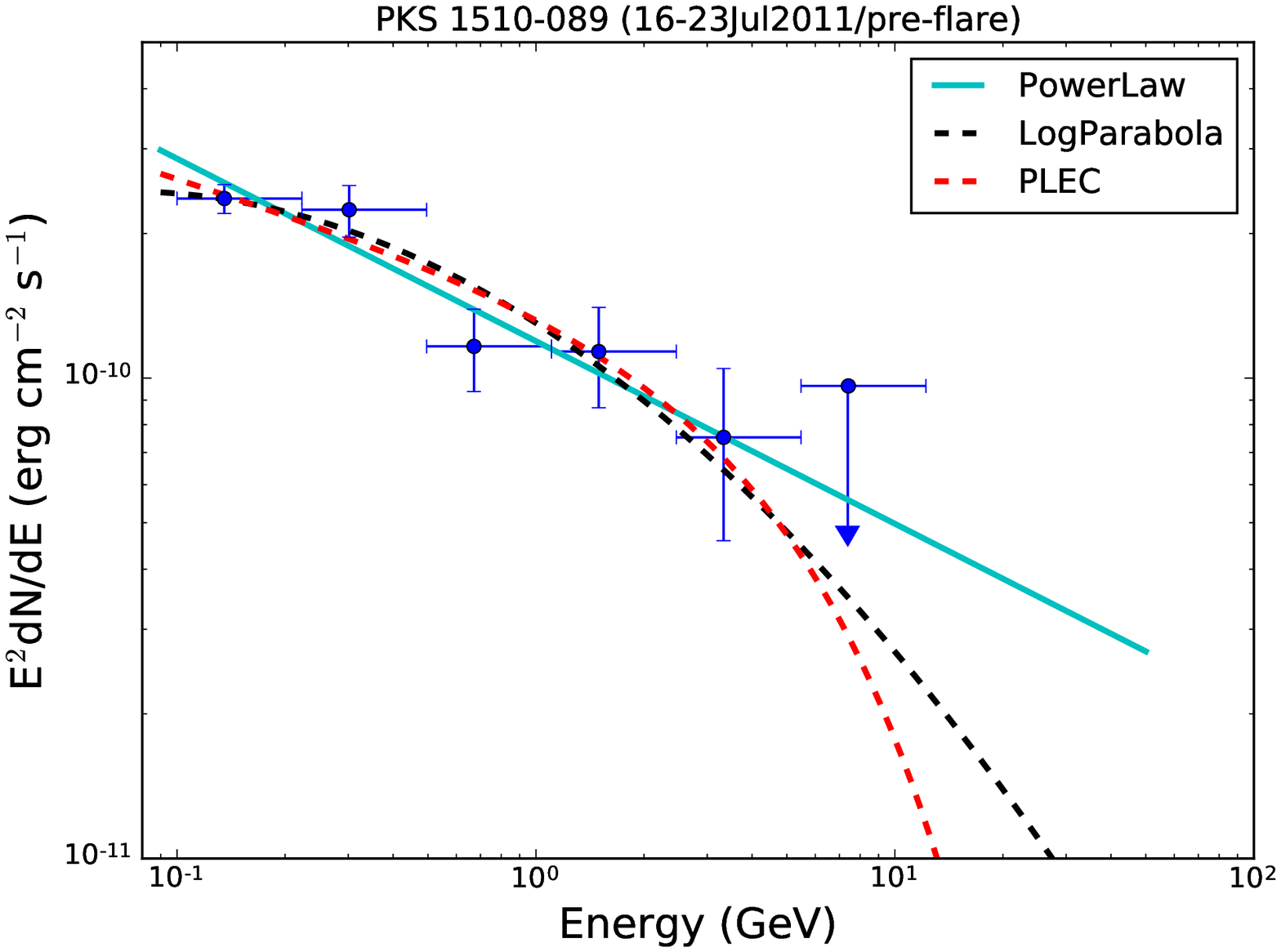}
\includegraphics[scale=0.35]{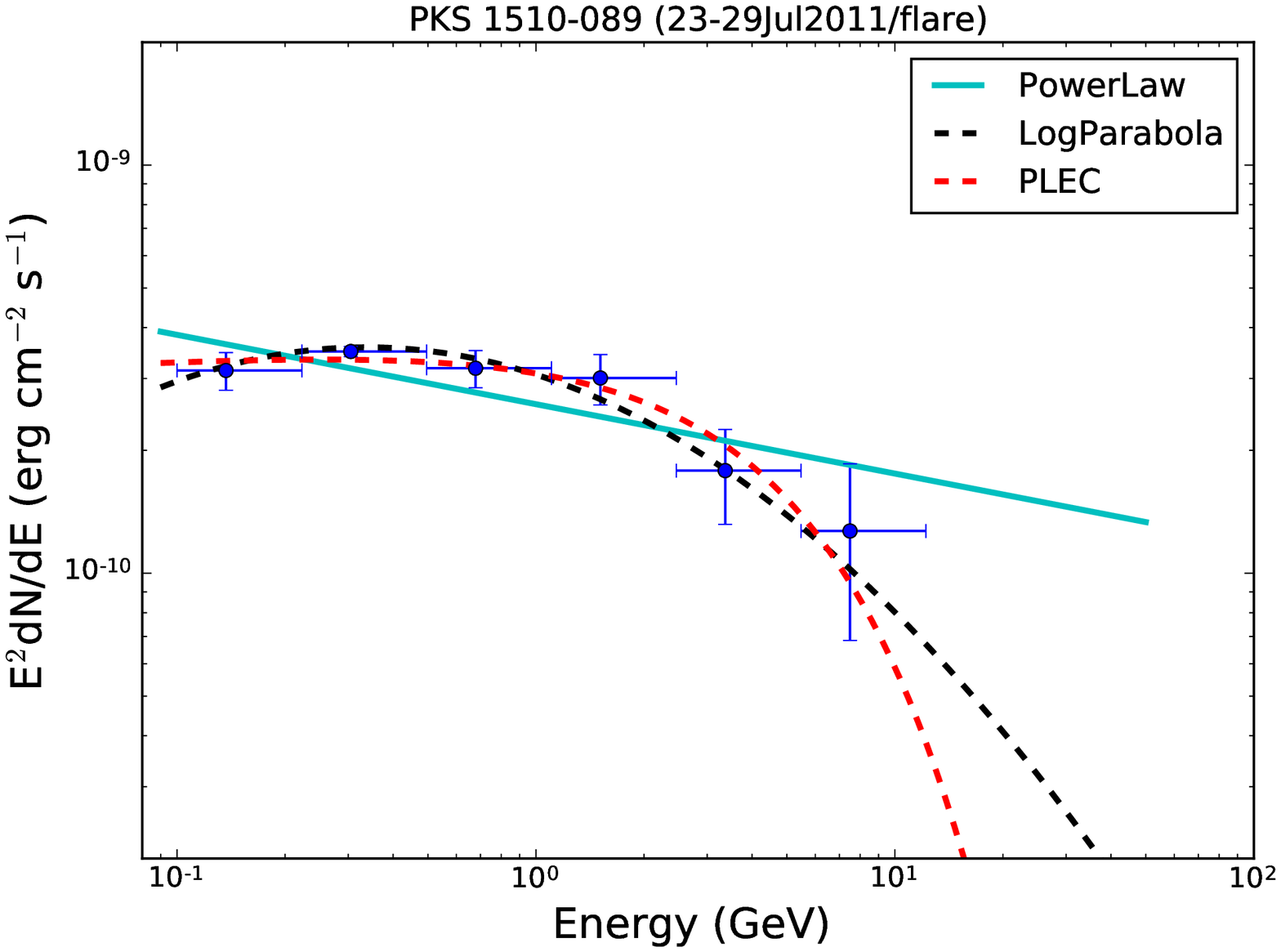}
\end{center}
\begin{center}
\includegraphics[scale=0.35]{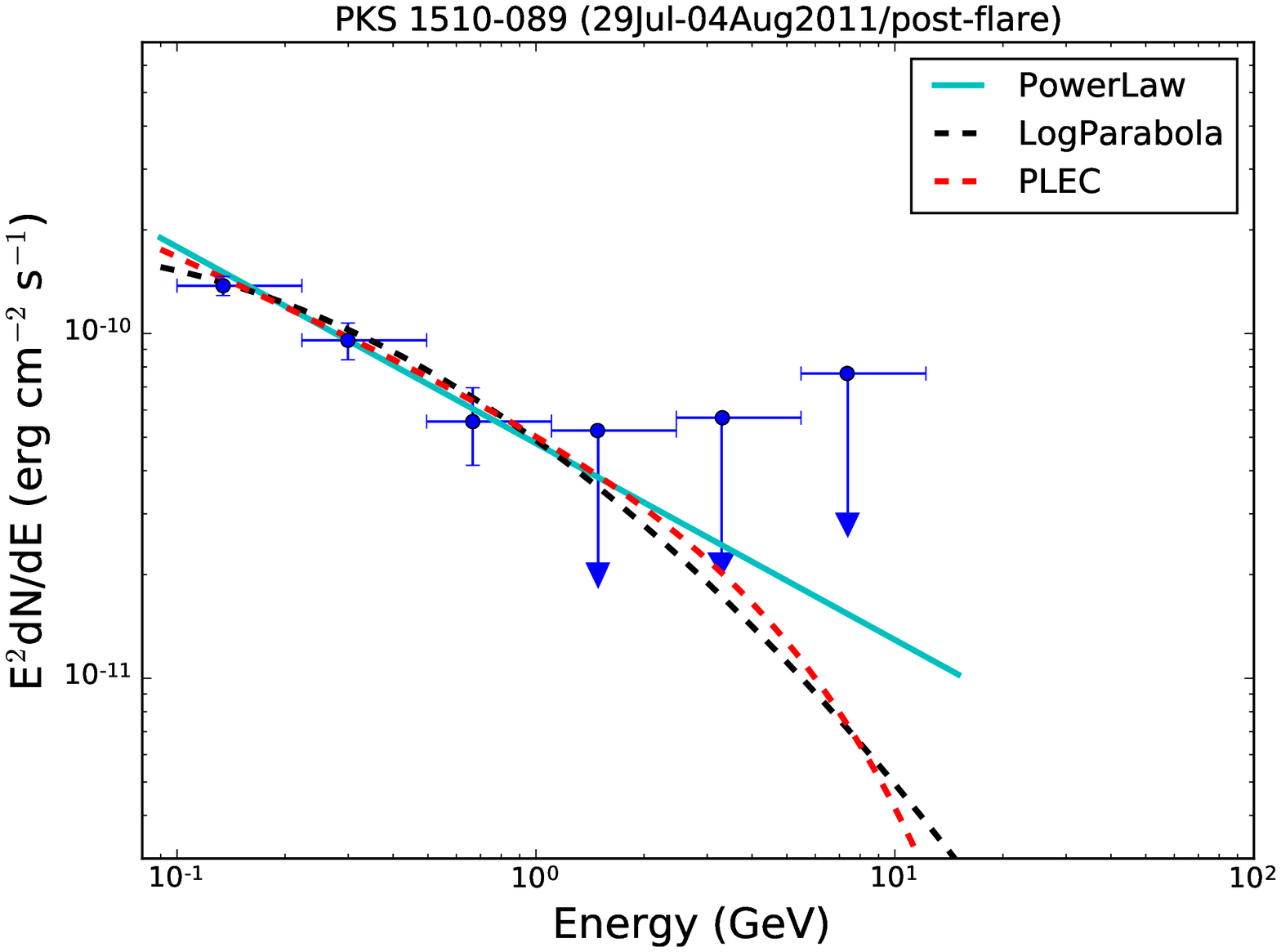}
\end{center}
\caption{Fermi-LAT SEDs during different activity states of flare-2(B) as defined in Fig.5 . PL, LP, PLEC models are
shown in cyan, black and red color and there respective parameters are given in the Table-10.}
\label{fig:flare-2(b)}
\begin{center}
\includegraphics[scale=0.35]{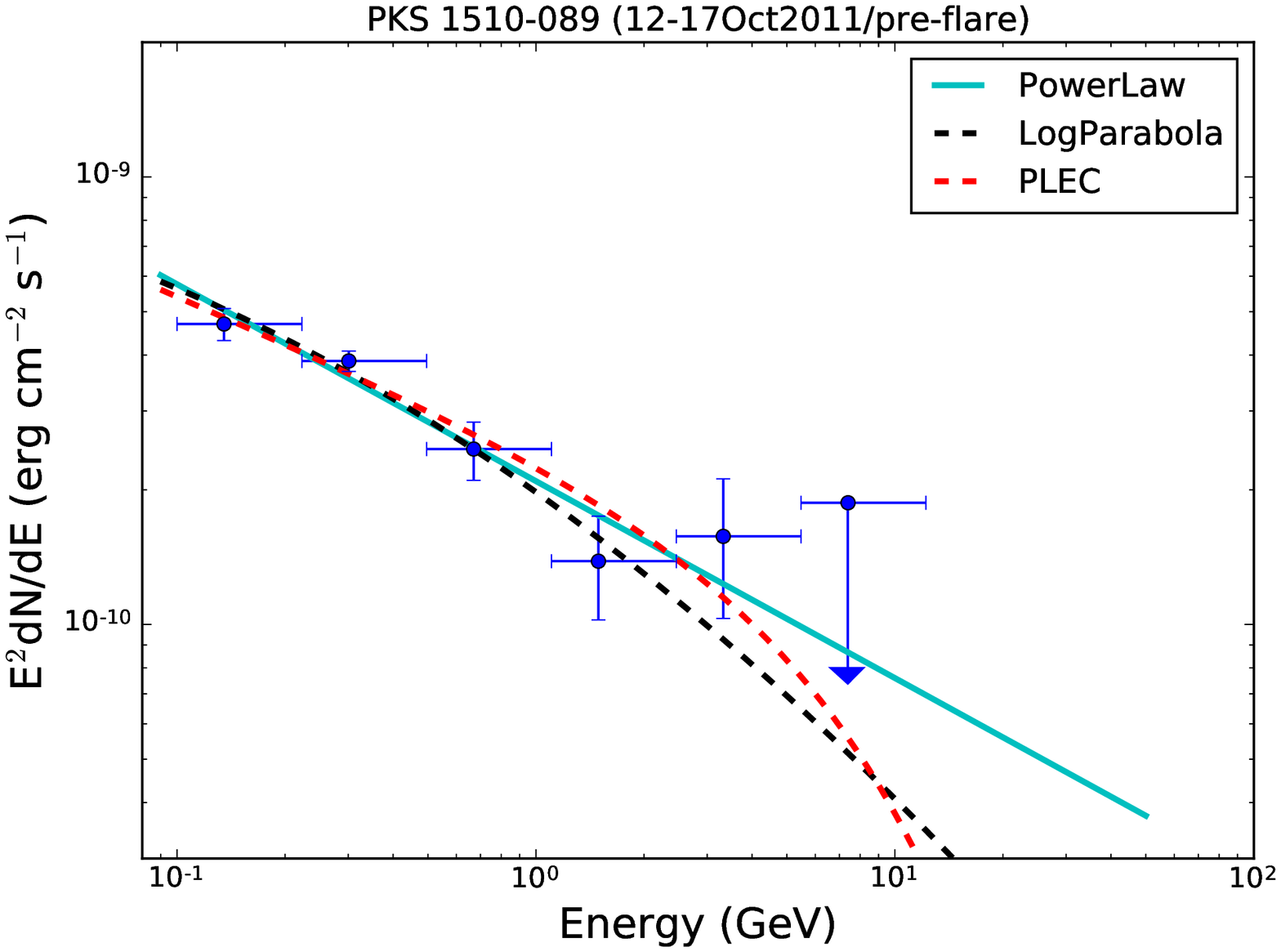}
\includegraphics[scale=0.35]{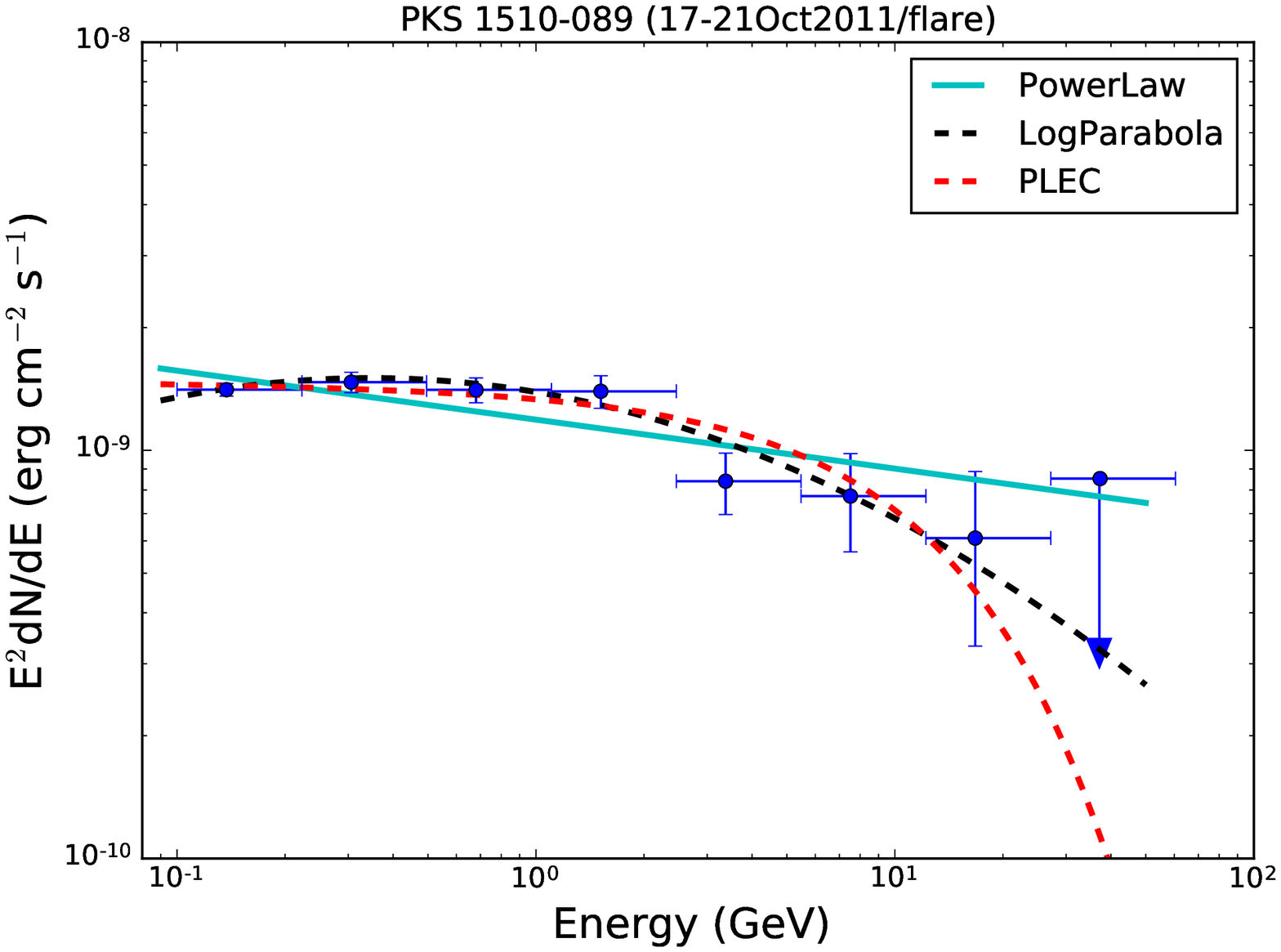}
\end{center}
\begin{center}
\includegraphics[scale=0.35]{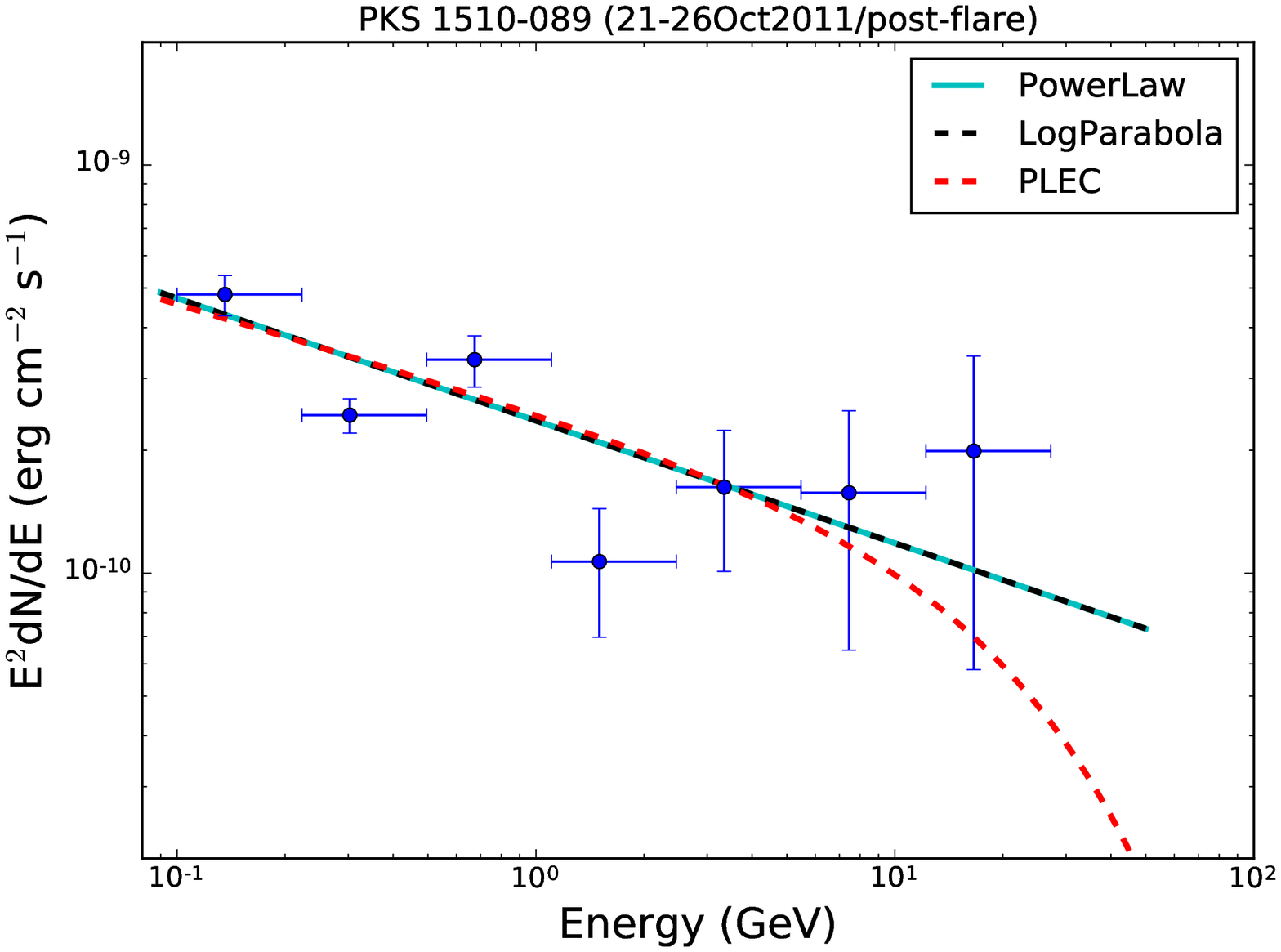}
\end{center}
\caption{Fermi-LAT SEDs during different activity states of flare-2(C) as defined in Fig.6 . PL, LP, PLEC models are 
shown in cyan, black and red color and there respective parameters are given in the Table-11. }
\label{fig:flare-2(c)}
\end{figure*}

\begin{figure*}[htbp]
\begin{center}
\includegraphics[scale=0.35]{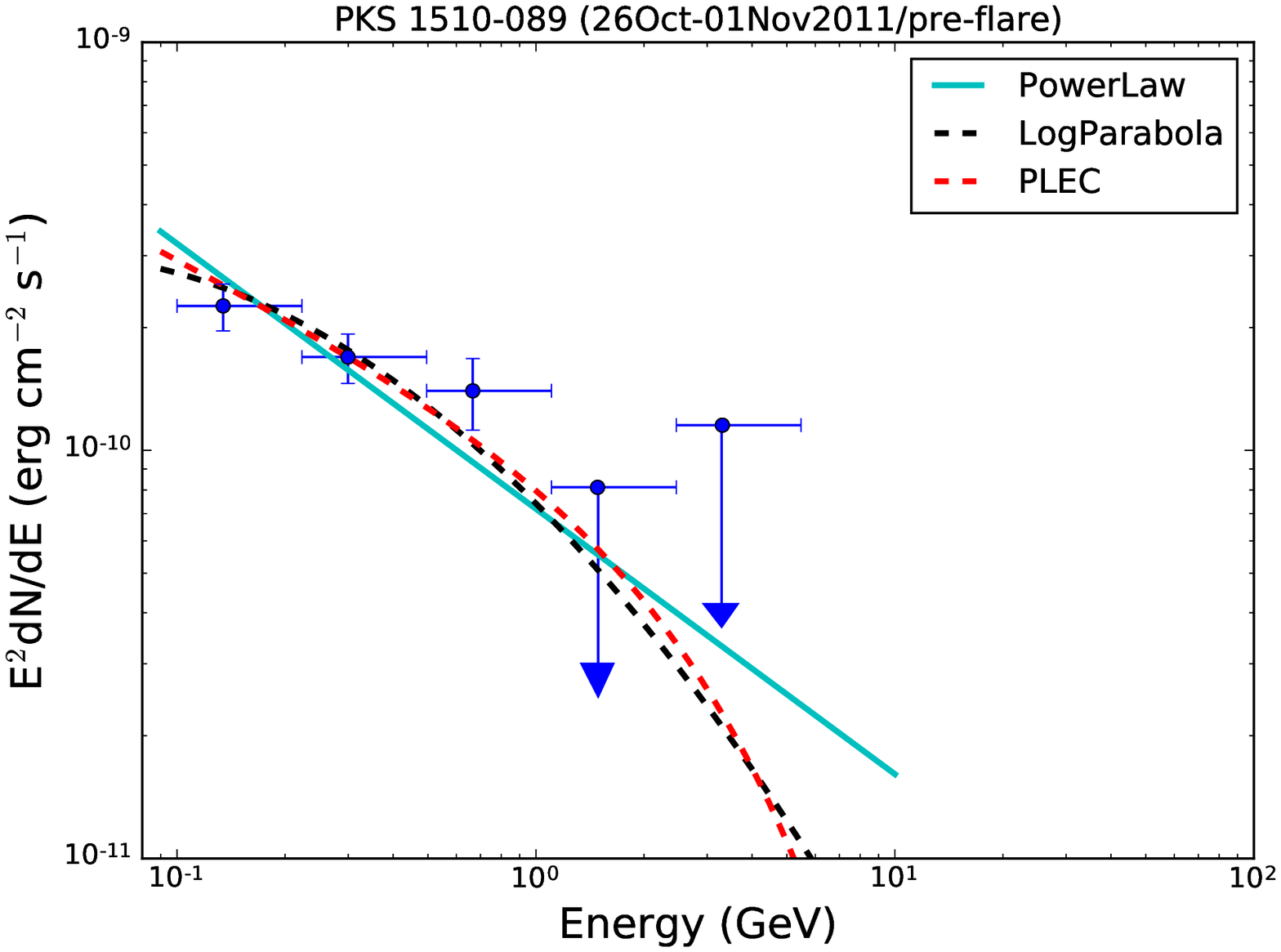}
\includegraphics[scale=0.35]{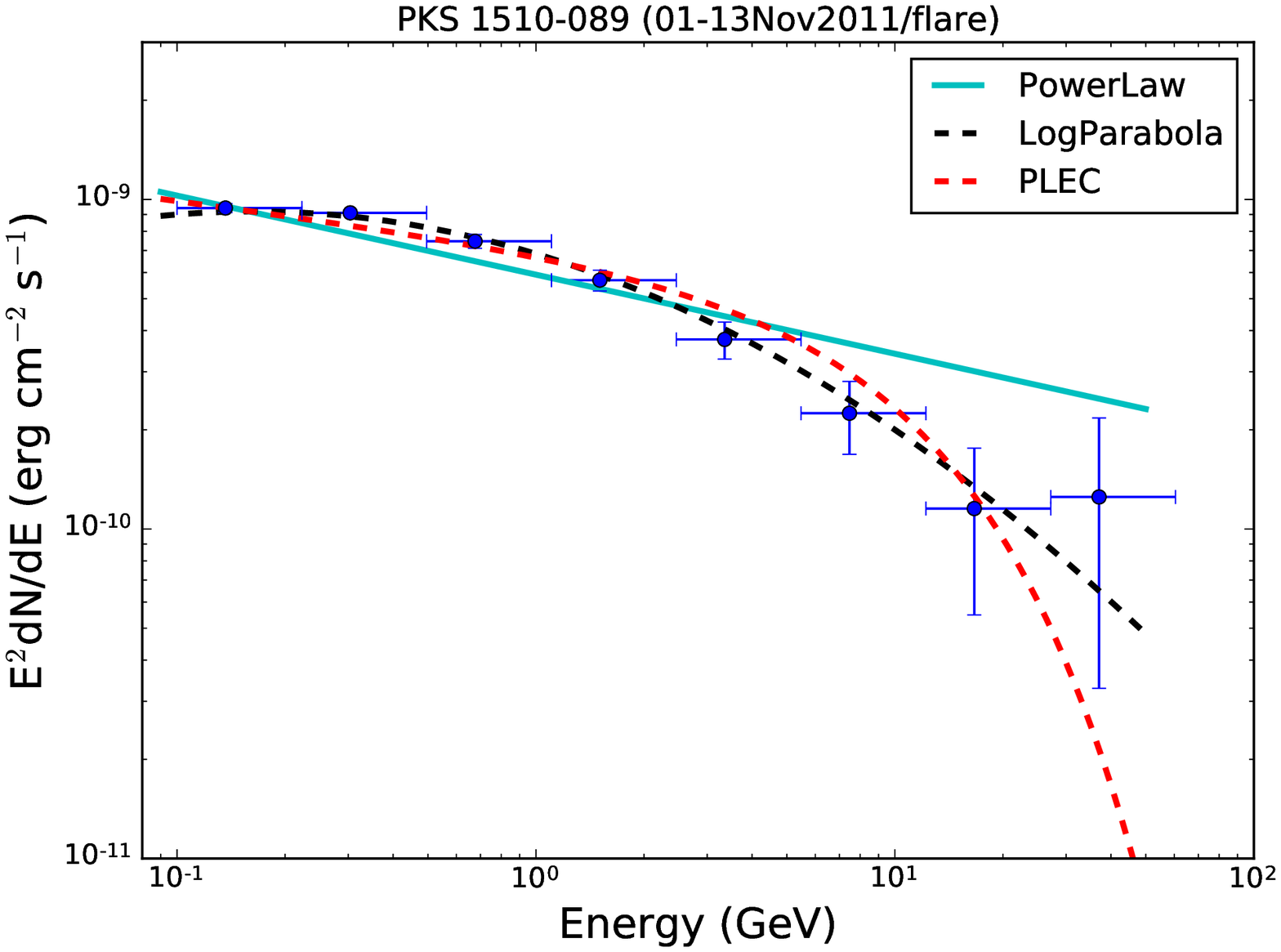}
\end{center}
\begin{center}
\includegraphics[scale=0.35]{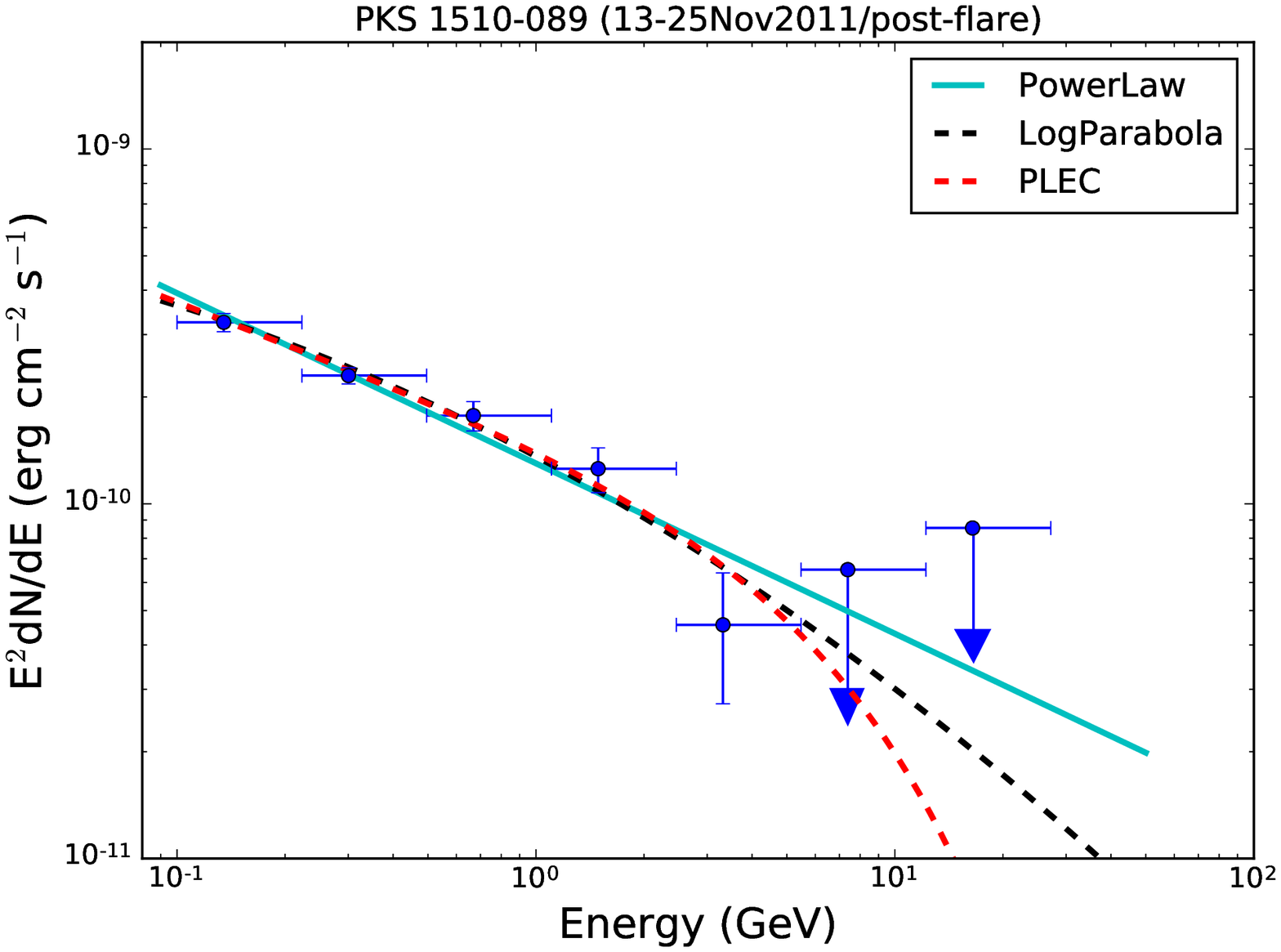}
\end{center}
\caption{Fermi-LAT SEDs during different activity states of flare-2(D) as defined in Fig.7 . PL, LP, PLEC models are 
shown in cyan, black and red color and there respective parameters are given in the Table-12. }
\label{fig:flare-2(d)}
\begin{center}
\includegraphics[scale=0.35]{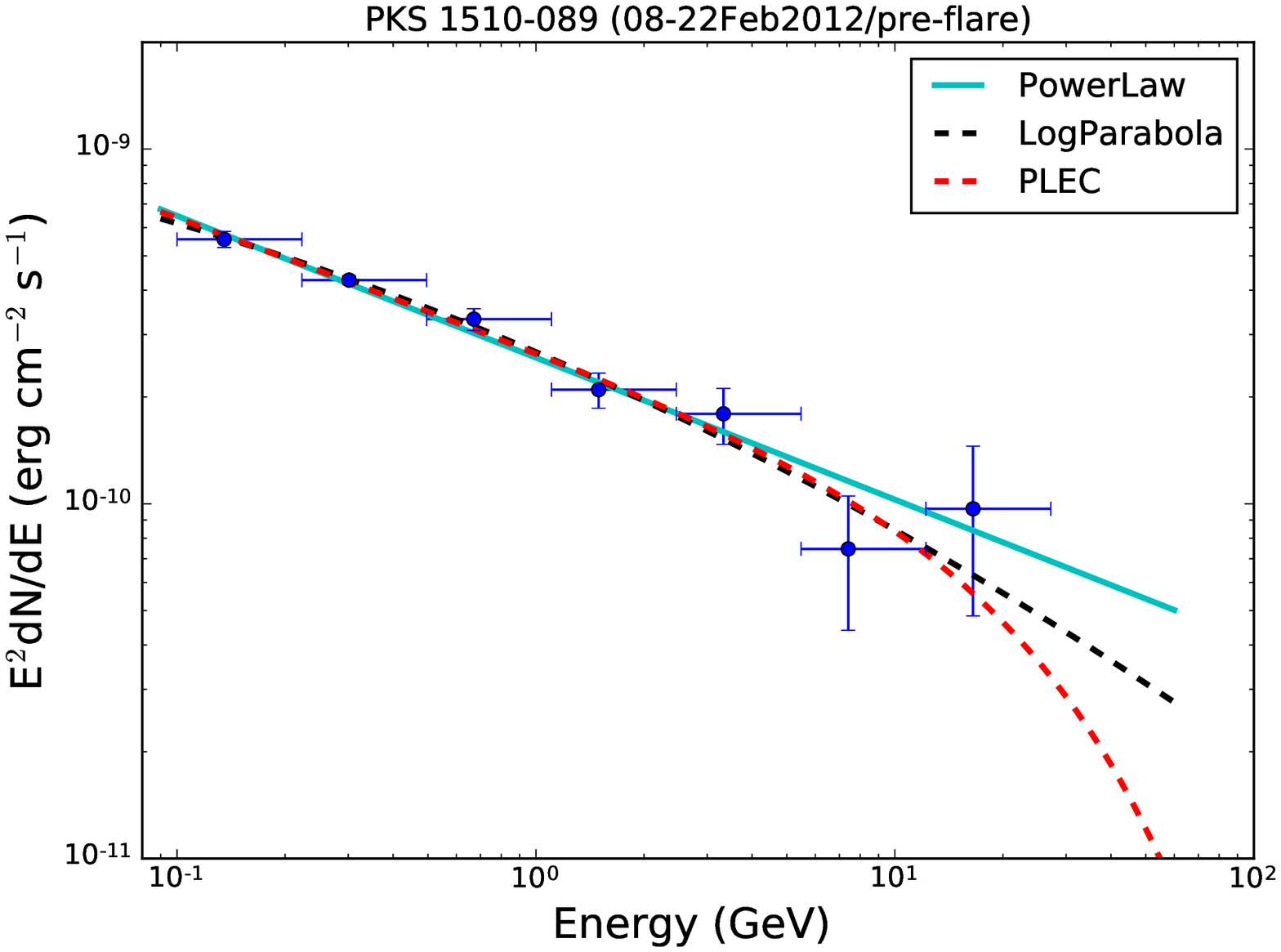}
\includegraphics[scale=0.35]{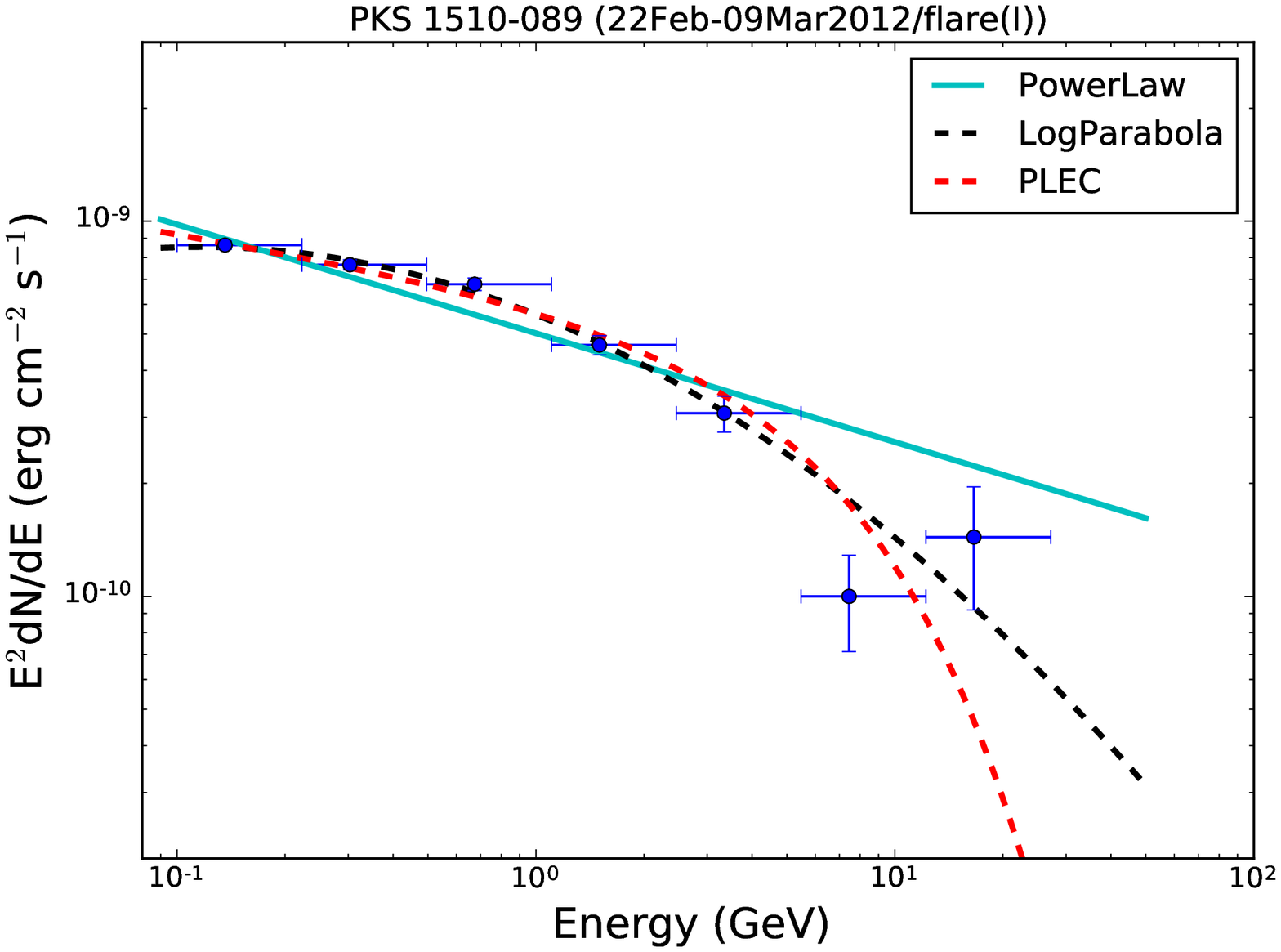}
\end{center}

\begin{center}
\includegraphics[scale=0.35]{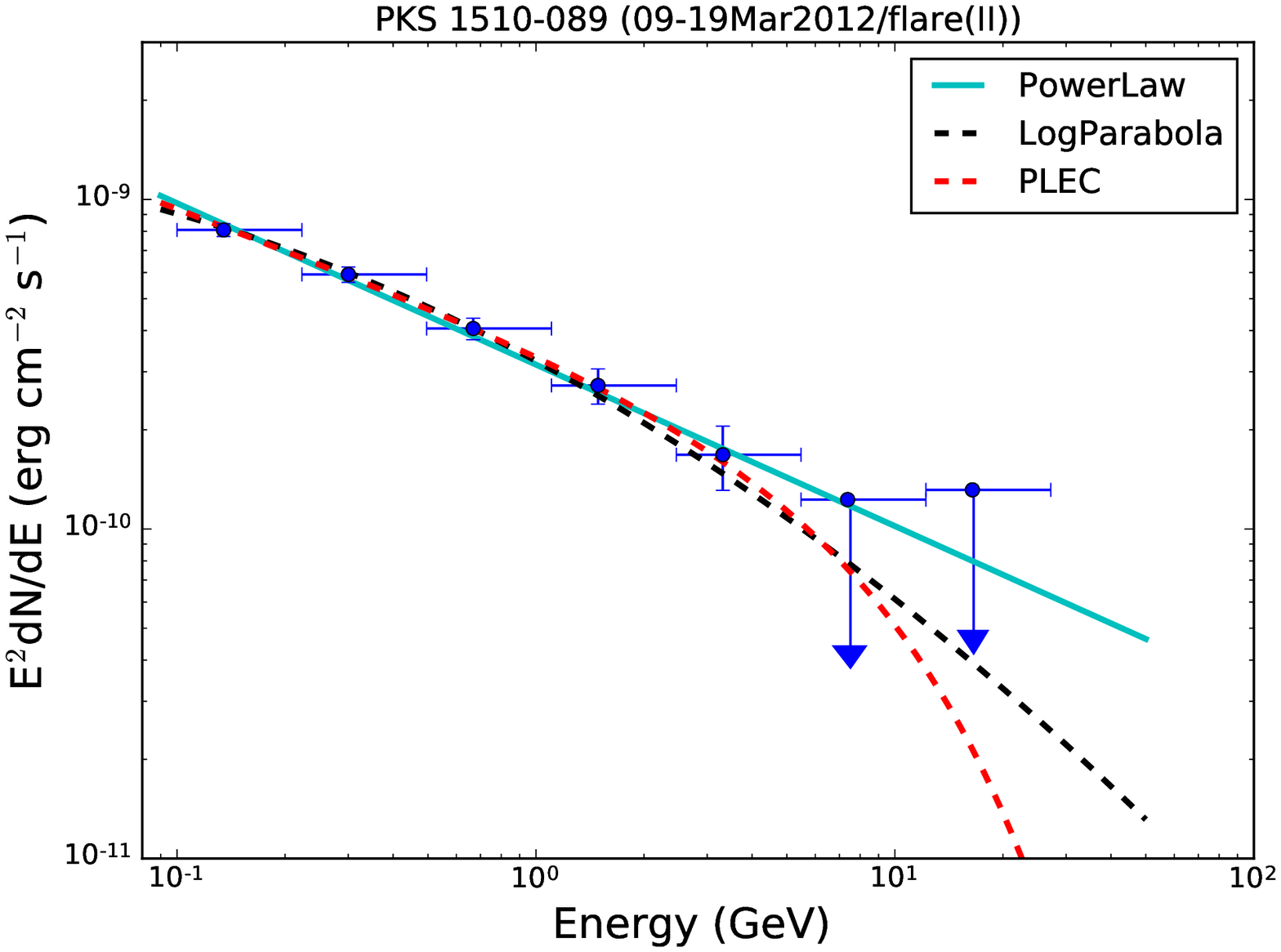}
\includegraphics[scale=0.35]{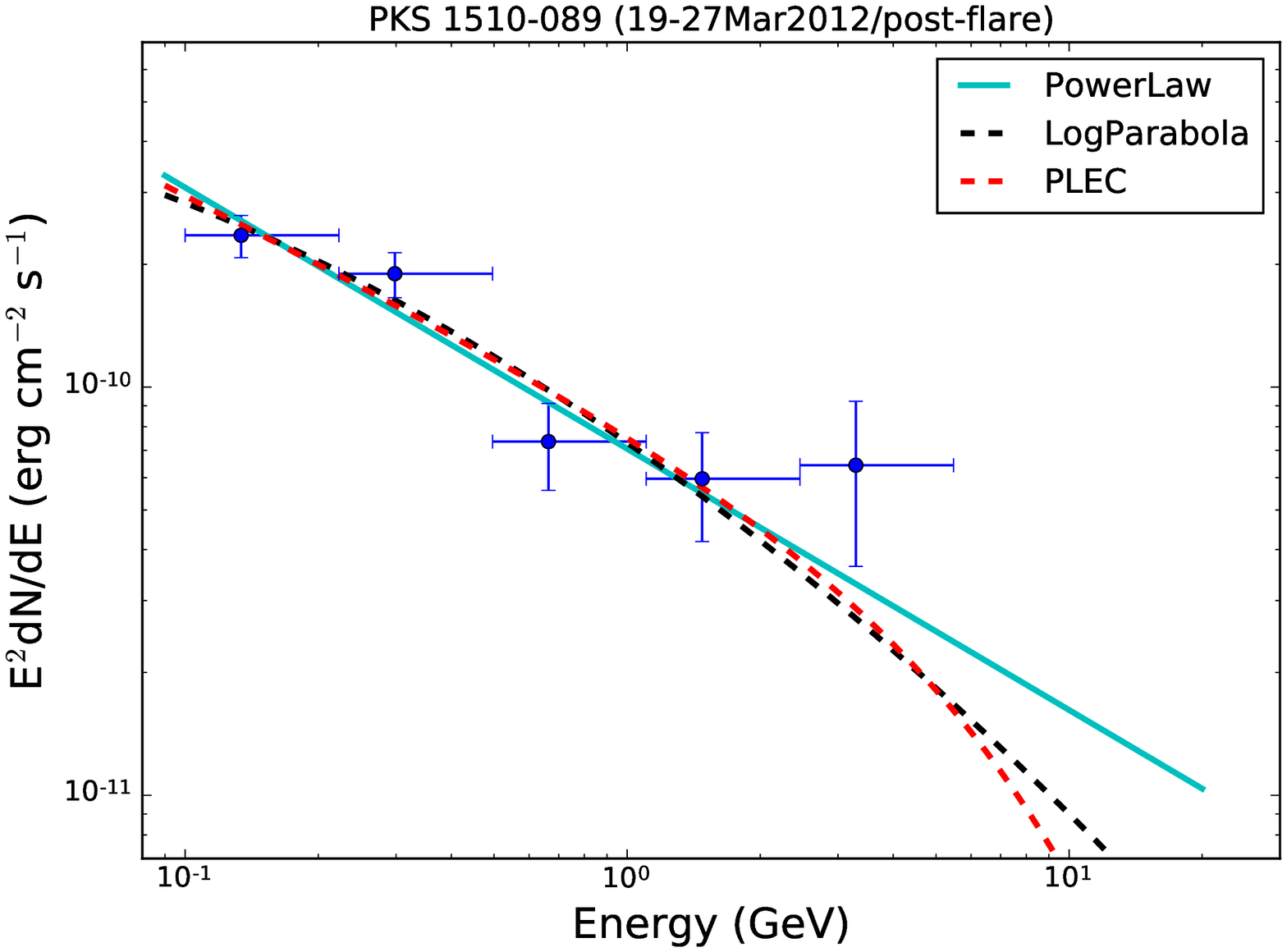}
\end{center}
\caption{Fermi-LAT SEDs during different activity states of flare-2(E) as defined in Fig.8 . PL, LP, PLEC models are 
shown in cyan, black and red color and there respective parameters are given in the Table-13. }
\label{fig:flare-2(e)}
\end{figure*}

\begin{figure*}[htbp]
\begin{center}
\includegraphics[scale=0.35]{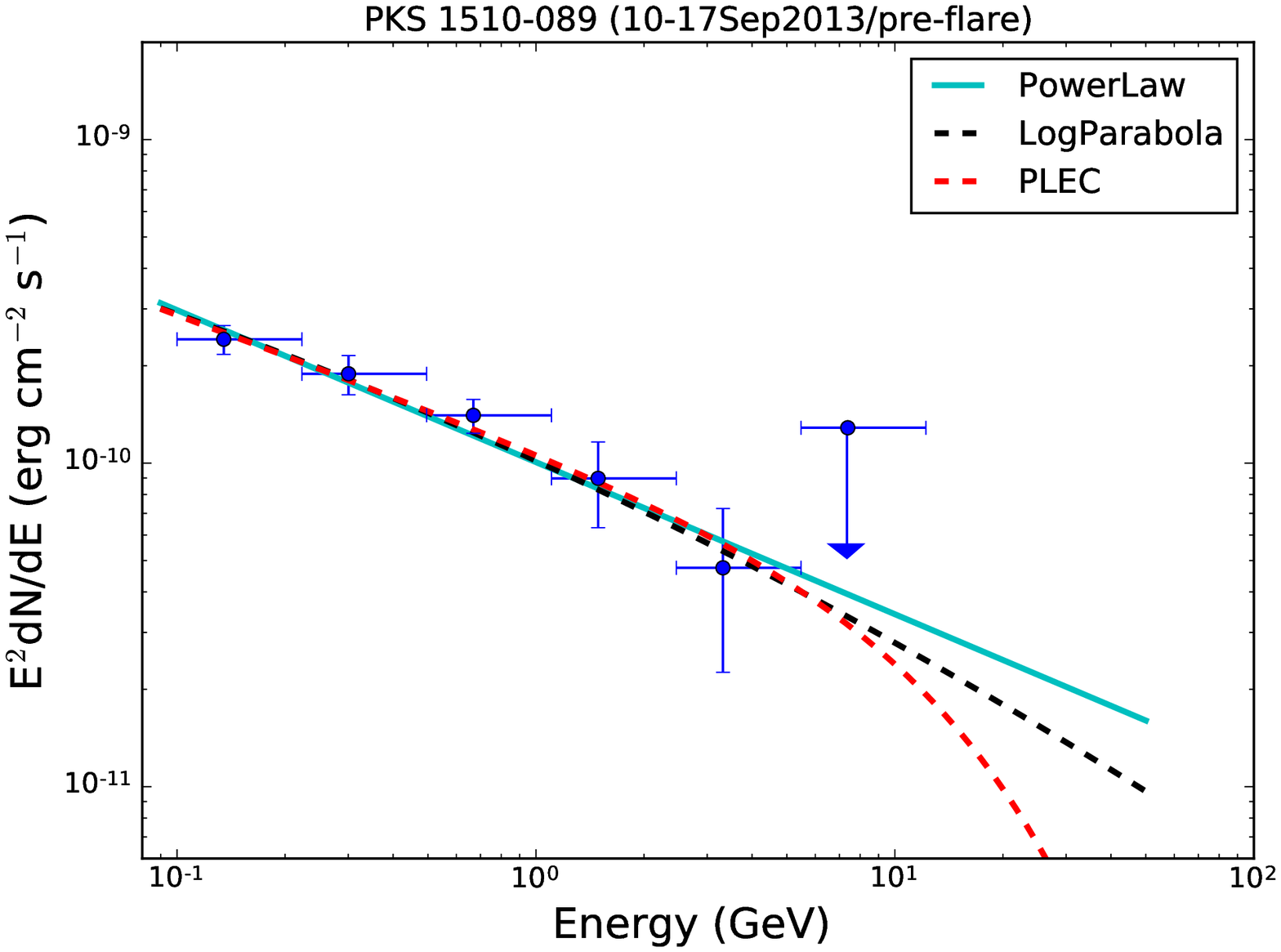}
\includegraphics[scale=0.35]{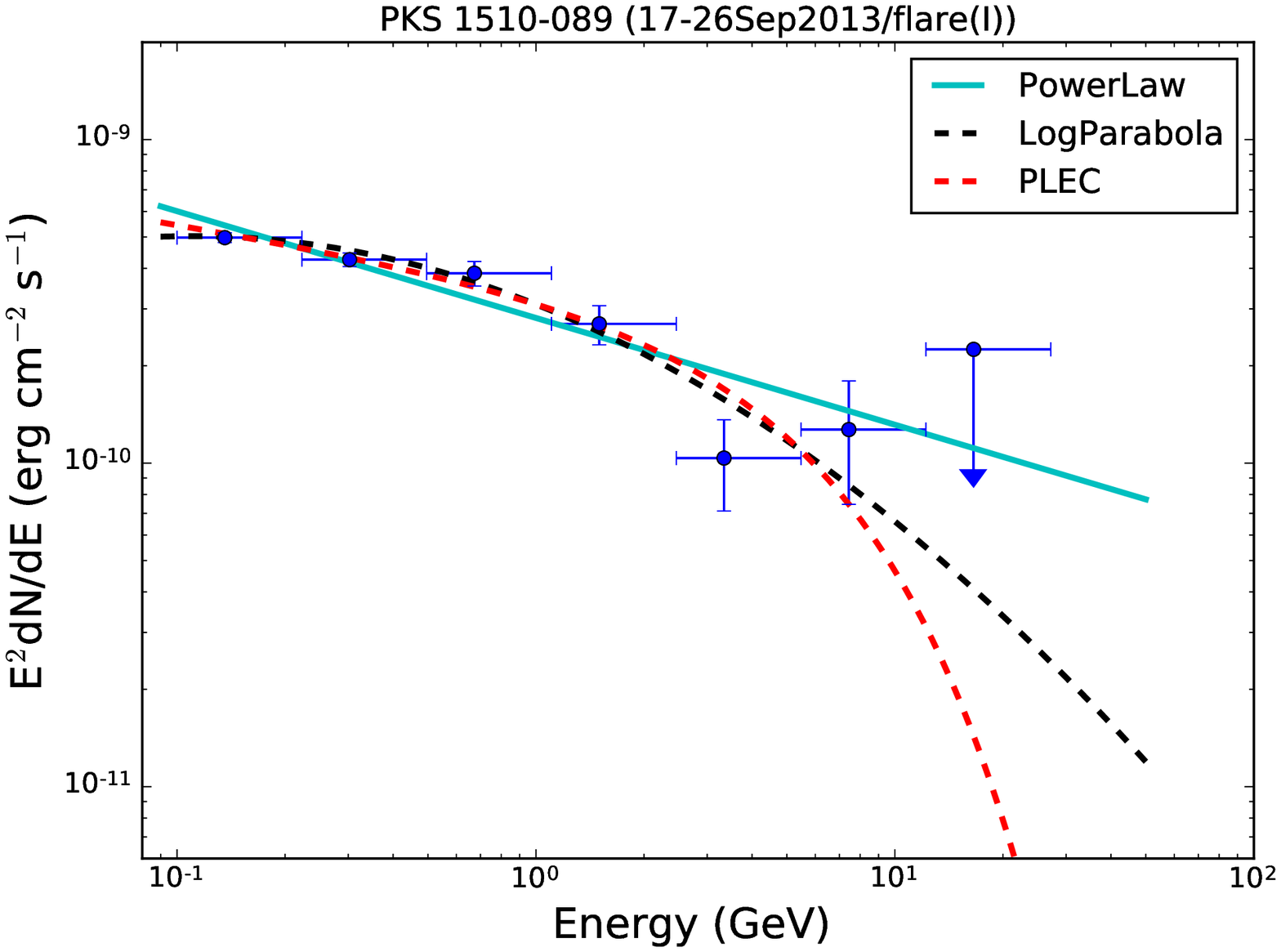}
\end{center}
\begin{center}
\includegraphics[scale=0.35]{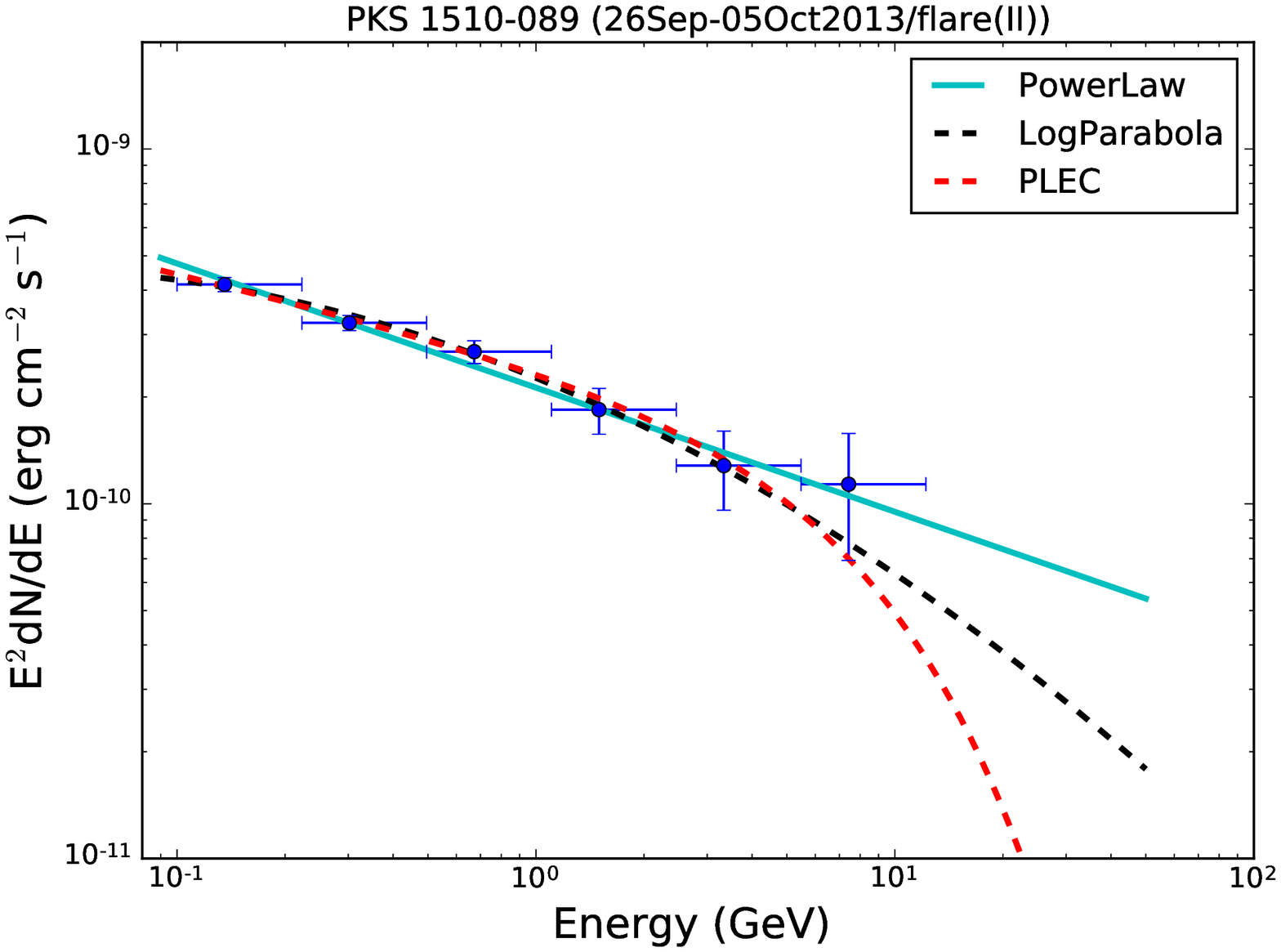}
\includegraphics[scale=0.35]{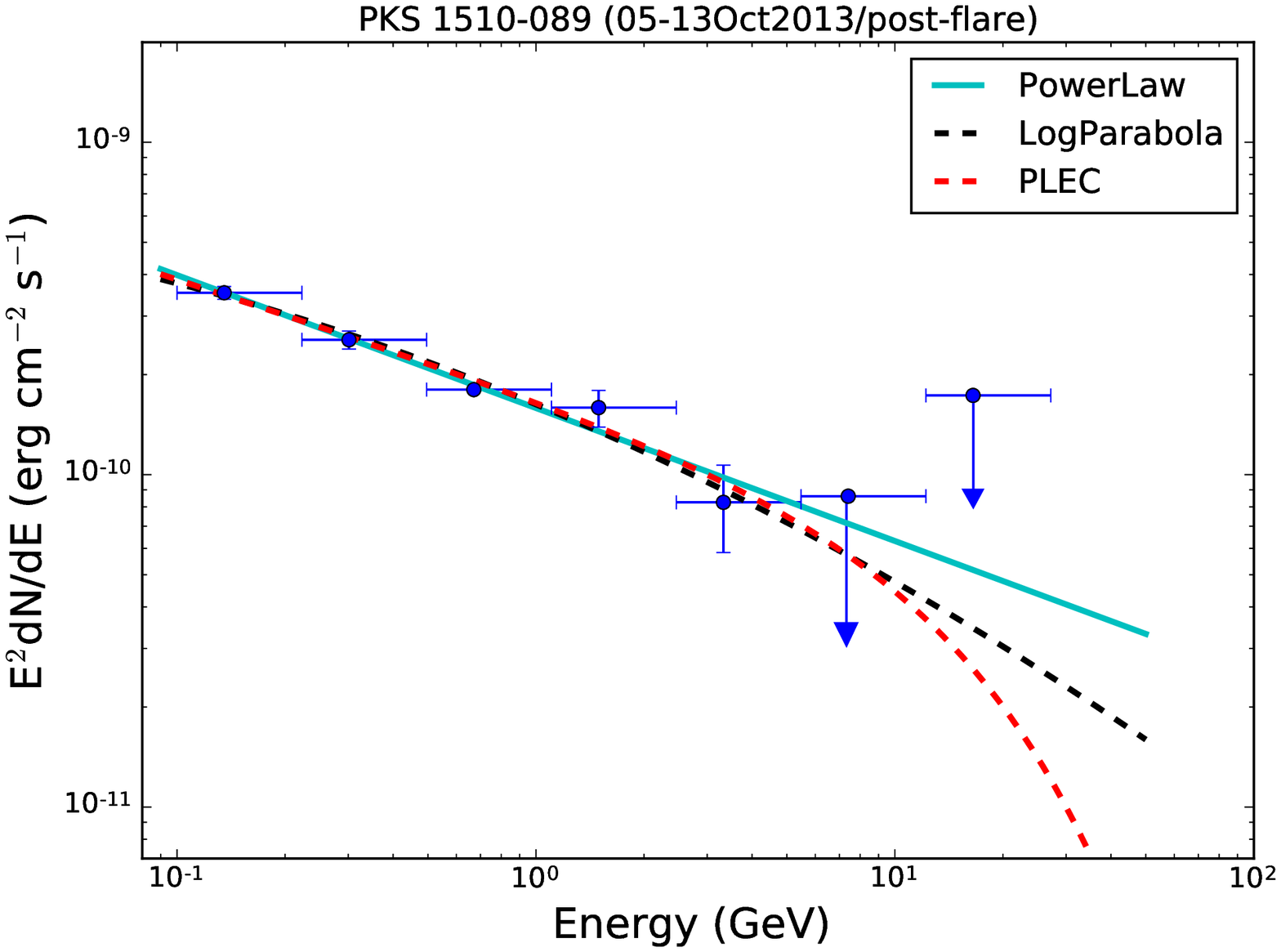}
\end{center}
\caption{Fermi-LAT SEDs during different activity states of Flare-3 as defined in Fig.9 . PL, LP, PLEC models are 
shown in cyan, black and red color and there respective parameters are given in the Table-14. }
\label{fig:flare-3}
\begin{center}
\includegraphics[scale=0.35]{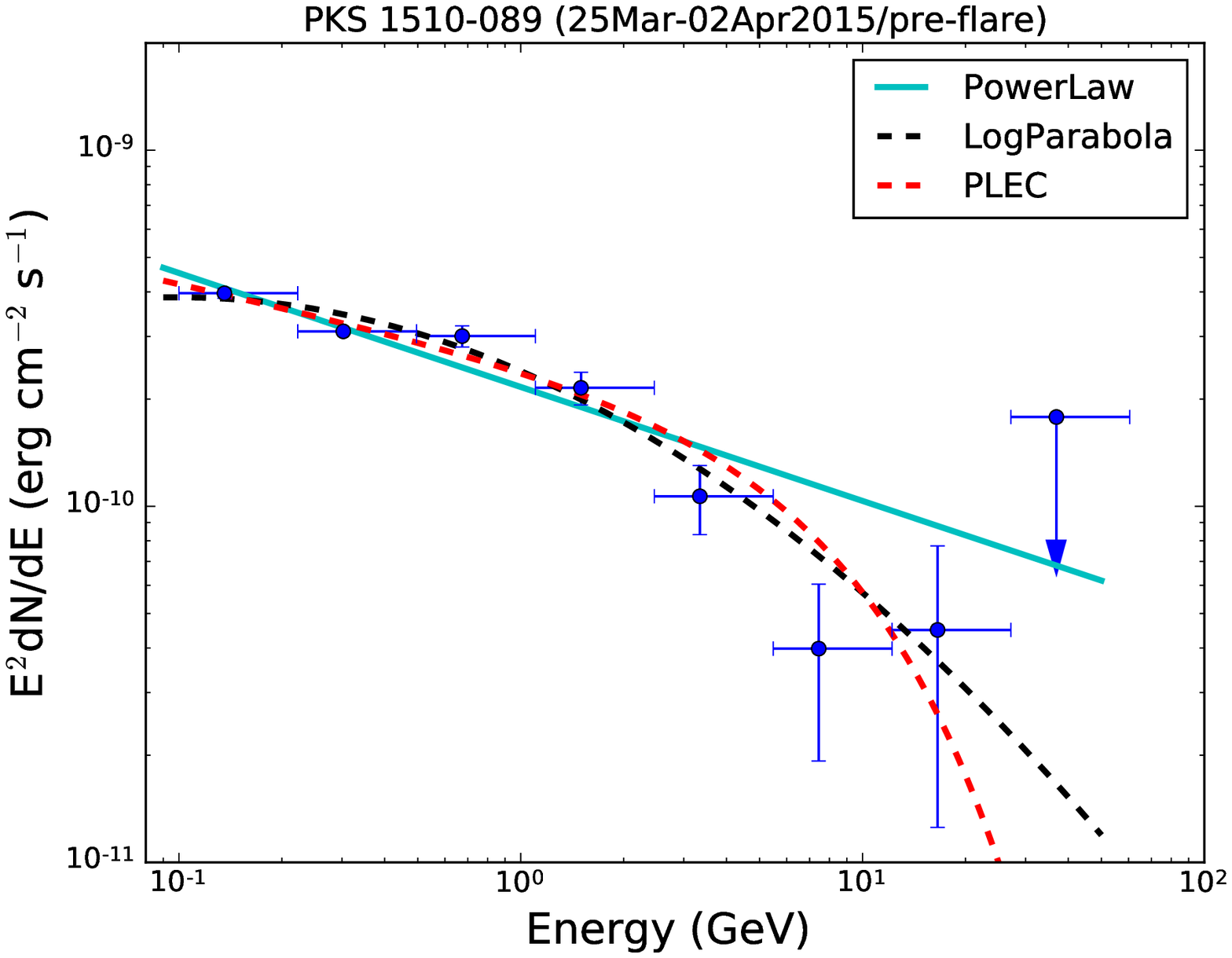}
\includegraphics[scale=0.35]{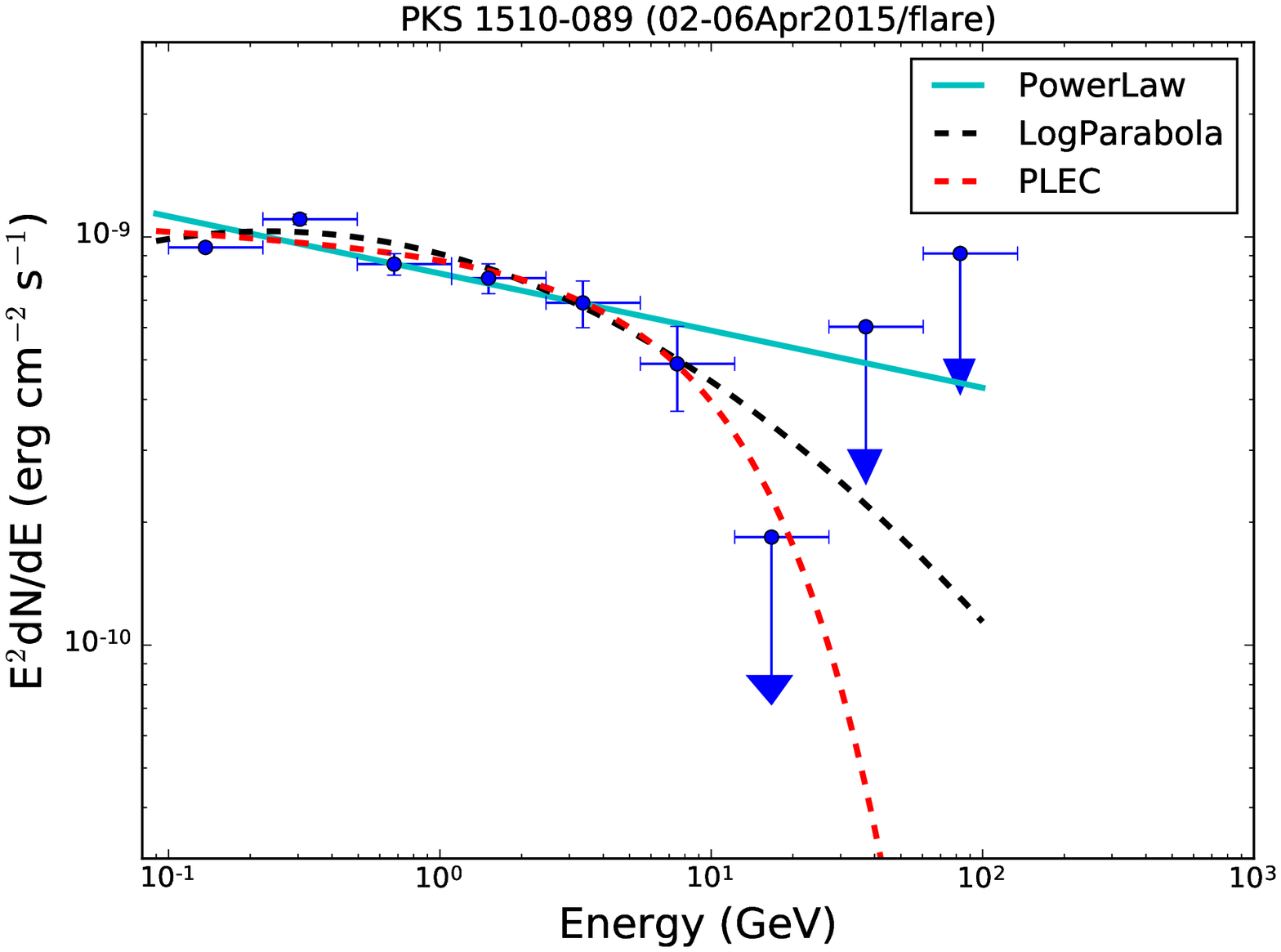}
\end{center}
\begin{center}
\includegraphics[scale=0.35]{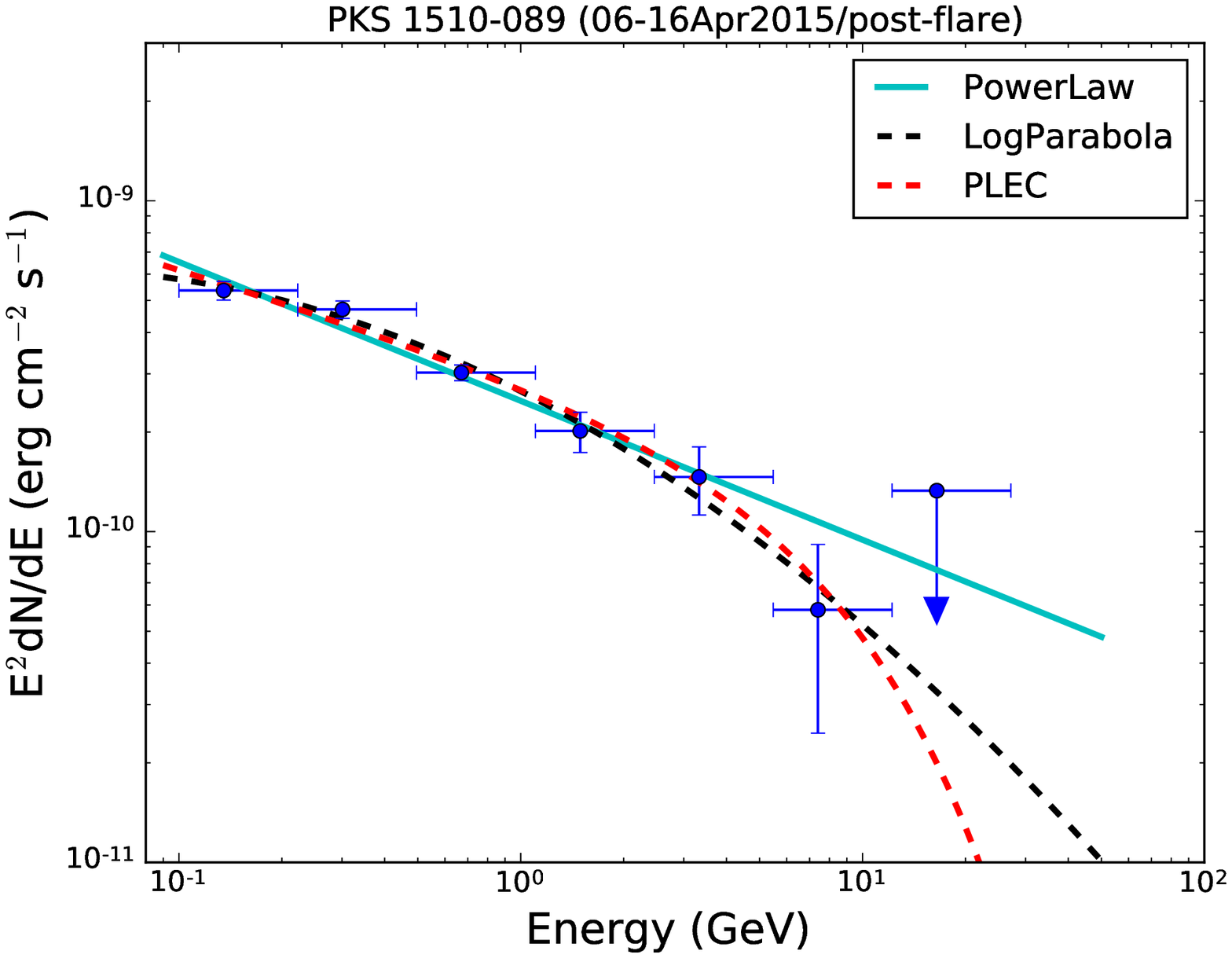}
\end{center}
\caption{Fermi-LAT SEDs during different activity states of flare-4(A) as defined in Fig.10 . PL, LP, PLEC models are
shown in cyan, black and red color and there respective parameters are given in the Table-15. }
\label{fig:flare-4(a)}
\end{figure*}

\begin{figure*}[htbp]
\begin{center}
\includegraphics[scale=0.35]{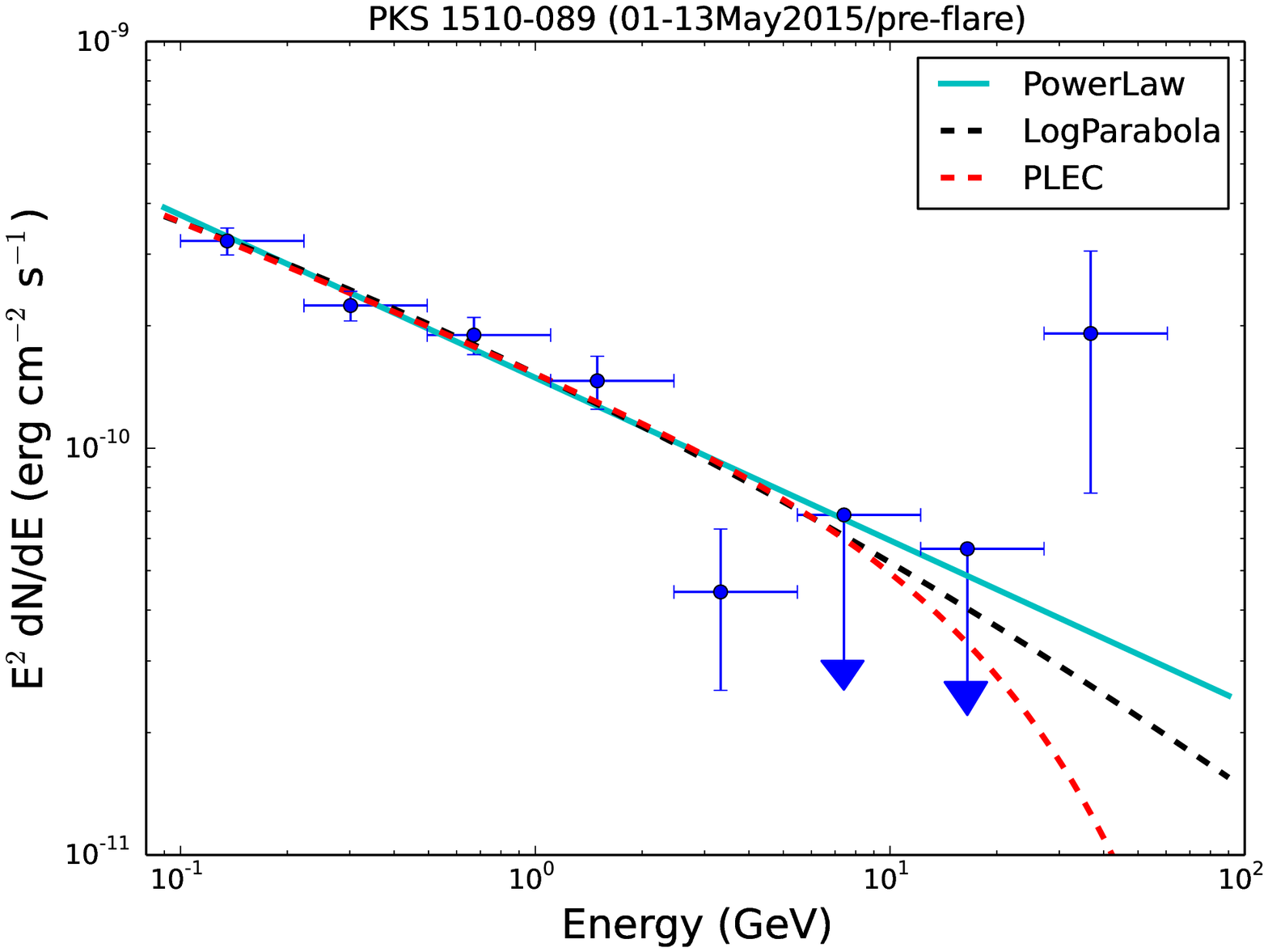}
\includegraphics[scale=0.35]{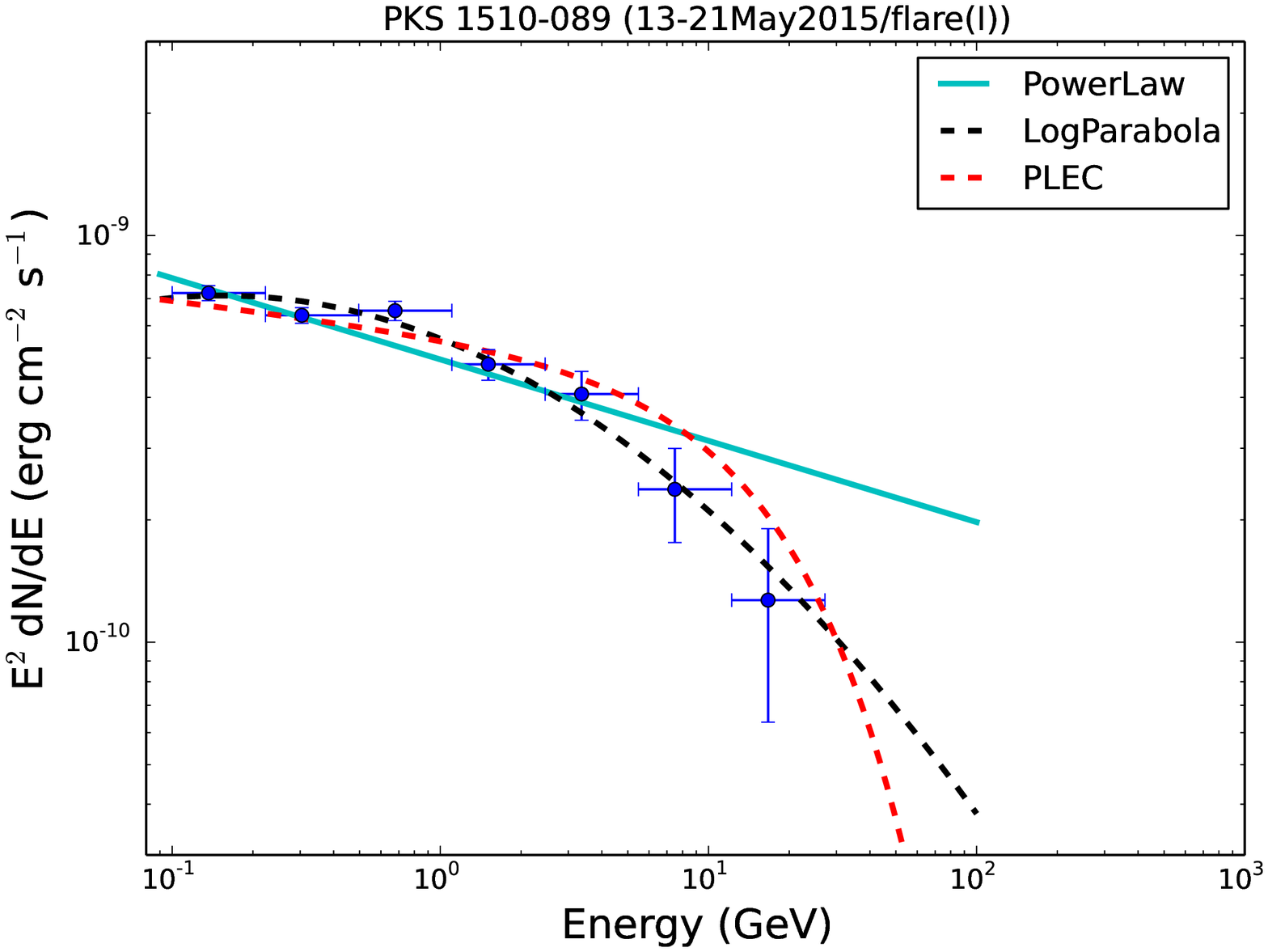}
\end{center}
\begin{center}
\includegraphics[scale=0.35]{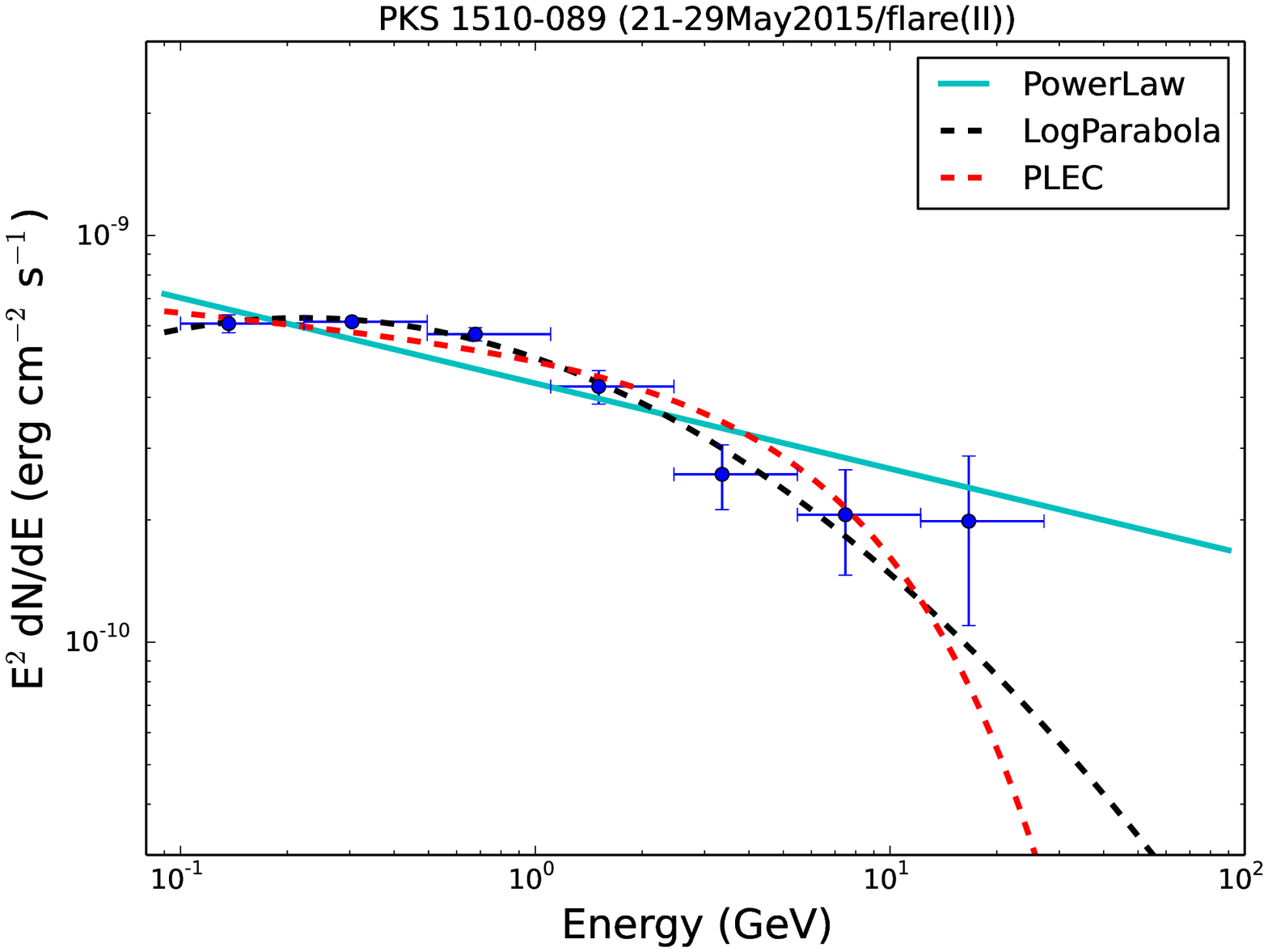}
\includegraphics[scale=0.35]{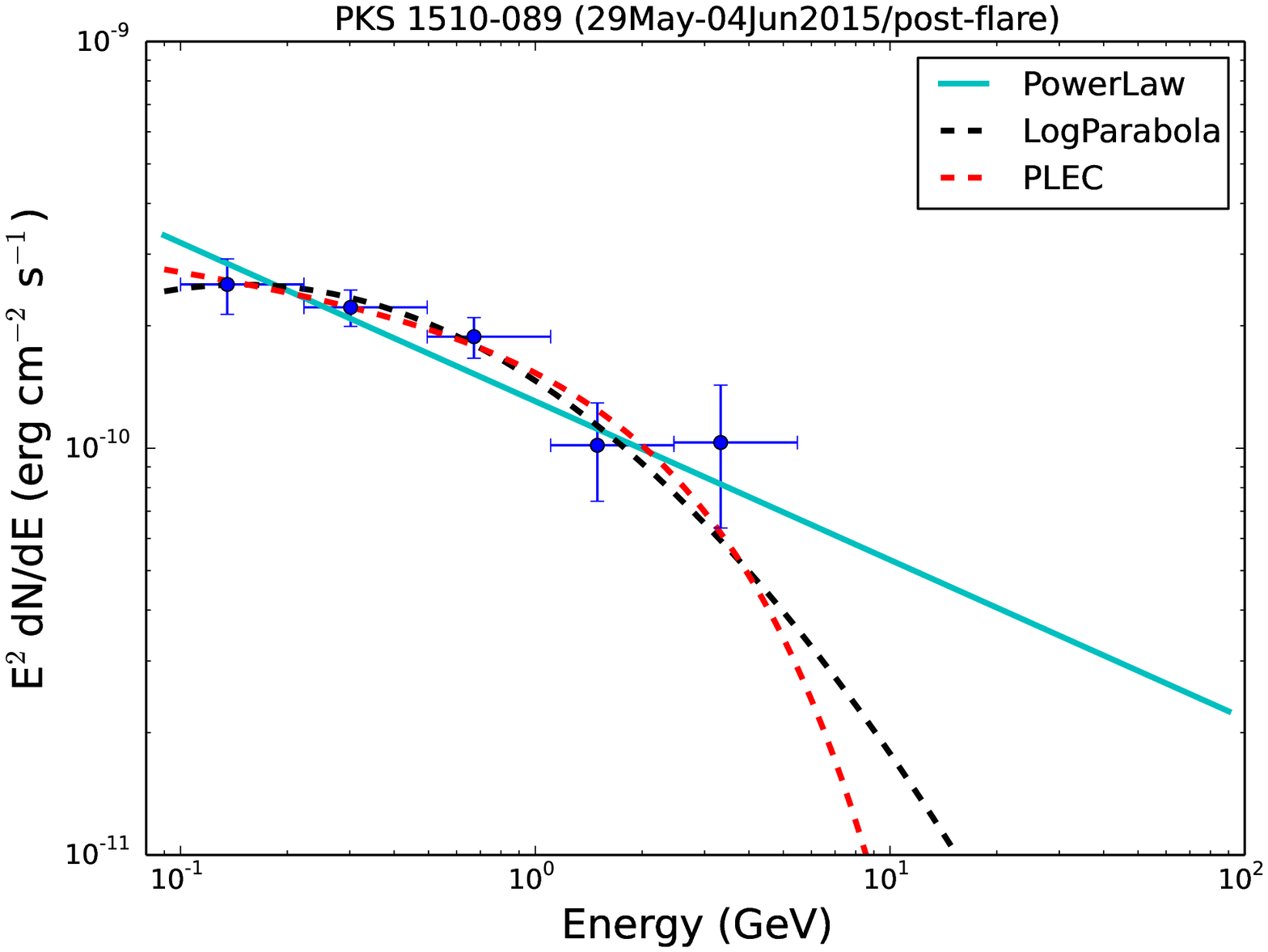}
\end{center}
\caption{Fermi-LAT SEDs during different activity states of flare-4(B) as defined in Fig.11 . PL, LP, PLEC models are 
shown in cyan, black and red color and there respective parameters are given in the Table-16. }
\label{fig:flare-4(b)}
\begin{center}
\includegraphics[scale=0.35]{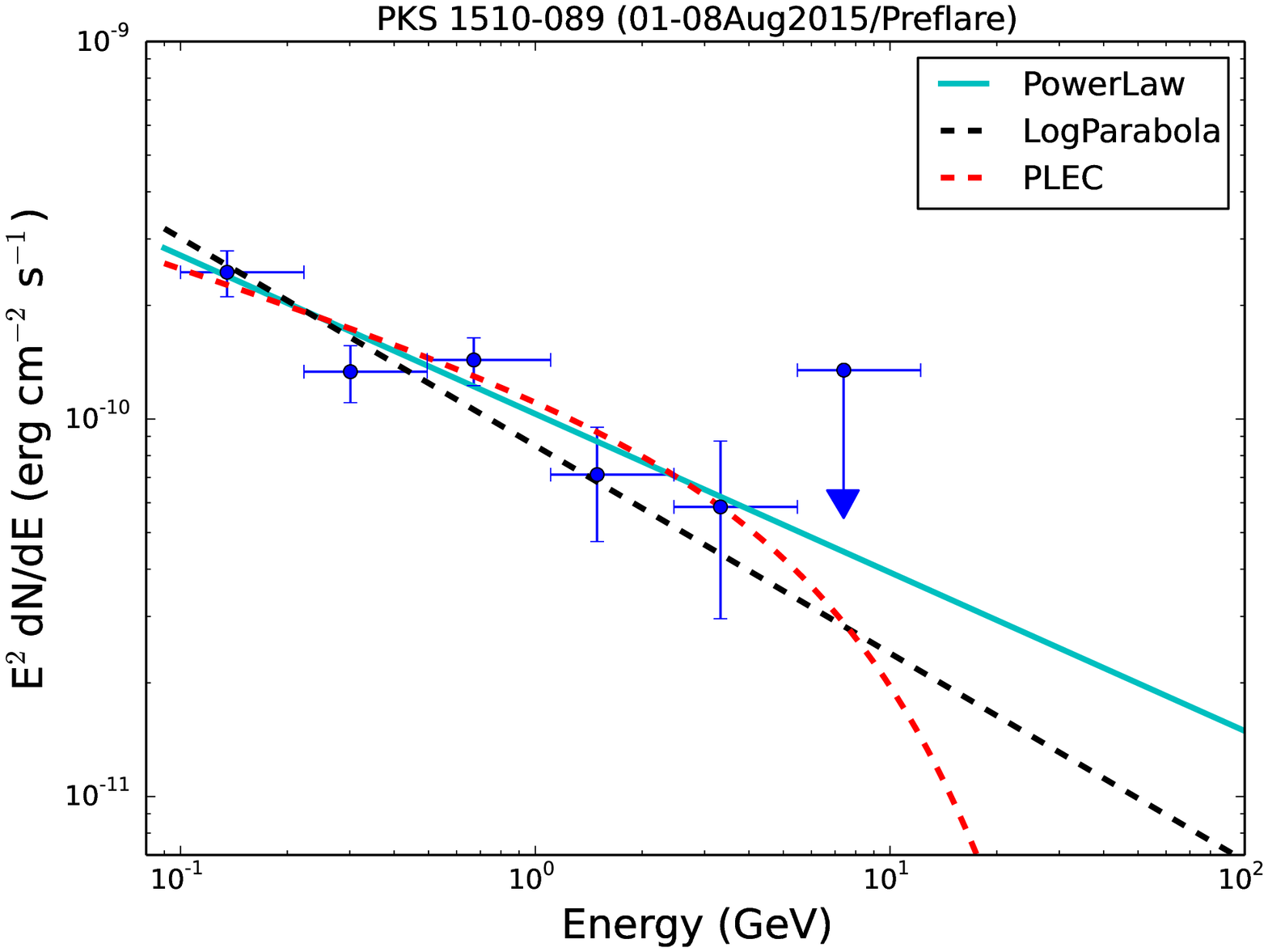}
\includegraphics[scale=0.35]{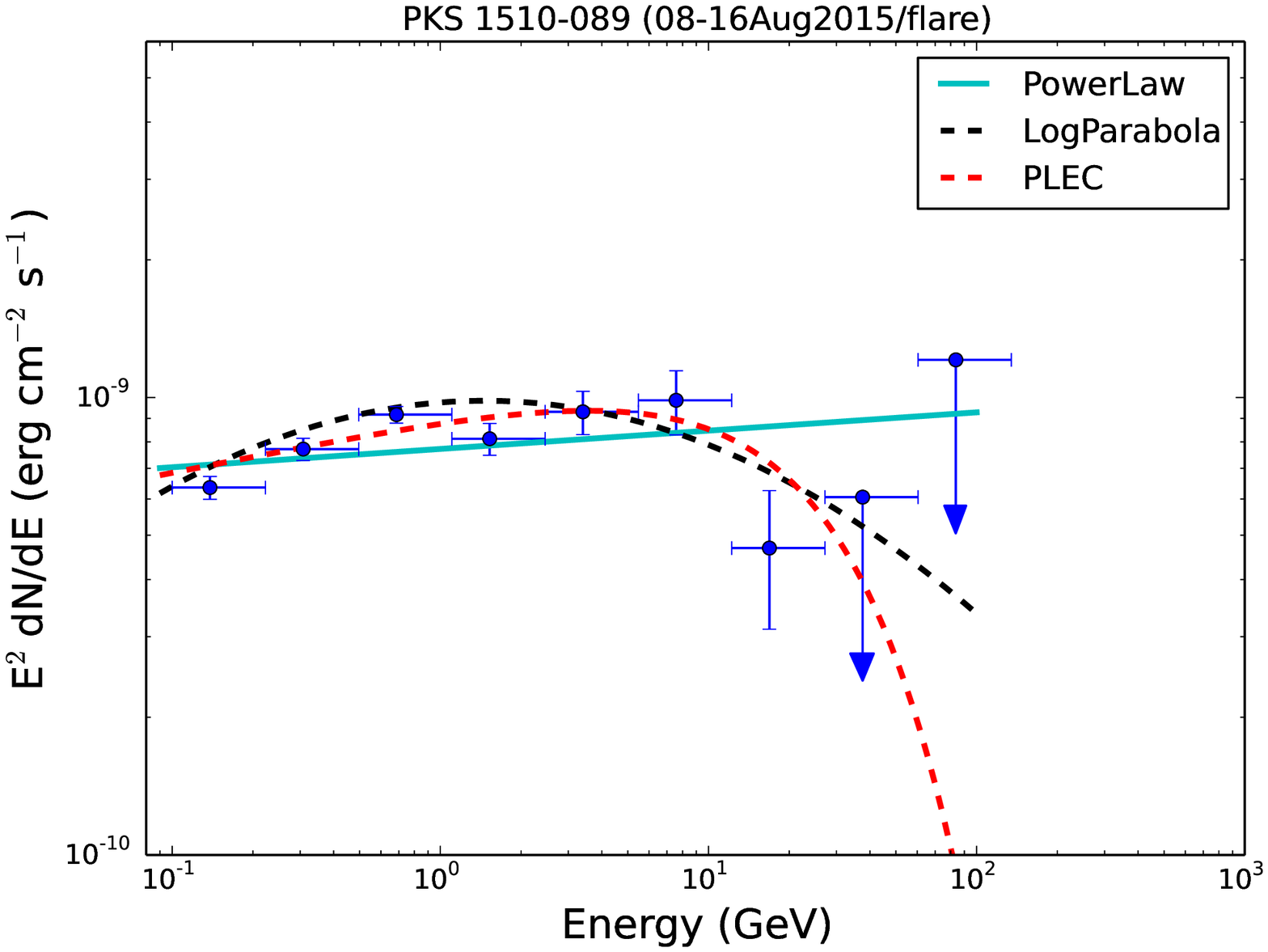}
\end{center}
\begin{center}
\includegraphics[scale=0.35]{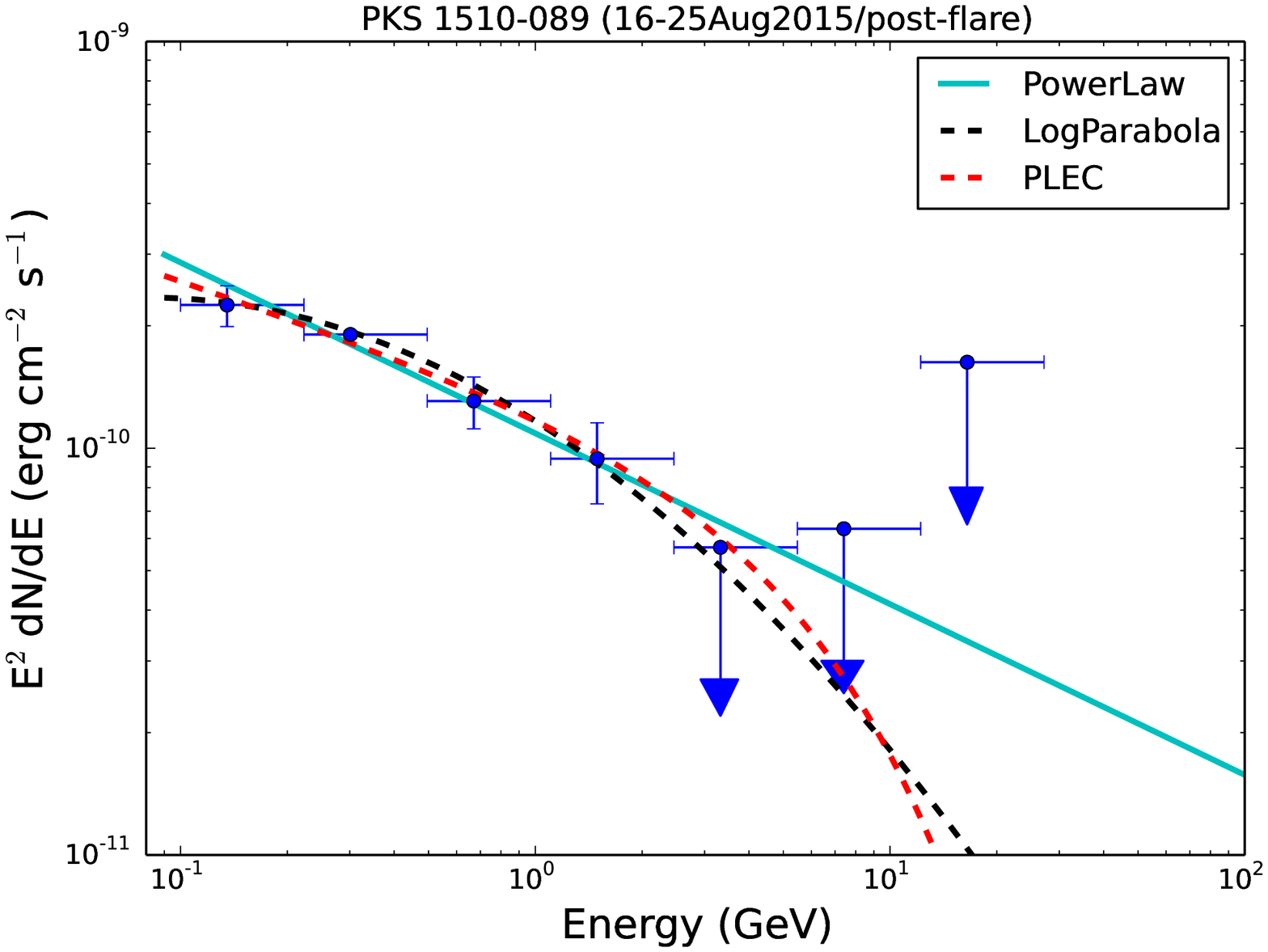}
\end{center}
\caption{Fermi-LAT SEDs during different activity states of flare-4(C) as defined in Fig.12 . PL, LP, PLEC models are
shown in cyan, black and red color and there respective parameters are given in the Table-17. }
\label{fig:flare-4(c)}
\end{figure*}

\begin{figure*}[htbp]
\begin{center}
\includegraphics[scale=0.35]{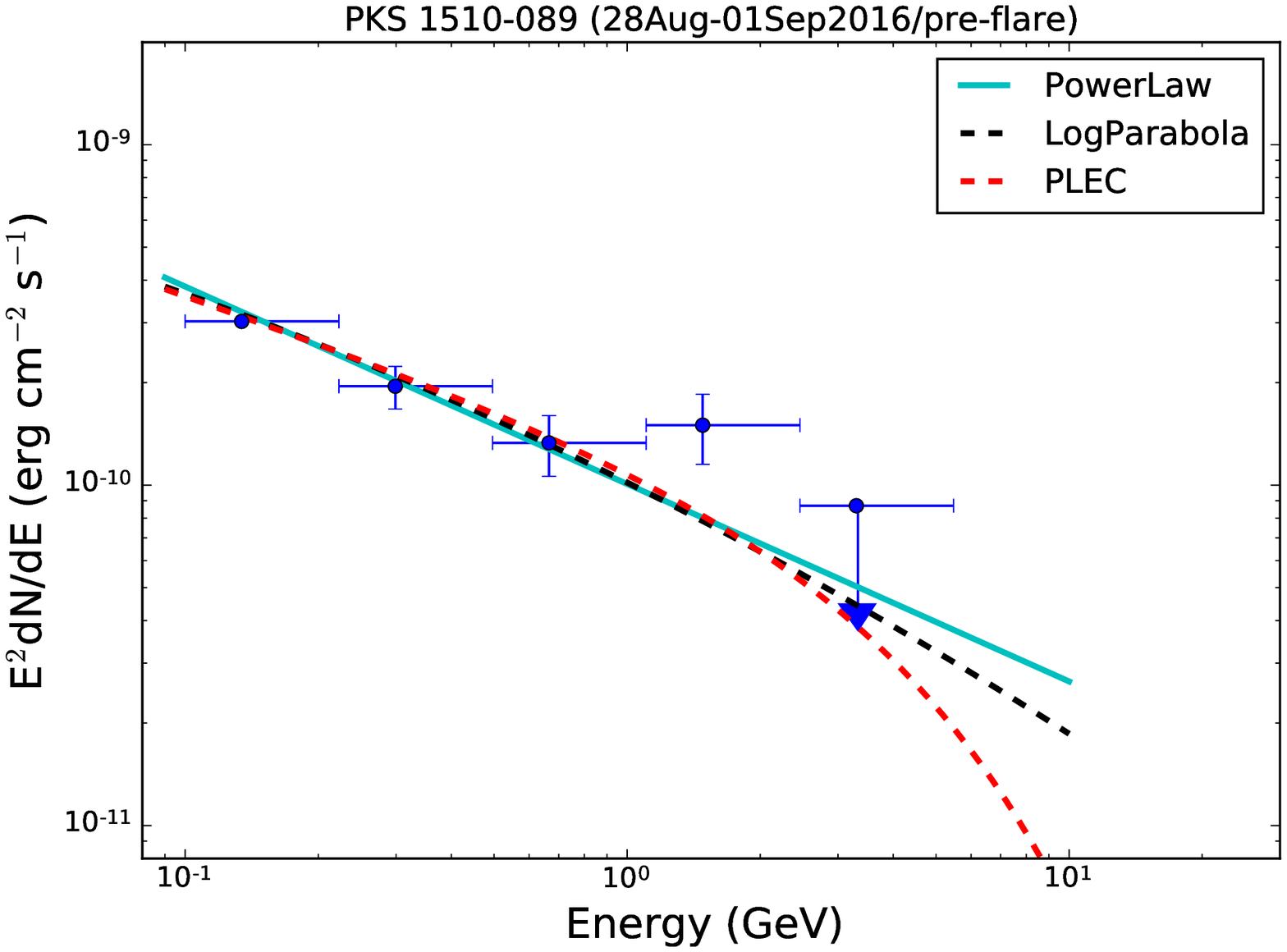}
\includegraphics[scale=0.35]{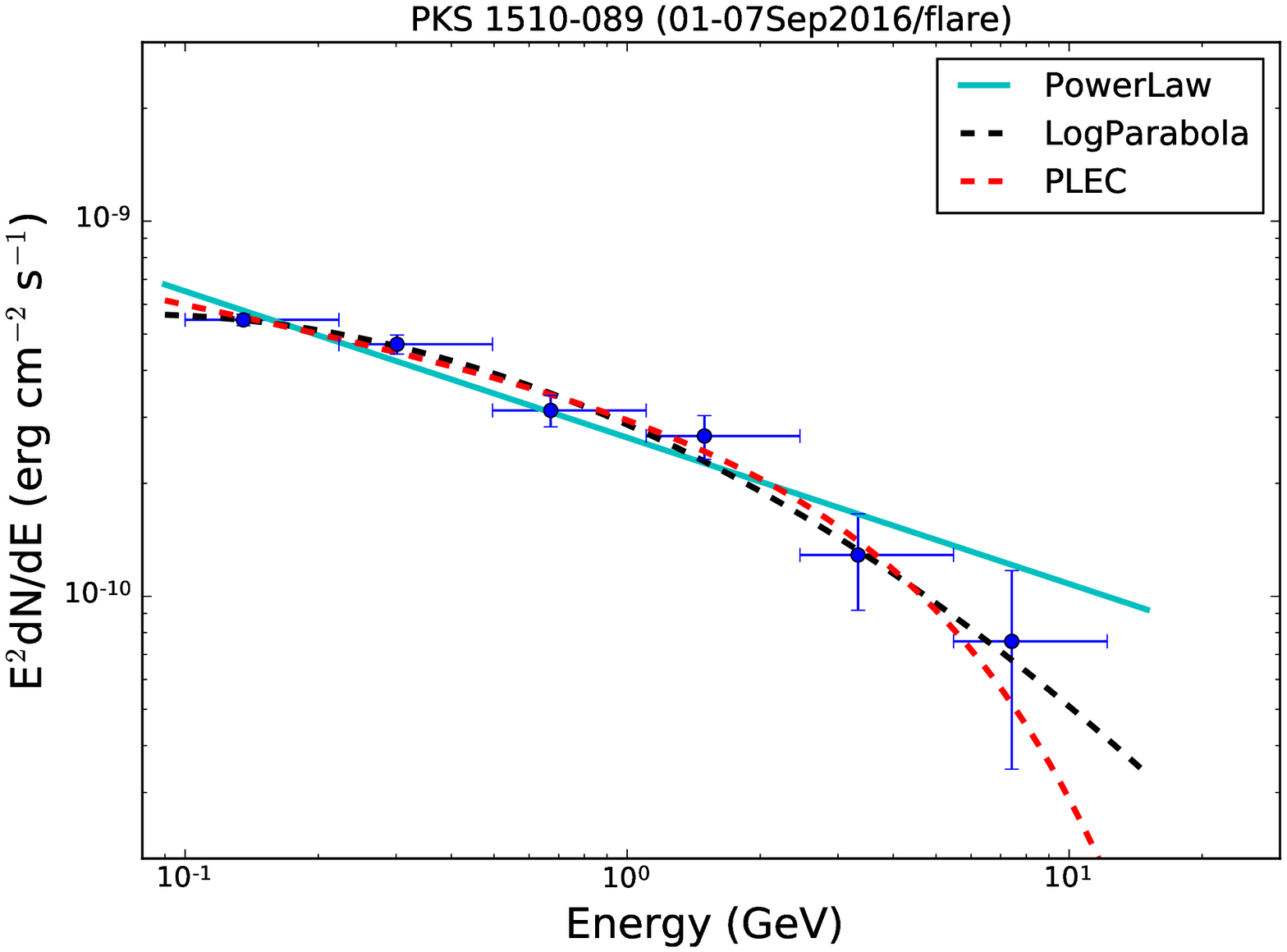}
\end{center}
\begin{center}
\includegraphics[scale=0.35]{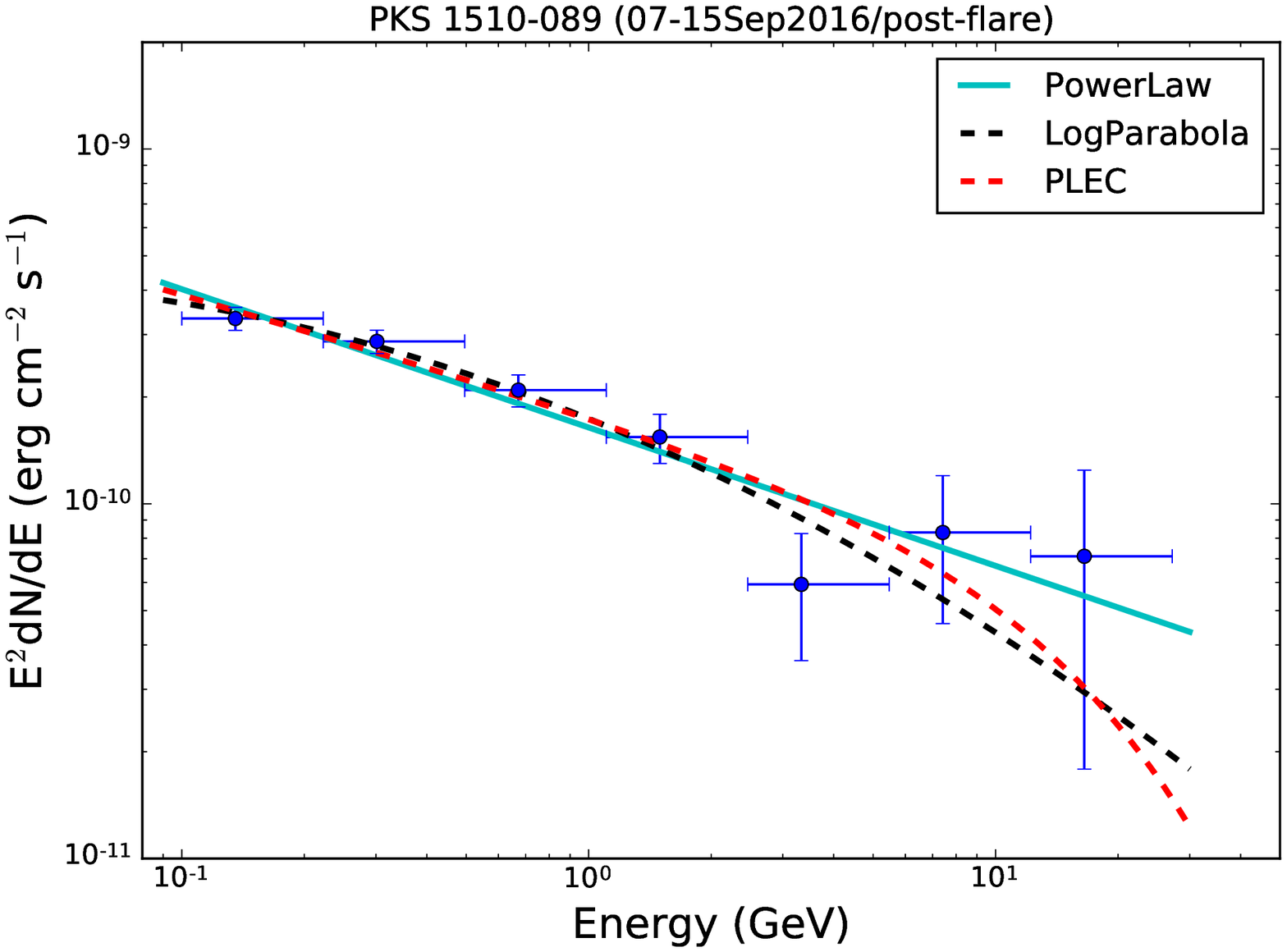}
\end{center}
\caption{Fermi-LAT SEDs during different activity states of Flare-5 as defined in Fig.13 . PL, LP, PLEC models are 
shown in cyan, black and red color and there respective parameters are given in the Table-18.}
\label{fig:flare-5}
\end{figure*}

\begin{figure*}[htbp]
\begin{center}
\includegraphics[scale=0.35]{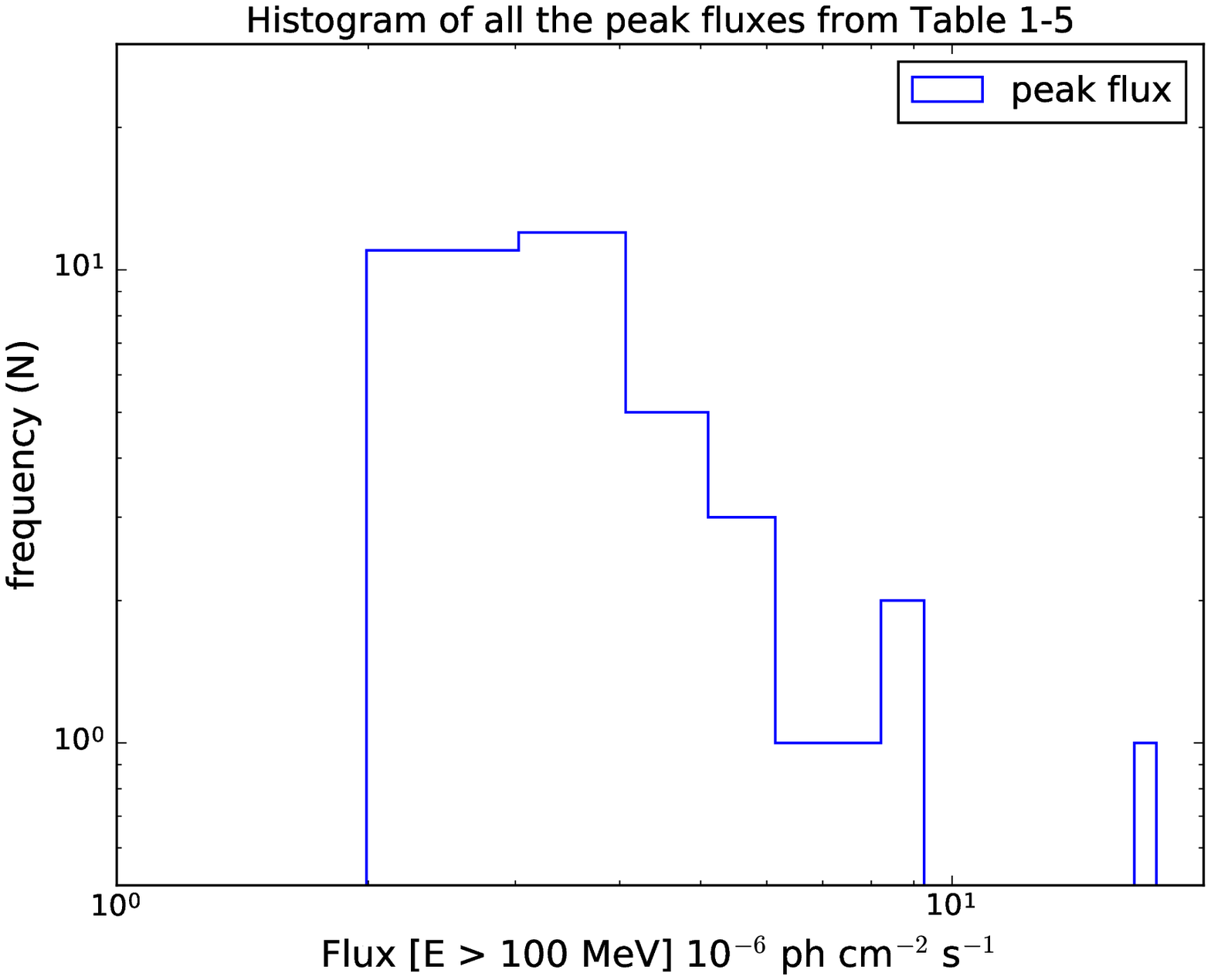}
\includegraphics[scale=0.35]{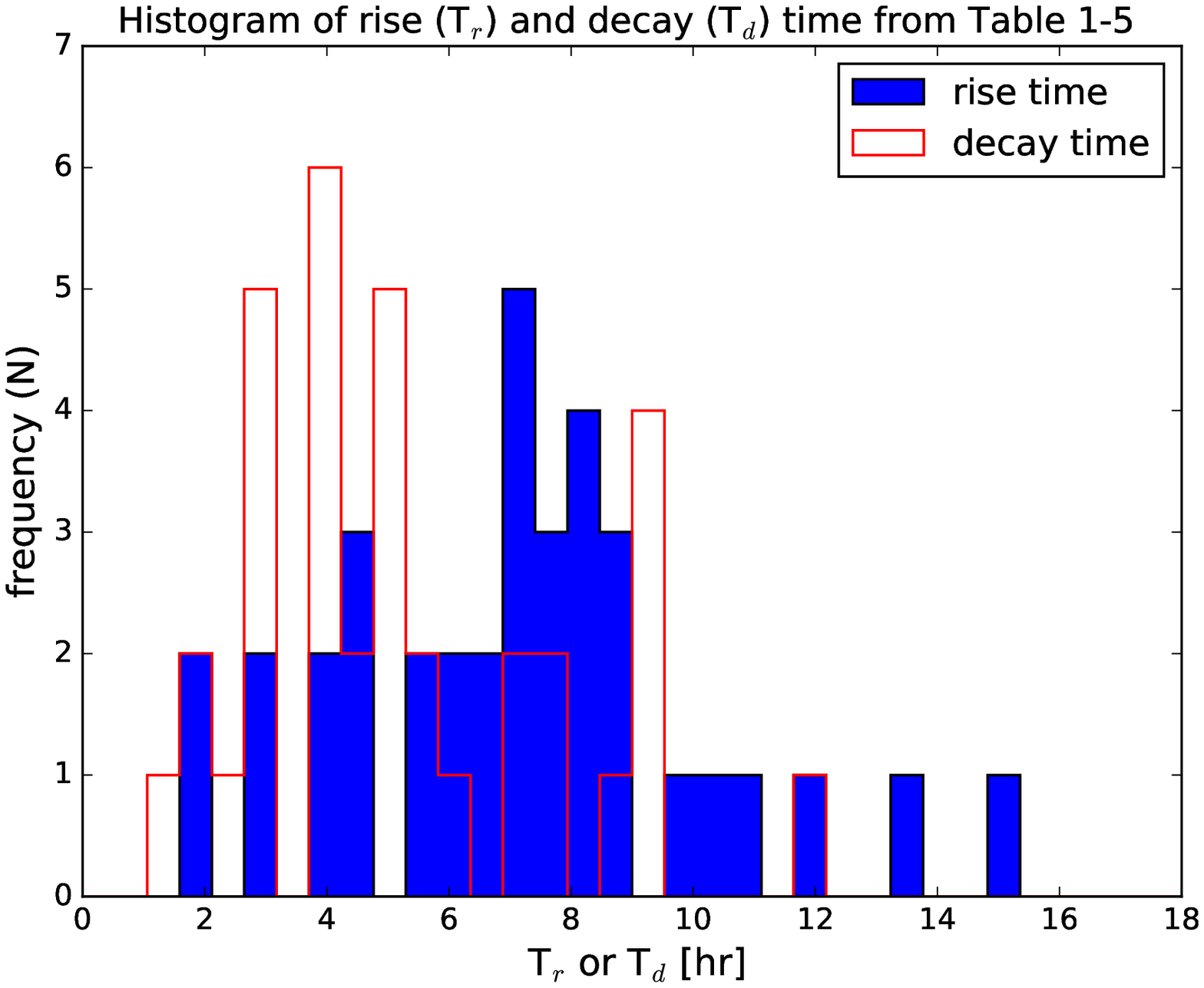}
\end{center}
\caption{Left panel: Histogram of peak fluxes from Tables 1-5. The mean flux is 3.54$\pm$0.08 and the standard deviation of the 
sample is 1.69. Right panel: Histogram of rise and decay time from Tables 1-5.
Their mean values are 6.04$\pm$0.22 hr and 3.88$\pm$0.16 hr respectively. The sample standard deviations are 2.40 hr and 2.20 hr.}

\begin{center}
\includegraphics[scale=0.35]{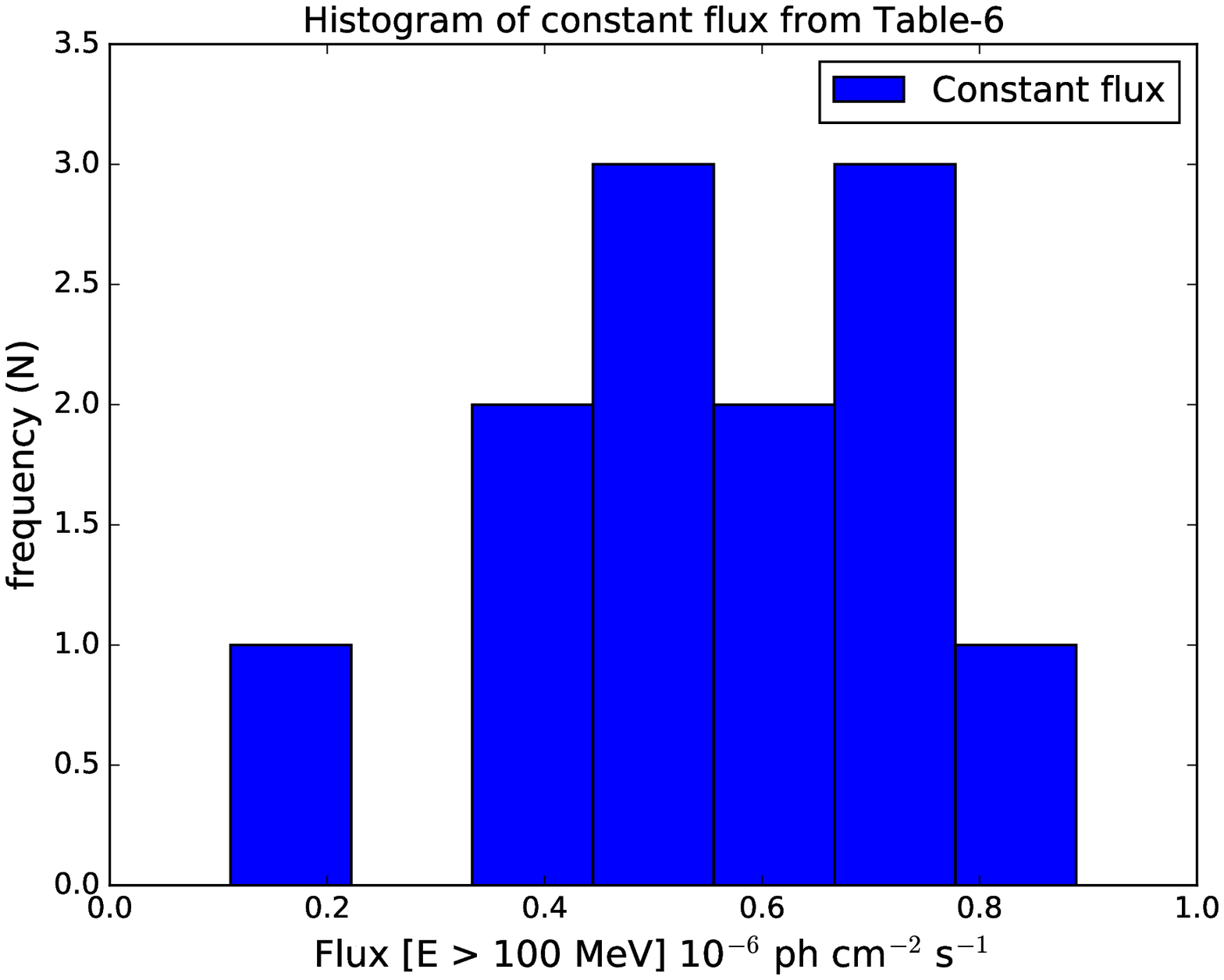}
\includegraphics[scale=0.35]{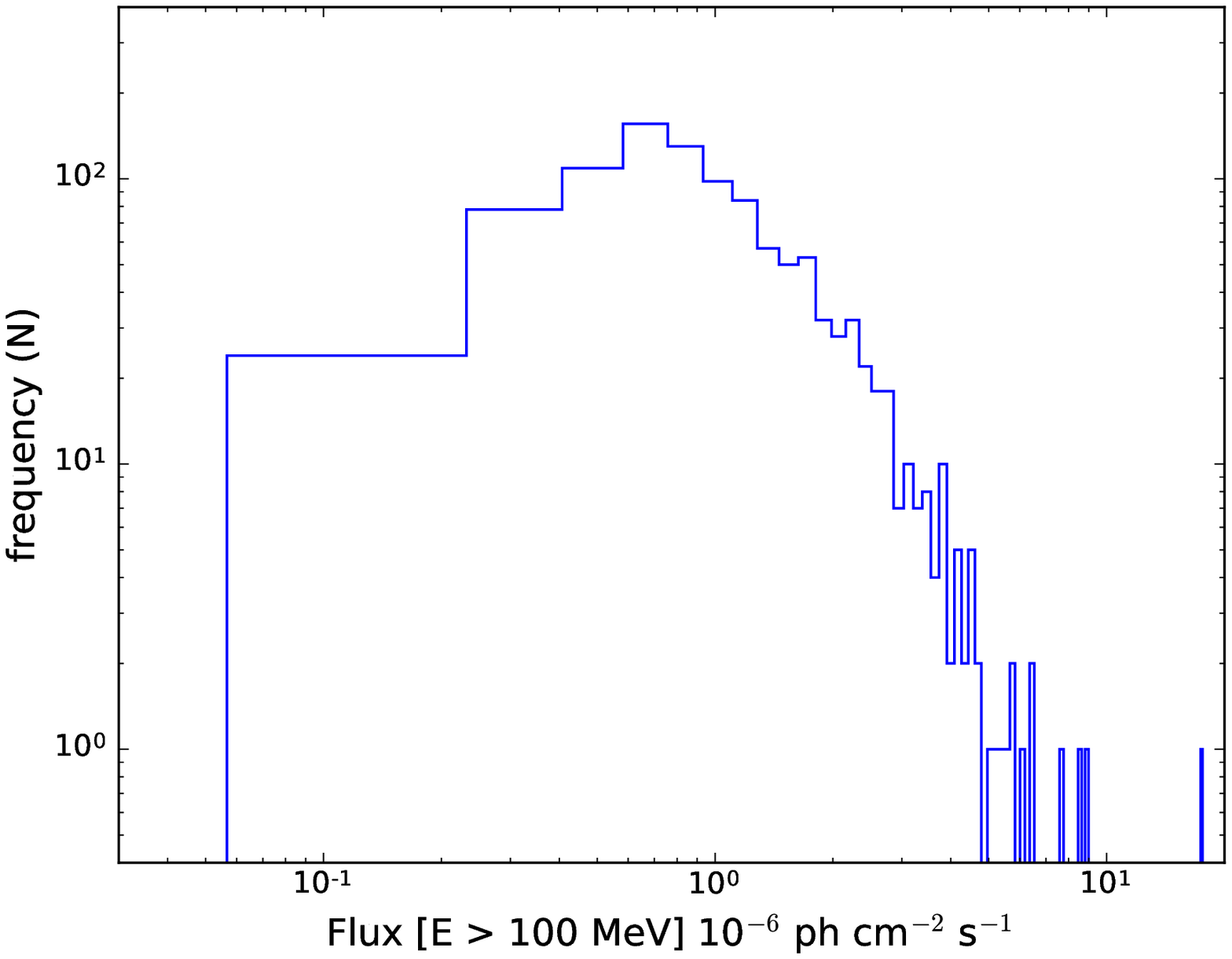}
\end{center}
\caption{Left panel: Histogram of constant flux from Table-6. The mean constant flux is found to be 0.51$\pm$0.01 and the
standard deviation is 0.20. Right panel: Histogram of all the flux data points.
The distribution is peaked, with slow rise up to peak and fast decay after that.}

\begin{center}
\includegraphics[scale=0.35]{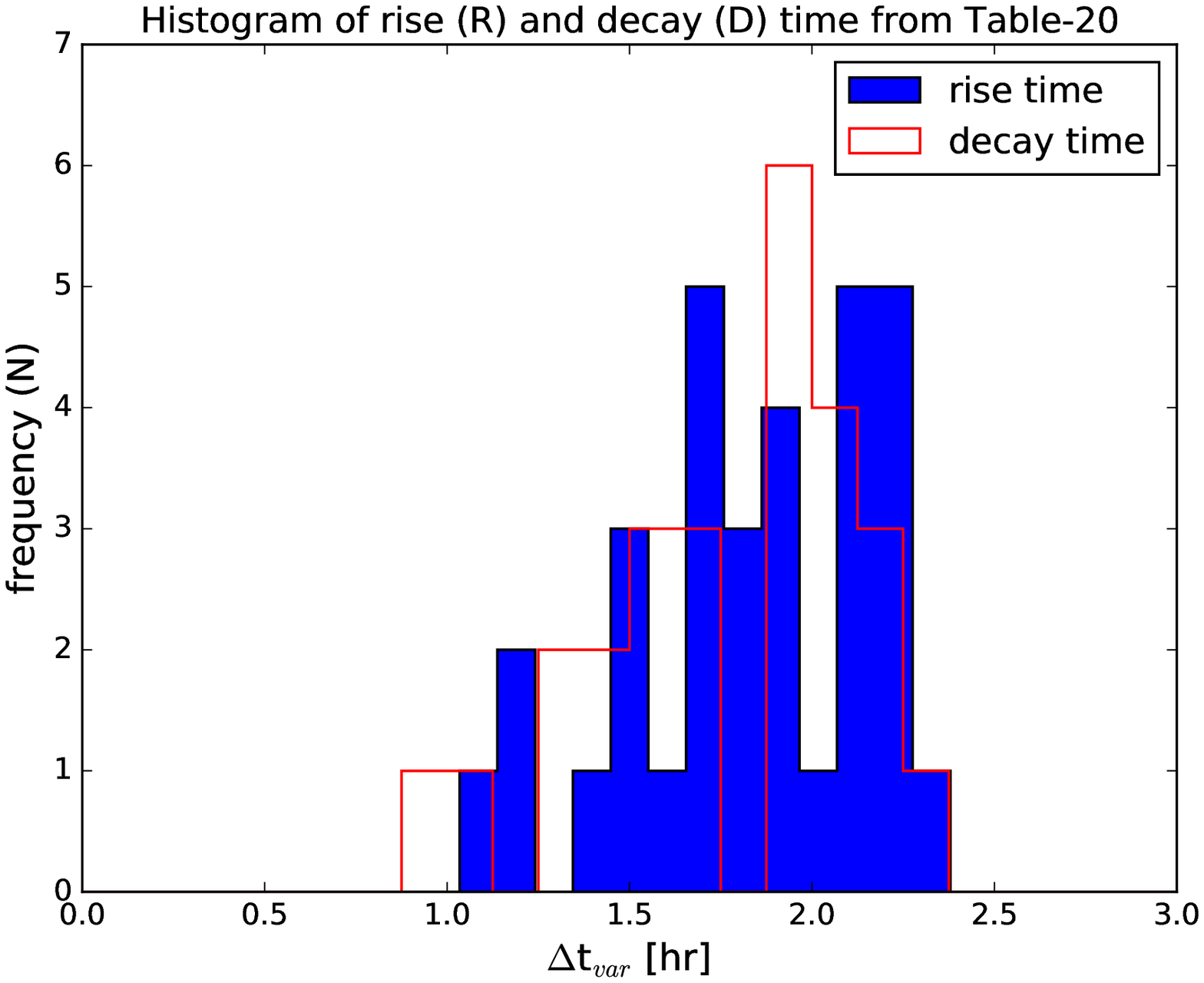} 
\end{center}
\caption{ Histogram of the rise and decay time from the fastest variability time, Table-20. They are distributed with mean of 
1.75$\pm$0.02 hr and 1.76$\pm$0.02 hr and standard deviation of 0.35 hr, 0.40 hr respectively.}
\end{figure*}

\bibliographystyle{plain}

\end{document}